\documentclass[manuscript]{aastex}
\shorttitle{Ices in the Large Magellanic Cloud}
\shortauthors{Oliveira et al.}

\usepackage{subfig}
\newcommand{\spitzer}{{\it Spitzer}}
\newcommand{\spec}{SAGE-Spec}

\begin{document}
\title{Ice chemistry in embedded young stellar objects in the Large Magellanic 
Cloud}

\author{J. M. Oliveira\altaffilmark{1}, 
J. Th. van Loon\altaffilmark{1},
C.-H. R. Chen\altaffilmark{2}, 
A. G. G. M. Tielens\altaffilmark{3},
G. C. Sloan\altaffilmark{4},
P. M. Woods\altaffilmark{5}, 
F. Kemper\altaffilmark{5}, 
R. Indebetouw\altaffilmark{2,6},
K. D. Gordon\altaffilmark{7},
M. L. Boyer\altaffilmark{7},
B. Shiao\altaffilmark{7},
S. Madden\altaffilmark{8},
A. K. Speck\altaffilmark{9},
M. Meixner\altaffilmark{7},
M. Marengo\altaffilmark{10}}

\altaffiltext{1}{School of Physical and Geographical Sciences, Lennard-Jones
Laboratories, Keele University, Staffordshire ST5 5BG, UK}

\altaffiltext{2}{Department of Astronomy, University of Virginia, P.O. Box 
400325, Charlottesville, VA 22904, USA}

\altaffiltext{3}{Leiden Observatory, P.O. Box 9513, NL-2300 RA Leiden, The 
Netherlands}

\altaffiltext{4}{Department of Astronomy, Cornell University, Ithaca, NY 14853, 
USA}

\altaffiltext{5}{Jodrell Bank Centre for Astrophysics, Alan Turing Building, 
School of Physics and Astronomy, University of Manchester, Oxford Road, 
Manchester M13 9PL, UK}

\altaffiltext{6}{National Radio Astronomy Observatory, 520 Edgemont Road,
Charlottesville, VA 22903, USA} 

\altaffiltext{7}{Space Telescope Science Institute, 3700 San Martin Drive, 
Baltimore, MD 21218, USA}

\altaffiltext{8}{Service d'Astrophysique, Commissariat \'{a} L'Energie Atomique 
de Saclay, 91191 Gif-sur-Yvette, France}

\altaffiltext{9}{Physics and Astronomy, University of Missouri, Columbia, MO 
65211, USA}

\altaffiltext{10}{Department of Physics and Astronomy, Iowa State University, 
Ames IA 50010, USA}

\begin{abstract} We present spectroscopic observations of a sample of 15 
embedded young stellar objects (YSOs) in the Large Magellanic Cloud (LMC). These 
observations were obtained with the {\it Spitzer Infrared Spectrograph} (IRS) as
part of the \spec\ Legacy program. We analyze the two prominent ice bands in the
IRS spectral range: the bending mode of CO$_2$ ice at 15.2\,$\mu$m and the ice 
band between 5 and 7\,$\mu$m that includes contributions from the bending mode of
water ice at 6\,$\mu$m amongst other ice species. The 5$-$7\,$\mu$m band is
difficult to identify in our LMC sample due to the conspicuous presence of PAH
emission superimposed onto the ice spectra. We identify water ice in the spectra
of two sources; the spectrum of one of those sources also exhibits the 6.8-$\mu$m
ice feature attributed in the literature to ammonium and methanol. We model the 
CO$_2$ band in detail, using the combination of laboratory ice profiles available
in the literature. We find that a significant fraction ($\ga$\,50\%) of CO$_2$ 
ice is locked in a water-rich component, consistent with what is observed for 
Galactic sources. The majority of the sources in the LMC also require a 
pure-CO$_2$ contribution to the ice profile, evidence of thermal processing. 
There is a suggestion that CO$_2$ production might be enhanced in the LMC, but 
the size of the available sample precludes firmer conclusions. We place our 
results in the context of the star formation environment in the LMC.

\end{abstract}


\section{Introduction}
The formation of stars in the high-redshift Universe occurred in a metal-poor 
environment. Much of what we know about the process of star formation is deduced
from observations in nearby, Galactic star-forming regions with essentially 
solar metallicity. However, many of the physical processes that affect star
formation and their observational diagnostics are expected to depend on 
metallicity, for instance cooling and gas-phase and grain surface chemistry. The
nearest templates for the detailed study of star formation under metal-poor 
conditions are the Magellanic Clouds (MCs). The Interstellar Media (ISM) of the 
Large and Small Magellanic Cloud (LMC and SMC) have significantly lower 
metallicities ($Z_{\rm LMC}\sim 0.4 $\,Z$_\odot$ and $Z_{\rm SMC}\sim 
0.2 $\,Z$_\odot$), than the solar neighborhood ISM ($Z_{\rm Gal}\sim $ 
Z$_\odot$). Only relatively recently are we able to study the details of the 
star formation process in the MCs \citep[for a review of recent results see][]
{oliveira09}. 

The star-formation process is a complex interplay of various chemo-physical 
processes. During the onset of gravitational collapse of a molecular cloud, 
sufficiently dense cores can only develop if the heat produced during the 
contraction can be dissipated. The most efficient cooling mechanisms are via 
radiation through fine-structure lines of carbon and oxygen (e.g., via the
emission lines of [C\,{\sc ii}] at 158\,$\mu$m and [O\,{\sc i}] at 63 and 
146\,$\mu$m), and rotational transitions in abundant molecules such as CO, 
O$_{2}$ and water \citep{goldsmith78}. These cooling agents all contain at least 
one heavy atom --- H$_{\rm 2}$ has no permanent dipole moment, and emission 
through its rotational transitions is very unlikely, making it an inefficient 
coolant. Besides also contributing to the thermal balance in the molecular cloud,
dust grains are crucial in driving cloud chemistry, as dust opacity shields 
molecules from radiation and grain surfaces enable chemical reactions to occur 
that would not happen in the gas phase. Surface chemistry is also thought to play
an important role in the formation of H$_{2}$ \citep[e.g.,][]{williams03} and 
H$_{2}$O \citep[e.g.,][]{hollenbach09}. Simple theoretical arguments suggest 
that, even over the range covered by the Milky Way and the MCs, cooling and 
chemistry rates and timescales are affected by metallicity \citep{banerji09}. Not
only is it crucial to understand the ISM chemistry at low metallicity, it is also
important to investigate the chemical evolution during the star formation 
process.

In cold molecular clouds, layers of ice form on the surface of dust grains. In
particular, when densities increase dramatically during the cold collapse phase,
most molecules accrete onto dust grain surfaces. In fact, as water, CO, and 
also CO$_{2}$ are completely frozen out of the gas phase in these environments, 
ices are substantial reservoirs of the available oxygen and carbon 
\citep[see review by][]{bergin07}. Even if CO seems to form exclusively in the
gas-phase, there is laboratory evidence that water ice is predominantly formed
by hydrogenation on icy grains \citep{ioppolo08}. Therefore, understanding ice 
chemistry is crucial to understanding the gas-phase chemistry and the physical 
conditions within molecular clouds. Water is by far the most abundant ice 
(typically $10^{-5}-10^{-4}$ with respect to H$_{2}$), followed by CO$_{2}$ and 
CO, with a combined abundance of 10$-$30\% with respect to water ice 
\citep{gibb04}. Other ice species have also been identified (methanol, ammonia, 
ammonium, methane and formaldehyde), albeit in much smaller concentrations 
\citep{vandishoeck04}. While water and CO can be observed from the ground, 
CO$_2$ can only be observed from space and its ubiquitous presence in a variety
of environments was one of the major discoveries of the {\it Infrared Space 
Observatory ({\it ISO})} \citep[see][ and references therein]{ehrenfreund03}.
 
It is still unknown whether metallicity affects ice chemistry. The most obvious 
effect of a lower metal content is a reduced availability of carbon and oxygen, 
the cornerstones of interstellar chemistry. Indeed the CO abundance in the 
diffuse ISM in the LMC is lower than that measured in Galactic environments 
\citep{fukui01}. In metal-poor clouds, it could be expected that ice abundances 
are generally depressed (i.e. all ices could be less abundant as a result of 
lower abundances of C and O), but it is possible that the abundance balance of 
ice species is also reshaped. The lower dust:gas ratio should make cloud cores 
more transparent to ambient UV light, possibly exposing ices to more radiative 
processing, and making complex species more abundant than in the Galaxy.

Interstellar ices have been detected in the envelopes of heavily embedded young 
stellar objects (YSOs) in the MCs. The first evidence of ices in the envelope of
a massive YSO in any extra-galactic environment was a serendipitous discovery by
\citet{vanloon05a}. The spectrum of IRAS\,05328$-$6827 in the LMC, obtained with
the Infrared Spectrometer \citep[IRS,][]{houck04} onboard the {\it Spitzer Space
Telescope} \citep[\spitzer,][]{werner04} and the Infrared Spectrometer And Array
Camera at the Very Large Telescope (ISAAC/VLT), shows clear absorption 
signatures of water ice at 3.1\,$\mu$m and CO$_{2}$ ice at 15.2\,$\mu$m. 
\citet{oliveira06} and \citet{shimonishi08} identified more embedded YSOs with 
ice signatures in the LMC. Three embedded YSOs have been recently identified in 
the SMC \citep{vanloon08}. There is a suggestion that the stronger UV radiation 
field \citep{welty06} and/or higher dust temperature \citep{aguirre03} could 
be responsible for a higher CO$_{2}$-to-water column density ratio observed 
towards a few YSOs in the LMC when compared to Galactic YSOs 
\citep{vanloon05a,shimonishi08}. These studies are however hampered by small 
sample size and limited spectral resolution and thus not yet conclusive. 

CO$_2$ ice is found to be abundant and ubiquitous in a variety of molecular
environments in the Galaxy, from quiescent clouds \citep{whittet07} to the 
envelopes of high- and low-luminosity YSOs \citep{gerakines99,pontoppidan08}. 
This molecule is generally believed to form exclusively via surface reactions, 
even though its formation mechanism is not fully understood 
\citep[for a recent discussion of this issue see][]{pontoppidan08}. 
The most prominent CO$_2$ solid state features are the 4.3-$\mu$m stretching 
and 15.2-$\mu$m bending modes \citep[e.g.,][]{ehrenfreund97}. Comparisons of
observed 15.2-$\mu$m profiles with laboratory spectra suggest that CO$_2$ ices 
are dominated by components that are either water- or CO-rich, rather than pure 
CO$_2$ \citep{gerakines99,pontoppidan08,whittet09}. For some YSOs, narrow 
substructure in the observed profiles provides evidence of thermal processing 
\citep[e.g.,][]{gerakines99,white09}. Therefore the CO$_2$ ice profile is a 
sensitive probe of its molecular environment.

Another ice band in the IRS range (5$-$7\,$\mu$m) is less well understood, with 
contributions not only from water ice at 6\,$\mu$m but also ammonia (NH$_3$), 
methanol (CH$_3$OH), formic acid (HCOOH) and formaldehyde (CH$_2$O) ices 
\citep{boogert08}. This superposition of multiple ice components is possibly 
responsible for the discrepancy between the column densities determined from the
3-$\mu$m (O$-$H stretching mode) and the 6-$\mu$m (O$-$H bending mode) water 
ice features, with the 6-$\mu$m column density systematically higher 
\citep{gibb04,boogert08}. There is an ice feature at 6.8\,$\mu$m attributed to 
ammonium (NH$_4^+$) \citep{boogert08}. This band also falls in a wavelength range
strongly affected by emission features generally attributed to polycyclic 
aromatic hydrocarbon (PAH) molecules. The libration mode of water ice is a very 
broad feature that peaks at 13\,$\mu$m and extends between 10$-$30\,$\mu$m; in 
practice this feature is very difficult to isolate both because of its width and
the fact that it is hard to disentangle from the more prominent silicate dust 
features.

In a series of papers we will investigate ice chemistry in the envelopes of 
massive embedded YSOs in the low metallicity environment of the MCs. The 
\spitzer\ Legacy survey of the LMC \citep[SAGE,][]{meixner06} identified large 
numbers of YSO candidates for the first time in any extra-galactic environment 
\citep{whitney08}. The follow-up Legacy program with the \spitzer-IRS includes a
variety of point and extended sources among which YSOs feature prominently 
\citep[\spec,][]{kemper09}. We selected the LMC sample described here from both 
the \spec\ program and the \spitzer\ archive, as those objects that have 
spectral energy distributions (SEDs) consistent with early embedded YSOs and
exhibiting in their spectrum the absorption feature associated with CO$_{2}$ 
ice, the most prominent ice band in the \spitzer-IRS spectral range. 

In this paper we present the first detailed analysis of CO$_2$ ice profiles for
massive YSOs in the LMC. The paper is organized as follows. First we introduce 
the sample and discuss the issues of continuum determination and constraining the
PAH spectrum. We then describe the observed CO$_2$ profiles in terms of 
laboratory components and measure column densities. We also investigate features 
in the 5$-$7\,$\mu$m range. Finally we interpret our results in terms of possible
effects of low metallicity.
 
\section{Observations and data reduction}

The \spitzer-IRS spectra of our target sample were obtained mostly as part of the
\spitzer\ \spec\ Legacy program (PID: 40159). For three targets, IRS spectra were
recovered from the archive (PIDs: 3591, 3505). \citet{kemper09} describe in 
detail the original target selection, observing strategy, and data reduction for 
the \spec\ program; we provide here only a brief summary. 

The IRS observations of the point sources in \spec\ were performed in staring
mode, using the Short-Low and Long-Low modules (SL and LL respectively). The 8- 
and 24-$\mu$m fluxes were used to set exposure times, aiming at signal-to-noise
ratios of $\sim$\,60 in SL and $\sim$\,30 in LL. Archival spectra mentioned 
previously were also obtained in staring mode with SL and LL. 

The spectra were reduced following standard reduction techniques. Flatfielded 
images from the standard \spitzer\ reduction pipeline were background subtracted 
and cleaned of ``rogue'' pixels and artifacts. Spectra were extracted 
individually for each data collection event (DCE) and co-added to produce one 
spectrum per nod position. The nods were combined to produce a single spectrum 
per order, rejecting ``spikes'' that appear in only one of the nod positions. 
Finally, the spectra of all the segments were combined including the two bonus 
orders, that are useful in correcting for discontinuities between the orders. 
Archival spectra of LMC objects were re-reduced by the \spec\ team to the same 
uniform standard as the \spec\ sources.

When comparing spectra obtained with different instruments, and of objects at 
different distances, it is important to consider potential differences in the 
physical scales that are sampled. The {\it Spitzer} IRS slits are 
$3.6^{\prime\prime}$ and $10.5^{\prime\prime}$ wide, respectively for the SL and
LL modes. In the cross-dispersion direction the typical 2-pixel scales sampled 
are $3.6^{\prime\prime}$ and $10.2^{\prime\prime}$, respectively.
At the distance of the LMC ($d\approx$\,50\,kpc), these approximately square 
regions on the sky correspond to 0.9\,$\times$\,0.9\,pc$^2$ and 
2.5\,$\times$\,2.5\,pc$^2$ at 6\,$-$\,7 and 15\,$\mu$m, respectively. 

By comparison, the {\it ISO} SWS apertures are 
$14^{\prime\prime}\times20^{\prime\prime}$ around 3 and 6\,$-$\,7\,$\mu$m, and 
$14^{\prime\prime}\times27^{\prime\prime}$ around 15\,$\mu$m. Observations 
obtained with these modules of massive-star forming regions in the Galaxy at
typical distances of $\sim$\,1.5\,kpc sample physical scales of 
$0.10\times0.15$\,pc$^2$ and $0.10\times0.20$ pc$^2$ at 6$-$\,7 and 15\,$\mu$m, 
respectively. Hence these SWS observations would sample scales roughly ten times 
smaller in typical Galactic regions than the {\it Spitzer} observations in the 
LMC. Observations obtained with {\it Spitzer} of low-mass YSOs in nearby 
complexes at typical distances $d\approx$\,150$-$\,350\,pc sample even finer 
scales, by a factor $\sim$\,200 compared to observations in the LMC.

\section{The massive YSO sample}

This study focuses on the least evolved, most embedded massive YSOs in the LMC.
Embedded YSOs can be selected based on their mid-IR colors. However, the colors 
and even broad-band SEDs of YSOs are in some instances similar to those of 
evolved objects like planetary nebulae and post-AGB stars. \citet{whitney08}
discuss in detail the YSO selection criteria using \spitzer\ data and 
contamination by evolved stars. One of the main goals of the \spec\ program is 
to use spectra to clarify the boundaries between different classes of objects in 
the color-color and color-magnitude space. For this study we have selected only 
objects whose nature is spectroscopically unambiguous, which also means that 
these objects are the most embedded and luminous.

We initially identified YSO candidates based on their very red continuum over the
5$-$38\,$\mu$m range. Then we narrowed the sample to include only those objects 
that showed an indication of the CO$_2$ ice feature at 15.2\,$\mu$m. CO$_2$ in 
the gas phase has been detected towards many AGB stars \citep[e.g.,][]
{justtanont98}; CO$_2$ ice however is only observed in YSO envelopes and 
quiescent molecular clouds (Section\,4.3) --- water ice is also observed in the 
envelopes of evolved stars \citep[e.g.,][]{dijkstra03}. Therefore the shape of 
the SED and the CO$_2$ ice feature unambiguously identify these objects as 
embedded YSOs. We did not make use of the 5$-$7\,$\mu$m ice complex for sample 
selection, because this wavelength range is dominated by PAH emission in our YSO 
sample (see below).

The final sample then includes 15 objects, 10 of which were classified as
``high-probability YSOs'' based only on their IRAC colors \citep{whitney08}. 
Tables\,\ref{targets1} and \ref{targets2} provide the positions and observed 
fluxes for the YSOs. Even if some objects are known IRAS or MSX sources, many 
objects had not been classified or studied prior to the \spitzer\ surveys. Van 
Loon et al. (2005a) analyzed the IRS and ground-based L-band spectra of 
IRAS\,05328$-$6827, which show ice signatures; we re-analyzed here its IRS 
spectrum. \citet{shimonishi08} obtained low-resolution {\it AKARI} near-IR 
spectra \citep[3$-$5\,$\mu$m,][]{murakami07} of three of the sources, which also 
showed ice features. Van Loon et al. (2005b) tentatively identified 
IRAS\,05246$-$7137 as an AGB or post-AGB object. More recently, \citet{whitney08}
classified this source as a YSO based on its IRAC colors. The object's IRS 
spectrum (showing strong silicate absorption, weak PAH emission, and a tentative 
ice detection) and IR luminosity support the YSO classification. MSX\,LMC\,464 
was tentatively identified as a possible H\,{\sc ii} region \citep{kastner08} 
based on 2MASS and MSX colors. Van Loon et al. (2009) describe the \spitzer\ 
MIPS-SED spectra \citep{rieke04} of IRAS\,05328$-$6827, MSX\,LMC\,1786, and 
IRAS\,05246$-$7137; they all show a cold dust continuum between 53$-$92\,$\mu$m 
consistent with a YSO status.

\section{Analysis of the IRS spectra and ice absorption features}

Figure\,\ref{spectra} shows the spectra of the 15 objects analyzed in this work,
covering the whole IRS range. The insets show the spectral regions associated
with the ice bands, CO$_2$ at 15.2\,$\mu$m and the 5$-$7\,$\mu$m complex. Objects
are displayed in decreasing flux intensity at 5.8\,$\mu$m. Many objects show
broad absorption at 10\,$\mu$m (and occasionally 18\,$\mu$m) attributed to
silicate dust grains; one object (IRAS\,05240$-$6809) shows silicate emission. 
PAH emission is also present in many objects in our sample: the prominent 6.2- 
and 11.3-$\mu$m PAH features are easily identified in at least two thirds of the 
spectra. Figure\,\ref{spectra} also shows that there are a variety of strength 
and profile shapes associated with the CO$_2$ ice and that few objects exhibit 
clear ice absorption at 6\,$\mu$m. In the next sections we will analyze these 
features in more detail, starting with the issue of continuum determination.

\subsection{YSO properties: SED fitting}

We have fitted the observed SEDs of the LMC YSOs using the on-line fitting 
tool\footnotemark\ developed  by \citet{robitaille07}. \footnotetext{SED fitter 
available at http://caravan.astro.wisc.edu/protostars/index.php.} This tool 
relies on a database of 20,000 pre-calculated models that are compared to the 
observed SEDs, and the best fit is selected using a $\chi^2$ minimization 
technique. \citet{robitaille06,robitaille07} explain these tools in more detail, 
and \citet{chen09} give examples of their use. A typical YSO model includes a 
central stellar photosphere, a flared accretion disk, and a rotationally 
flattened infalling envelope with bipolar cavities, characterized by a total of 
14 parameters. The balance of the different model components determines the 
object's evolutionary status. 

To constrain the observed SEDs we made use of the IRS spectrum discussed here
as well as all the photometric data available for each object 
(Tables\,\ref{targets1} \& \ref{targets2}). The SEDs of 10 objects are described 
by \citet{whitney08}; for the remaining objects the results of the SED fits will 
be discussed in detail by Chen et al. (in preparation).  The objects in the 
sample have estimated YSO luminosities in the range 
$\sim$\,5\,$-$\,50\,$\times$\,10$^3$\,L$_{\odot}$ and masses between $\sim$\,10 
and 25\,M$_{\odot}$. Taking into account their observed luminosity, the objects 
in our sample are the LMC counterparts of the luminous Galactic YSOs such as 
those studied by \citet{gerakines99} and \citet{gibb04} with {\it ISO}. In terms 
of evolutionary stages the YSOs are all in Stage I, i.e they are still deeply 
embedded in their envelopes, as expected from the presence of ice absorption 
features. The Stage I classification is equivalent to the Class\,I as defined by
\citet{lada87}, but instead of relying on the spectral index of the SED, it is 
based on modelled SED quantities, in this case the accretion rate of the envelope
normalized to the stellar mass,
{\it \.M}$_{\rm env}$/$M_{\star}$\,$>$\,10$^{-6}$\,yr$^{-1}$ 
\citep{robitaille06}.

\subsection{Dust continuum, PAH emission and silicate features}

The mid-IR spectra of embedded YSOs exhibit a wealth of solid-state and gas 
features super-imposed on a dust continuum: silicate absorption/emission 
features at 10 and 18\,$\mu$m, PAH emission features from 5 to 19\,$\mu$m,
molecular hydrogen and forbidden line emission across 5\,$-$\,35\,$\mu$m
and finally the ice signatures we are interested in here. In our sample, 11 out
of 15 objects show prominent PAH emission (look for the conspicuous 11.3-$\mu$m
feature in Figure\,\ref{spectra}). As we want to measure ice optical depth, we 
need both to determine the continuum level and account for any emission or 
absorption features that fall in the desired spectral ranges. We fitted continua
between 5 and 20\,$\mu$m, focusing on obtaining good continuum determinations 
around 5\,$-$\,7 and 15\,$\mu$m.

Determining a global IR continuum for YSOs is not trivial. This is because the 
lines and bands listed above merge together, meaning that few true continuum
points exist across the spectrum. As a consequence continuum fits rarely offer 
unique solutions. The CO$_2$ ice band at 15.2\,$\mu$m is relatively narrow and 
falls in the region between the 11.3$-$12.7\,$\mu$m and 17-$\mu$m PAH complexes 
and between the silicate absorption bands. Therefore, it is possible and 
practical to use a local continuum determination, for instance a low-order 
polynomial fitted to the local pseudo-continuum in the neighborhood of the ice 
feature. 

This is not the case for the 5$-$7\,$\mu$m ice band. First, there is prominent
PAH emission in this region, namely the 6.3-$\mu$m feature and the 7.7-$\mu$m 
complex, as well as unresolved H$_{2}$ emission at 5.51 (S7), 6.11 (S6), and 
6.91\,$\mu$m (S5). \citet{spoon02} show that relatively weak water ice signatures
at 6\,$\mu$m are difficult to separate from the PAH emission contribution. 
Secondly, this range falls at the edge of the IRS spectrum, where the spectra are
noisier. This means that it is very difficult if not impossible to find bona-fide
continuum points to anchor a local pseudo-continuum determination. 

PAHFIT \citep{smith07} provides a convenient way of constraining the dust
continuum and extinction as well as accounting for PAH and gas emission. The
dust continuum is modelled with up to eight thermal dust components represented 
by modified blackbodies of a range of temperatures; dust extinction is produced 
by adopting extinction profiles similar to those of Milky Way dust. PAH emission 
features are represented by individual and blended Drude profiles --- depending 
on whether PAH complexes are resolved into their multiple components --- with 
fixed widths and central wavelengths. Molecular and atomic emission lines 
(e.g., H$_{2}$ pure rotational lines, forbidden line emission from 
[S {\sc iii}], [Ne {\sc ii}], and [Ne {\sc iii}] etc) are modelled as Gaussian 
profiles; their widths and central wavelengths are allowed to change to account
for the fact that these lines are not resolved in the IRS low-resolution 
spectra. To help PAHFIT constrain the continuum we added the IRAC 3.6- and 
4.5-$\mu$m photometric points to the IRS range. The very broad water libration 
mode at 13\,$\mu$m is treated as a component of the continuum in this approach. 
Once all the PAHFIT components are accounted for, we are left only with the ice 
contributions. 

Figure\,\ref{PAHFIT} shows fits to the 11 YSOs in the sample with PAH emission, 
using PAHFIT. Total fits are shown, as well as color-coded individual 
contributions of dust continuum, PAH emission and molecular and atomic line 
emission. For each object we show their spectrum and the continuum-subtracted
spectrum to highlight the PAH contribution. We found that PAHFIT fits the spectra
well from 5$-$17\,$\mu$m. PAH emission is unequivocal in Figure\,\ref{PAHFIT}. 
In terms of general appearance, the continuum-subtracted spectra of objects like 
MSX\,LMC\,464, SAGE1C\,J050354.56$-$671848.9, and IRAS\,05452$-$6924 are 
reminiscent of the spectral components between 5 and 14\,$\mu$m attributed to 
neutral PAHs, PAH clusters, and PAH cations \citep{tielens08}.

At longer wavelengths, we find that PAHFIT has some trouble fitting the red side
of the 17-$\mu$m PAH complex that also includes H$_2$ emission at 17.0\,$\mu$m
(examples are IRAS\,04514$-$6931 and IRAS\,F04532$-$6709 in 
Figure\,\ref{PAHFIT}). In particular at about 18\,$\mu$m there seems to be more 
emission than can be produced by PAHFIT: there are no known PAH emission features
between 17.9 and 18.9\,$\mu$m. It is also not due to spurious ``features'' 
related to diffuse gas in the LMC (i.e. [S {\sc iii}] at 18.7\,$\mu$m, as 
mentioned by \citealt{buchanan06}). The 17-$\mu$m complex is not as well studied 
as other, shorter-wavelength PAH features. Even though identified with 
{\it ISO}-SWS, it was only \spitzer-IRS that allowed a more detailed analysis of 
this complex \citep[see][ for a review]{tielens08}. It is described as a broad 
plateau that can extend from 16 to 19\,$\mu$m, but observations show a varied 
substructure. \citet{kerckhoven00} and \citet{peeters04} have used large molecule
databases to show that a huge range of merged profile structures can  be achieved
by combining individual PAH emission contributions. Thus it is not surprising 
that the fits are not perfect in this region. None of this affects the continuum 
underlying the ice bands. 

We note that a detailed analysis of PAH emission is beyond the scope of this 
paper, but we verified that PAHFIT is an adequate tool to use in the context of 
a YSO study. While the spectral contributions underlying the CO$_2$ ice feature 
are well constrained, that is not the case for the 5$-$7\,$\mu$m ice complex. 
Even using PAHFIT to account for emission features, it is very difficult to 
isolate this ice band: the only objects for which this was possible have no 
detectable PAH emission (see below).

\begin{figure}
\begin{center}
\includegraphics[scale=0.92]{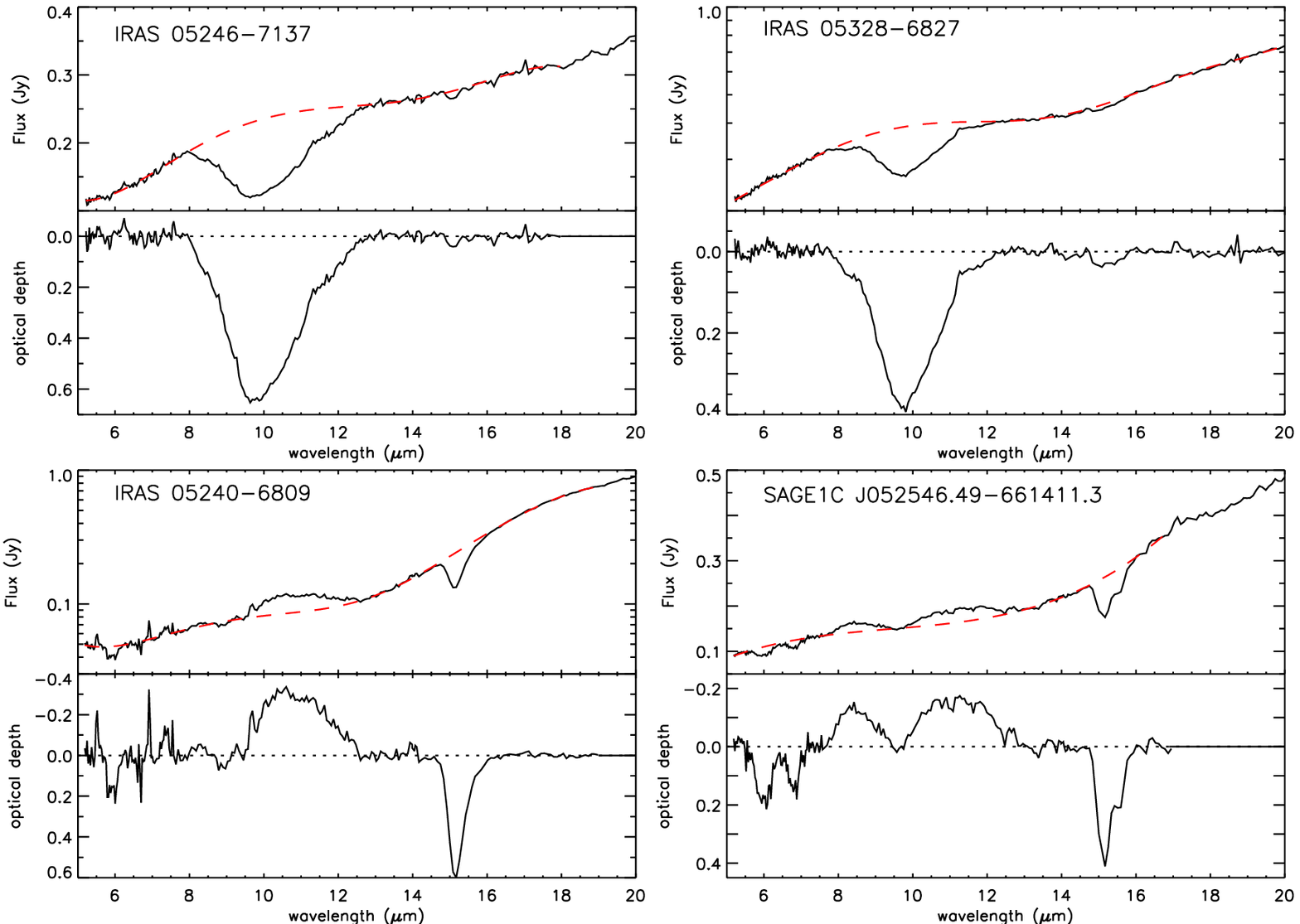}
\caption{Continuum fits to YSO spectra without PAH emission. For each object the
total continuum fit is shown (red dashed line, low-degree polynomial) as well as
the optical depth spectra. The two objects at the top have strong silicate 
absorption and weak CO$_2$ ice; the two objects at the bottom exhibit either a 
self-absorbed or emission silicate feature and strong CO$_2$ ice.
\label{polyfit}}
\end{center}
\end{figure}

The four remaining objects do not show obvious PAH emission in their spectrum; we
thus adopt the traditional approach of fitting a low-order polynomial to the 
continuum. Continuum points are selected to avoid the ice bands at 5\,$-$\,7 
and 15.2\,$\mu$m and the silicate feature at 10\,$\mu$m. The 18-$\mu$m silicate
feature and the water libration mode are very broad, and we do not try to extract
them from the spectra; instead we treat them as contributing to the local 
``pseudo-continuum''. Figure\,\ref{polyfit} shows the continuum fits for these 
objects. The two objects at the top (IRAS\,05246$-$7137 and
IRAS\,05328$-$6827) show strong silicate in absorption and weak CO$_2$ ice
absorption. SAGE1C\,J052546.49$-$661411.3 and IRAS\,05240$-$6809 show silicate 
emission and self-absorption respectively, even though the nature of the 
silicates is difficult to constrain without a detailed study. In general the 
silicate optical depth is not well constrained as the shape of the continuum 
underlying it is somewhat uncertain. Both these objects show ice absorption at 
15.2 and 6\,$\mu$m. IRAS\,05240$-$6809 also shows H$_2$ emission at 5.51, 6.91, 
and 9.66\,$\mu$m. These two apparently PAH-free objects are the only YSOs for 
which we were able to successfully isolate the 5\,$-$\,7\,$\mu$m ice band.

For the 11 PAH-rich YSOs we use the PAHFIT output to extract the optical depth
of the ice features, for the remaining objects we use the spectrum after
subtracting a polynomial continuum. We compared the optical depth spectra 
obtained when using the PAHFIT and the polynomial continua, for these four 
objects. This allowed us to constrain how the choice of continuum affects the
measured column densities for the ice features (Section\,4.3.3)

\subsection{CO$_2$ ice at 15.2\,$\mu$m}

It has been shown in laboratory experiments that the CO$_2$ ice profile at 
15.2\,$\mu$m is a sensitive probe of the environment in which the ice resides, 
both in terms of physical conditions like temperature 
\citep[it traces strong heating,][]{ehrenfreund98,gerakines99,white09} and 
also in terms of molecular environment
\citep[it can indirectly pinpoint other molecules in the ice matrix,]
[]{ehrenfreund97} as is discussed in the next section. On one hand this makes 
CO$_2$ an interesting ice species to study, but on the other it implies that 
laboratory studies of this solid are very complex as the whole molecular system 
needs to be considered, not just CO$_2$ ice in isolation. Realistic
astrophysical environments are also impossible to replicate. Thus we cannot 
model the CO$_2$ profiles directly using experiments. The next best thing is a 
phenomenological approach that attempts to model the observed profiles as a 
combination of ``components'' that have been well studied in the laboratory.

CO$_2$ does not form in any sizable abundance in the gas phase in the
diffuse interstellar medium or molecular clouds \citep[e.g.,][]{herbst86}, but 
it is not yet fully understood which of a number of possible grain-surface 
reactions is responsible for the abundance and ubiquity of CO$_2$ ices 
\citep[for a recent discussion on CO$_2$ formation mechanisms see][]
{pontoppidan08}. It was initially thought that this molecule would form during
strong UV photolysis of mixed ices containing H$_2$O and CO \citep[e.g.,][]
{dhendecourt86}, conditions easily found in the envelopes of luminous embedded 
YSOs. However, observations of abundant CO$_2$ also in the quiescent ISM 
\citep{whittet07,whittet09} indicate that {\it additional} radiative processing,
besides that provided by cosmic-ray induced photons and ambient radiation field 
photons, is not required for CO$_2$ formation. 

H$_2$O ice is the most abundant ice species, followed by CO and CO$_2$
\citep{gibb04}. These distinct ice species are often observed in the same 
sightline and are present on the same grain mantles possibly in a layered, 
``onion''-like structure. In particular, two independent components have been 
identified on icy mantles: H$_2$O-dominated and CO-dominated components, also
referred to as polar and apolar components, respectively 
\citep{tielens91,ehrenfreund03}. A possible scenario is that H$_2$O ice 
(with traces of CO, NH$_3$, and CH$_4$) forms easily and widely in molecular 
clouds, as it requires relatively low gas density and only moderately low 
temperature. At higher densities grains accrete mainly apolar molecules like CO 
(and IR inactive molecules like O$_2$ and N$_2$, see below); this rapid freezeout
creates a mantle of almost pure CO on top of a water-rich layer 
\citep{tielens91,ehrenfreund97}. 

Laboratory experiments show that 15.2-$\mu$m CO$_2$ profiles in ice matrices 
dominated by either H$_2$O or CO have very distinct properties \citep[e.g.,]
[ see also Figure\,\ref{smooth_ices}]{ehrenfreund97}. The comparison of 
laboratory and astronomical CO$_2$ ice profiles shows that the latter are well 
represented by a combination of polar (H$_2$O-dominated) and apolar 
(CO-dominated) components \citep{gerakines99,whittet07,pontoppidan08,white09,
whittet09}. The polar component usually dominates, accounting for a large 
fraction of the total CO$_2$ column density \citep{gerakines99,pontoppidan08,
whittet09}. Thus, even though doubts remain about its formation mechanism, these 
observations indicate that viable CO$_2$ formation routes must be present both 
in H$_2$O- and CO-dominated ice mantles \citep{pontoppidan06,pontoppidan08}.

For some Galactic YSOs, the CO$_2$ ice profiles show additional substructure. 
The fact that this structure is absent in the CO$_2$ profiles in
quiescent regions \citep{whittet07,whittet09} suggests that thermal processing 
could play a role in the production of this additional CO$_2$ component. 
\citet{ehrenfreund98,ehrenfreund99} showed that when heated the spectra of 
laboratory mixtures of H$_2$O, CH$_3$OH (methanol), and CO$_2$ in equal parts 
develop a complex structure due to annealing and crystallization \citep[see also]
[]{white09}. The samples become very inhomogeneous and at temperatures of 
typically 100\,K CH$_3$OH and CO$_2$ segregate and inclusions of pure CO$_2$ are
produced. As a result the profiles develop the double-peaked structure 
that is characteristic of pure-CO$_2$ ice \citep{ehrenfreund97}. A red shoulder 
at 15.3\,$\mu$m also appears, very likely due to the interaction of CO$_2$ and 
CH$_3$OH molecules. Several studies have found that the annealed ice component
combined with the dominant water-rich component fits the observed CO$_2$ ice 
profiles well \citep{gerakines99,white09,zasowski09}. Recently 
\citet{pontoppidan08} suggested that for the case of low-luminosity YSOs, where 
the relatively high temperature needed for the segregation process may not be 
reached, the pure-CO$_2$ component arises from the distillation of the CO:CO$_2$ 
component through evaporation of CO at lower temperatures ($<$\,50\,K). 

Bringing all this information together, we  have the following schematic scenario
\citep{gerakines99,pontoppidan08}. CO$_2$ is formed plentifully on icy grain 
surfaces in quiescent molecular clouds, mainly in an H$_2$O-rich medium  but with
a contribution of a CO-rich mantle in agreement with observations. This complex 
ice mantle is heated up by the contracting YSO, and what happens next depends on
the temperature profile of the dusty envelope. In the environments of massive 
luminous YSOs, high enough temperatures can be reached ($\sim$\,100\,K in 
laboratory experiments, 50$-$80\,K in collapsing dusty envelopes) so that 
annealing and segregation give rise to co-existing H$_2$O-rich and CO$_2$-rich 
($\sim$ equal parts H$_2$O and CO$_2$) components \citep{gerakines99}. This 
heating process would sublimate the CO-rich component\footnotemark. In the
envelopes of lower luminosity YSOs, only moderate heating takes place 
($\la$\,50\,K); the CO gradually desorbs leaving behind a pure-CO$_2$ component 
\citep{pontoppidan08}. In the latter case, H$_2$O-rich, CO-rich and pure-CO$_2$ 
components co-exist in icy mantles. How these two scenarios are played out in a 
YSO environment would depend on the luminosity and evolutionary stage of the 
object, and the temperature profile across the dusty envelope. 

Recent studies have favored combining the water-rich component exclusively with 
either annealed ices \citep{zasowski09} or CO- and/or CO$_2$-dominated ices 
\citep{pontoppidan08} to model the observed profiles. Both approaches fit the 
data well, and it is not clear which of these is more appropriate. We avoid 
making a priori decisions on the nature of the profiles observed towards the LMC
sources. We model the CO$_2$ ice profiles in our sample in both ways: {\it{i}}) 
by combining H$_2$O-rich, CO-rich, and pure-CO$_2$ components; and {\it{ii}}) by 
combining a H$_2$O-rich component with an annealed component --- the same 
approach is used by \citet{gerakines99} and \citet{white09}. The next sections 
describe the laboratory ice profiles used in the analysis and the resulting
profile fits. 

\footnotetext{Sublimation temperatures are of the order of 20$-$30, 50$-$90, and 
100$-$120\,K, respectively for CO, CO$_2$, and H$_2$O depending on the ice 
mixtures in which the species reside 
\citep{gerakines99,vanbroekhuizen06,fraser01}.}

\subsubsection{Laboratory ices}

The laboratory ice profiles used in this study come from public databases from
\citet[][ Leiden database: polar, apolar, and pure CO$_2$ ices]{ehrenfreund97} 
and \citet[][ annealed ices]{white09}, which Table\,\ref{lab} lists. In the
Leiden database, particle shape corrections are applied to the laboratory 
profiles using different grain models \citep{ehrenfreund97}; following other 
authors \citep{gerakines99,pontoppidan08} we use a Continuous Distribution of 
Ellipsoids (CDE, each grain shape equally probable) that seems to work best for 
CO$_2$ profiles. According to \citet{gerakines99} no such particle shape 
corrections are appropriate for the highly inhomogeneous annealed ices.

These databases are incredibly rich, and many parameters can be tuned (laboratory
temperatures, fractions of ices in mixtures, etc). Instead of using the full 
databases and just testing which combination of lab profiles fits each spectrum 
best \citep[as done in][]{gerakines99}, we simplify the problem
\citep{pontoppidan08}, and reduce whenever possible the number of free variables 
--- i.e. temperatures are kept fixed except in the case of the processed ices 
where thermal effects are being probed, and relative concentrations are only 
changed when it affects the morphology of the profiles in a unique way. This
should make possible trends easier to detect. The lower resolution of the 
\spitzer\ LL mode (when compared to previously published analysis) also justifies
this simplification, as profile details are lost (see below). For the same reason
we only adopt the three dominant components used in the analysis of 
\citet{pontoppidan08} (H$_2$O-rich, CO-rich, and pure CO$_2$). In the rest of 
this section we describe the profiles used.

\begin{figure}
\begin{center}
\includegraphics[scale=0.75]{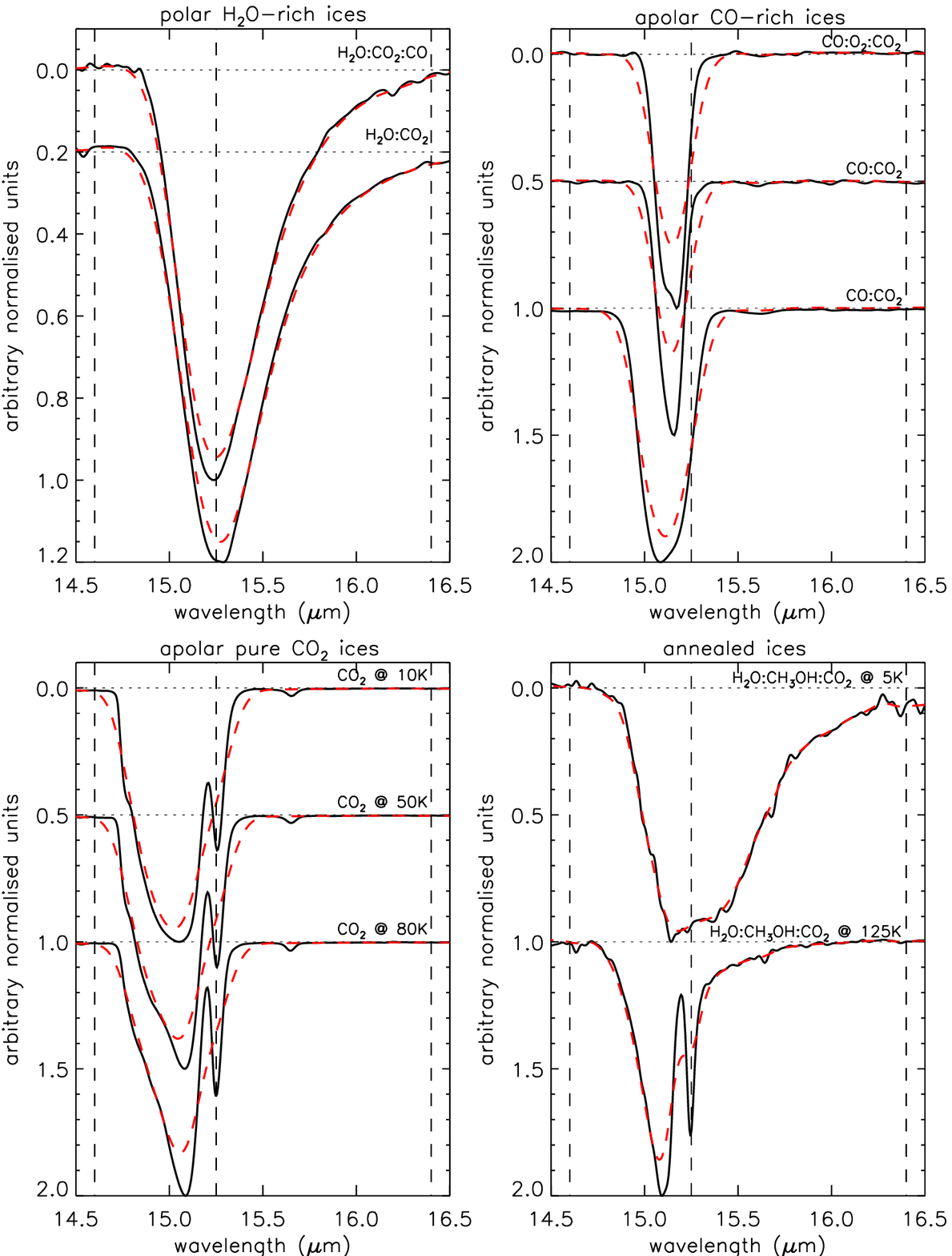}
\caption{Representative laboratory ice profiles for the different types of
mixtures considered here: H$_2$O-rich, CO-rich, pure-CO$_2$ and annealed ices.
The profiles used are listed in Table\,\ref{lab}. For the CO-rich ices the
following compositions are shown: CO:O$_2$:CO$_2$\,=\,100:50:16, 
CO:CO$_2$\,=\,100:16 and CO:CO$_2$\,=\,100:70. Dashed (red) lines are the 
laboratory profiles smoothed to match the resolution of the \spitzer\ IRS LL 
mode (see text). The double-peaked structure of the pure-CO$_2$ and annealed 
ices is smoothed out to such an extent that it is no longer clearly 
identifiable. The vertical lines delimit the red and blue wing of the profile 
(see text).\label{smooth_ices}}
\end{center}
\end{figure}

For polar (water-dominated) ices, two mixtures are available in the databases:
H$_2$O:CO$_2$\,=\,100:14 at 10\,K and H$_2$O:CO$_2$:CO\,=\,100:20:3 at 20\,K. As
it can be seen from Figure\,\ref{smooth_ices} (top left) these two mixtures 
produce very similar spectra, with only a minor shift in the peak position; 
still we consider them both for completeness sake. A 4$^{\rm th}$-order 
polynomial is used to remove the contribution of the H$_2$O libration mode at 
13\,$\mu$m from the spectra of all water-rich mixtures used.

For apolar mixtures of CO and CO$_2$ several types of mixtures are available, 
either just CO:CO$_2$ or including IR inactive molecules, CO:O$_2$:CO$_2$ and 
CO:N$_2$:CO$_2$. IR inactive molecules like O$_2$ and N$_2$ are formed in the gas
phase and accrete onto the cold grains at higher densities. They have no innate 
IR transitions, but in an ice matrix they interact with other molecules, subtly 
changing the shape of strong profiles \citep{ehrenfreund97}. For the same
CO$_2$:CO concentration ratio, the presence of O$_2$ in the ice matrix makes the
CO$_2$ ice profile slightly wider (Figure\,\ref{smooth_ices} top right, top two 
pairs of spectra). However, the exact amount of O$_2$ in the mixture is not well
constrained (profiles from spectra where the amount of O$_2$ is 10, 20 or 50\% of
CO are indistinguishable). Therefore we keep the O$_2$ concentration at 50\% of 
that of CO. The presence of N$_2$ does not change the profiles in a noticeable 
way \citep[N$_2$ is a ``silent'' matrix component,][]{ehrenfreund97} so we do not
include those laboratory profiles. 

With increasing CO$_2$ fraction in the ice mixtures, the CO$_2$ profiles become 
progressively more asymmetric, with a more pronounced blue wing and also 
blue-shifting the peak position (Figure\,\ref{smooth_ices} top right, bottom two 
pairs of spectra). These changes in the profile shape are monotonic with 
CO$_2$:CO ratio; thus we can interpolate between available laboratory spectra 
and ratios \citep[see][]{pontoppidan08}. Available spectra have CO$_2$:CO ratios
as listed in Table\,\ref{lab}. The temperature of the mixture (generally 10, 20 
and 30\,K) also has no unique effect on the shape of the profile for any of the 
mixtures mentioned so far; we adopt the profiles for 10\,K. 

Laboratory spectra of pure-CO$_2$ ices are available at three temperatures (10, 
50, and 80\,K); with increasing temperature the profile becomes narrower, and the
component at 15.25\,$\mu$m becomes more pronounced (Figure\,\ref{smooth_ices} 
bottom left). We use the laboratory spectra for the 
H$_2$O:CH$_3$OH:CO$_2$\,$\sim$\,1:1:1 mixture from \citet{white09} as 
representative of the annealed ice component. At lower temperature those 
profiles are rather broad and structureless; as the samples are gradually heated
the characteristic peak splitting becomes pronounced ($\ga$\,100\,K), and the 
profile becomes considerably narrower (Figure\,\ref{smooth_ices} bottom right). 
Profiles are available at temperatures from 5 to 135\,K in 5-K steps 
(Table\,\ref{lab}). We should point out that laboratory temperatures quoted 
throughout this paper correspond to lower temperatures in conditions typical of 
dense molecular clouds \citep{ehrenfreund98,boogert00}. 

A crucial thing to realize is that previous studies made use of spectra with 
higher spectral resolving power. Observations performed with the {\it ISO}-SWS 
\citep[e.g.,][]{gerakines99} and the \spitzer\ IRS SH mode \citep[e.g.,][]
{pontoppidan08,zasowski09,white09} have $R$\,$\sim$\,1000 and $R$\,$\sim$\,600, 
respectively; for the IRS LL mode at 15.2\,$\mu$m $R$\,$\sim 90$. We take this 
into account by convolving the laboratory ice profiles described previously with 
a Gaussian profile at each wavelength element with a width appropriate to 
simulate the lower resolution of the LL mode; Figure\,\ref{smooth_ices} shows the
effect of the resolution on the profiles. The convolved profiles for pure CO$_2$ 
and annealed ices at higher temperature (lower two panels) show that the 
substructure in the profile that is normally associated with thermal processing 
is smoothed out at the LL resolution. The double-peaked structure is {\em no 
longer clearly visible}. Thus we are not able to diagnose processing in our 
spectra as readily and to the same level of detail as in previous studies. 

Figure\,\ref{smooth_ices} also shows which particular components contribute to
specific parts of the profile. For illustration purposes, we divide the profile 
into a blue and a red wing (respectively short- and long-wards of 15.2\,$\mu$m). 
The H$_2$O-rich and annealed mixtures at low temperature contribute to both wings
of the profile and the red wing is much more extended than the blue wing ($\sim$ 
1.1 versus 0.45\,$\mu$m). The CO-rich, pure-CO$_2$, and annealed mixtures at high
temperature contribute almost exclusively to the blue wing of the profile --- the
pure CO$_2$ profiles extend the furthest into the blue.
 
\subsubsection{Laboratory fits to the CO$_2$ profiles}

The IRS spectra in our sample were compared to the laboratory profiles using a
$\chi^2$ minimization technique 
\citep[Levenberg-Marquardt least-squares minimization,][]{more77}, making use of
the IDL\footnotemark\ routines \footnotetext{Interactive Data Language.} 
developed by \citet[][ MPFIT\footnotemark]{markwardt09}. This section discusses 
the resulting fits.\footnotetext{http://purl.com/net/mpfit.} 

\begin{figure}
\begin{center}
\includegraphics[scale=0.8]{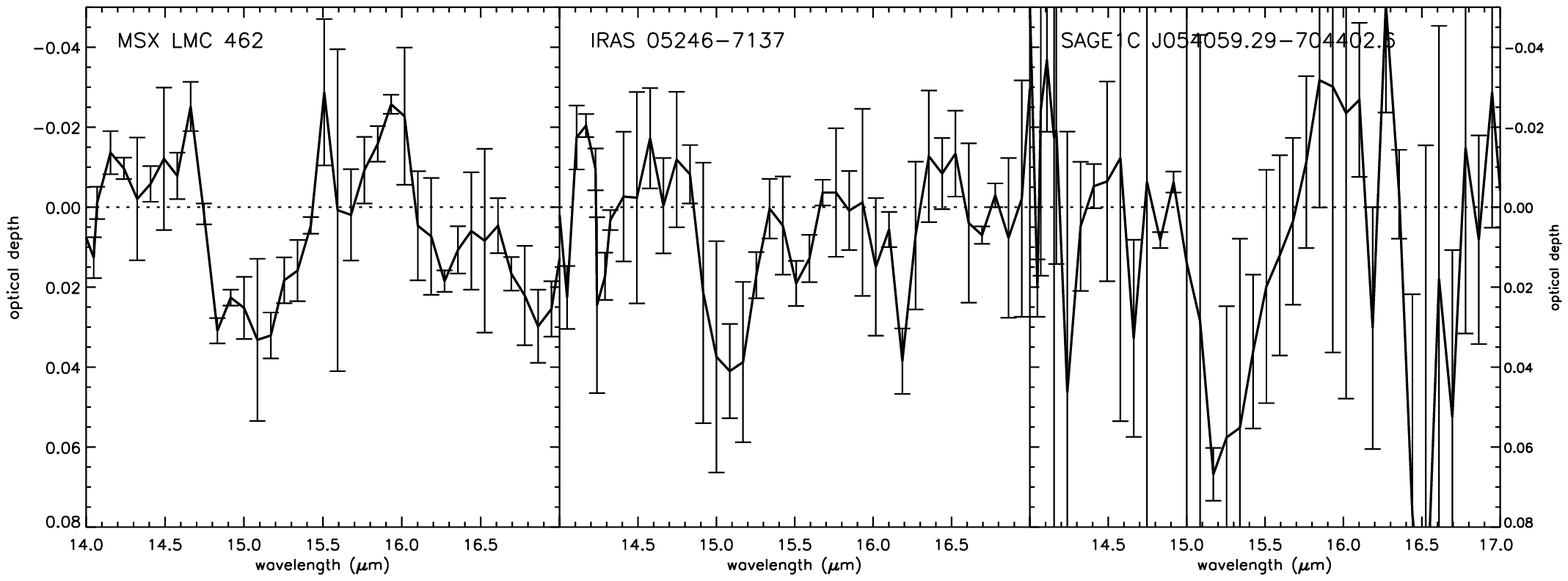}
\caption{CO$_2$ features for the three YSOs for which no fits with laboratory 
spectra were performed.\label{nofit}}
\end{center}
\end{figure}

Each observed CO$_2$ profile is modelled in two ways as described previously. 
Figure\,\ref{ice_fits} shows the resulting fitted profiles. Table\,\ref{fit_data}
lists relevant fitted parameters as well as the total CO$_2$ column density (see 
next section) and the percentages (in terms of column density) for each 
component. For three YSOs (MSX\,LMC\,464, IRAS\,05246-7137, and 
SAGE1C\,J054059.29$-$704402.6) the optical depth spectra were not deemed of good 
enough quality to fit meaningful models. Figure\,\ref{nofit} shows their CO$_2$ 
profiles, and Table\,\ref{fit_data} lists the column densities.

The spectra of the remaining 12 YSOs in the sample are well fitted by the
combinations of the lab profiles. Some objects exhibit features at 
$\sim$\,15.6\,$\mu$m (e.g., SAGE1C\,J052546.49$-$661444.3 and MSX\,LMC\,1786) 
and $\sim$\,16.2\,$\mu$m (IRAS\,04514$-$6931) that none of the model components 
can reproduce (Figure\,\ref{smooth_ices}). There is a [Ne\,{\sc iii}] emission 
line at 15.56\,$\mu$m and a weak PAH feature at 15.9\,$\mu$m. Therefore it is 
possible that these features are artifacts related to these two emission lines, 
either from the object itself or from its sky background.

The fitted profiles show that we are not able to distinguish between the 
different H$_2$O- and CO-rich mixtures used: similarly good fits are possible if 
the H$_2$O:CO$_2$ component is replaced by the H$_2$O:CO$_2$:CO mixture or the 
CO:CO$_2$ is replaced by CO:O$_2$:CO$_2$. There is also only a negligible 
difference in the fits when using the three available temperatures for 
pure-CO$_2$ ice. Accordingly, we make no such distinctions in 
Table\,\ref{fit_data}, referring only to H$_2$O-rich (polar), CO-rich 
(``apolar1'') and pure-CO$_2$ (``apolar2'') components. When the 
polar\,+\,annealed combination is used, laboratory temperatures $\ga$\,100\,K are
required, consistent with thermally processed mixtures in which CO$_2$ ice is 
segregated. 

For seven YSOs the two strategies provide equally good results in terms of 
$\chi^2$ minimization. For two YSOs the polar\,+\,apolar modeling fits their 
spectra better while for three other objects the polar\,+\,annealed combination 
performs better. \citet{gerakines99} found that the combination of 
polar\,+\,annealed components fitted the data of their Galactic sample better,
and suggested that this indicated that ice processing has occurred. We believe 
that for our data we cannot make this particular discrimination. The laboratory
components are obviously very different, but the loss of spectral detail 
at the LL resolution and the combination of multiple components limits what we 
can discriminate. 

In most cases, the two methods agree well on the amount of water-rich ice
required. The majority of YSOs have a large fraction of CO$_2$ locked into 
H$_2$O-rich ice: for eight of the twelve fitted objects the polar component 
accounts for $\sim$\,45\,$-$\,75\% in column density. This is consistent with 
what is observed for Galactic YSOs \citep[e.g.,][]{pontoppidan08,gerakines99}. 
For the other four objects either no water-rich ice is required or the two
strategies do not agree on how much is necessary. The polar component provides 
the dominant contribution to the red wing of the CO$_2$ profile 
(Figure\,\ref{smooth_ices}). When contrast in this region is low (as is the case 
for SAGE1C\,045228.65$-$685451.3, IRAS\,05452$-$6924, and IRAS\,05328$-$6827), it
becomes more difficult to determine the contribution of the polar component. We 
test this effect by adding noise to one of our best profiles and re-fitting it: 
the profile looks narrower, and the amount of the polar CO$_2$ ice is not well 
constrained. The spectrum of SAGE1C\,J052546.49$-$661444.3 is of good quality, 
but it is also dominated by the feature at $\sim$\,15.52\,$\mu$m.

If we consider the polar\,+\,apolar fits, a contribution from pure-CO$_2$ ice is
present for all but two fitted YSOs; it typically accounts for 20\,$-$\,50\% of 
the total column density. Similarly, the CO-rich component is present in 10 
profiles. Unlike \citet{pontoppidan08}, we have difficulty in constraining the 
CO$_2$:CO concentration ratio for the CO-rich component in the profiles 
individually. However, there seems to be a tendency in the sample for high 
CO$_2$:CO ratios. If we consider the polar\,+\,annealed fits, a contribution from
processed ices is needed in all cases. 

\subsubsection{CO$_2$ ice column density}

Table\,\ref{fit_data} lists computed column densities adopting a band strength 
A\,=\,1.1\,$\times$\,10$^{-17}$\,cm molecule$^{-1}$ \citep{gerakines95}; CO$_2$ 
column densities are in the range 0.45\,$-$\,16\,$\times$\,10$^{17}$\,cm$^{-2}$. 
Uncertainties listed in the table are statistical only. Different choices in 
defining the underlying continuum can change the measured column densities by 
$\sim$\,15\%.

We can compare our new CO$_2$ column density measurements for
IRAS\,05240$-$6809, SAGE1C\,J052546.49$-$661411.3, and
SAGE1C\,J051449.41$-$67221.5 with measurements performed with the low-resolution
spectrograph ($R$\,$\sim$\,20) on board {\it AKARI} for the 4.3-$\mu$m CO$_2$ 
stretching mode \citep{shimonishi08}. Their measurements suffered from very 
large uncertainties (at least 50\%, see Table\,\ref{fit_data}); both sets of 
measurements are consistent within the quoted uncertainties, however our 
measurements bring the estimated CO$_2$ column densities down considerably. 

For those three objects, \citet{shimonishi08} also obtained water ice column 
densities --- for IRAS\,05240$-$6809 they do not tabulate a value for the column
density, but we estimate it roughly from their Figure\,1, 
$\sim$\,40\,$\times$\,10$^{17}$\,cm$^{-2}$ with an assumed error of 25\%. 
Van Loon et al. (2005b) measured the column density of water ice in 
IRAS\,05328$-$6827; we re-estimate this column density using their spectrum and
obtain 3.26\,$\pm$\,0.27\,$\times$\,10$^{17}$\,cm$^{-2}$. For these four YSOs 
(IRAS\,05240$-$6809, SAGE1C\,J052546.49$-$661411.3, SAGE1C\,J051449.41$-$67221.5
and IRAS\,05328$-$6827) we combine our CO$_2$ measurements with the H$_2$O 
measurements. \citet{shimonishi08} computed CO$_2$ and water column densities
for a further three objects. Thus there are now seven YSOs in the LMC with column
densities measured for both ice species. There is a strong positive correlation 
between the CO$_2$ and H$_2$O column densities for Galactic YSOs; in the next 
section we compare the measurements for YSOs in the LMC and the Galaxy.

\subsection{The 5$-$7\,$\mu$m ice band}

The 6- and 6.8-$\mu$m ice bands dominate the 5$-$7\,$\mu$m range in YSOs. Even 
though first identified by \citet{puetter79}, their carriers are still not
conclusively identified. The main contributor to the 6-$\mu$m component is the 
bending mode of water \citep[e.g.,][]{tielens84}. However, the column densities
for the 3- and 6-$\mu$m water bands were found to be discrepant, with the 
6-$\mu$m band often much too deep \citep[e.g.,][]{keane01}. Other ice species 
like formic acid (HCOOH), formaldehyde (CH$_2$O), and ammonia (NH$_3$) may 
contribute to the depth of the 6-$\mu$m feature \citep[e.g.,][]
{knez05,boogert08,zasowski09}. Further, the 6-$\mu$m bending mode might be 
sensitive to the presence of both CO$_2$ and CO in the ice matrices 
\citep{oberg07,bouwman07}, which in some sources could account for some of the 
excess absorption at 6\,$\mu$m \citep{knez05}. The ice component at 6.8\,$\mu$m 
is now believed to be due mostly to ammonium (NH$_4^+$), with a smaller
contribution from methanol (CH$_3$OH) \citep{boogert08}.

In Section\,4.1 we discussed how PAH and H$_2$ emission have also important
spectral signatures in the 5\,$-$7\,$\mu$m spectral range, in particular the 
strong 6.3-$\mu$m PAH feature. \citet{spoon02} have shown that weak water ice 
features at 6\,$\mu$m are effectively masked in the presence of a strong PAH 
emission spectrum (their Figure\,1). As a large fraction of our sample shows
significant PAH emission, this effect limits our ability to identify ice 
absorption in this spectral range. A quick inspection of the LMC sample of 
\citet{seale09} reveals that the majority of the objects that show CO$_2$ ice 
also show prominent PAH emission.

\citet{boogert08} and \citet{zasowski09} recently analyzed IRS high-resolution 
spectra of Galactic YSOs and found little or no sign of significant PAH 
emission. However these samples are mainly of low-luminosity objects. Thus PAH 
emission, if excited by the central object, would not dominate. In the 
higher-luminosity {\it ISO} sample discussed by \citet{gibb04} some objects show
PAH emission, even though not as prominent as in our LMC sample. 

In spite of the fact that there is a global dearth of PAH emission in metal-poor
galaxies \citep{madden06}, we believe that the reason why PAH emission is so 
readily observed in YSOs in the LMC has to do with the issue of spatial 
resolution. As discussed in Section\,2, observations YSOs in the LMC probe a 
larger region of the objects' environment compared to Galactic sources. Extended 
PAH emission can then easily appear superimposed onto the spectra of icy mantles.
A Galactic example of such ``source confusion'' in star forming environments is 
Mon\,R2: its {\it ISO}-SWS spectrum shows contributions from both an embedded 
YSO with ice and silicate absorption (Mon\,R2\,IRS\,1) and an ultracompact 
H\,{\sc ii} region (Mon\,R2\,IRS\,2) \citep[][ their Fig\,4]{spoon04}. Such 
problem is also well documented in the analysis of the spectra of ultra-luminous 
infrared galaxies \citep[e.g.,][]{spoon02,spoon04}.

\begin{figure}
\begin{center}
\includegraphics[scale=0.8]{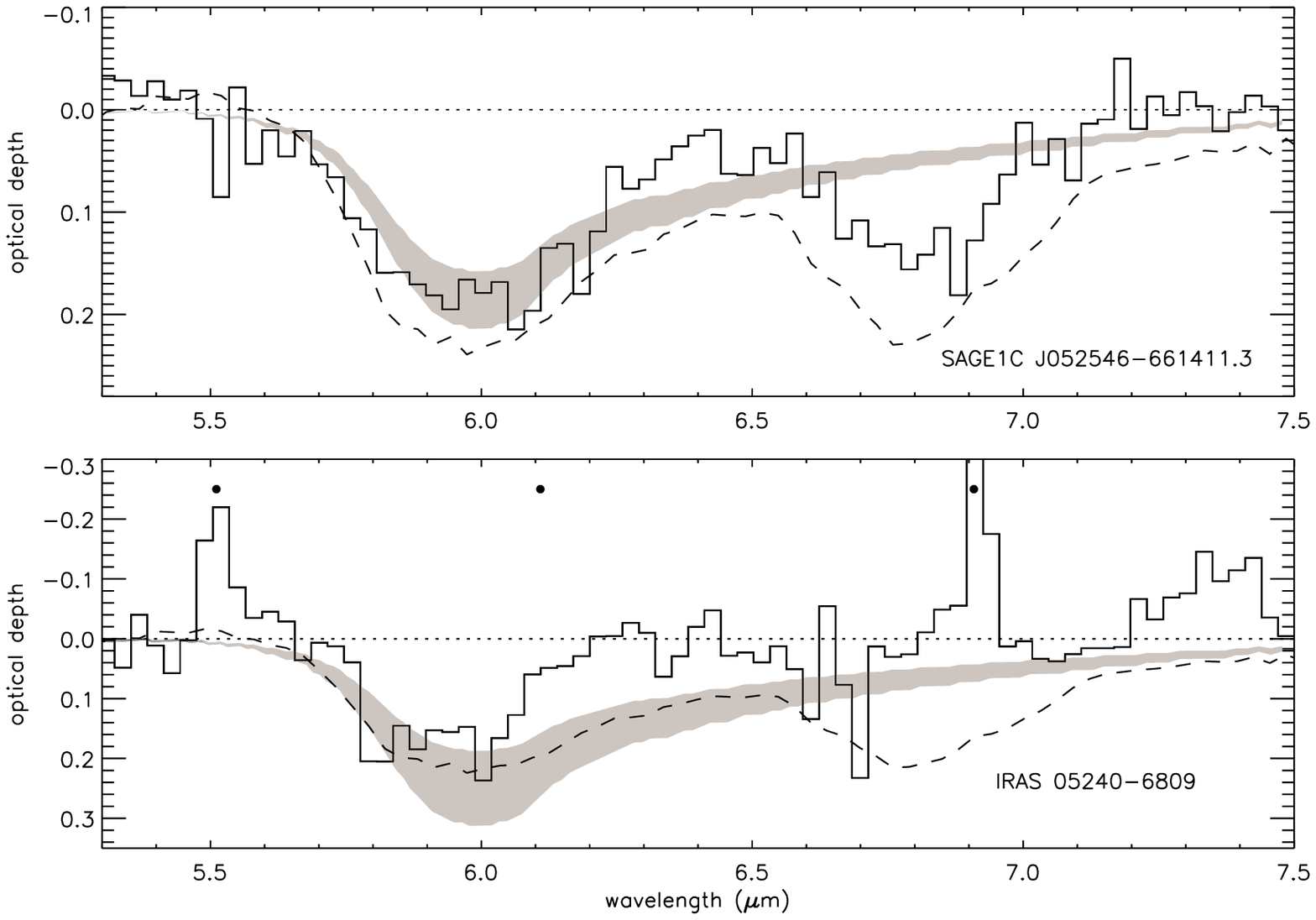}\\
\caption{\label{water_ice} Optical depth in the 5\,$-$\,7\,$\mu$m region, 
respectively for SAGE1C\,J052546.49$-$661411.3 and IRAS\,05240$-$6809. The 
filled circles indicate the position of unresolved H$_2$ lines, of which two are
clearly present in the spectrum of IRAS\,05240$-$6809. We also show typical 
laboratory ice models for pure water \citep{hudgins93}, scaled to match the water
column densities measured using the 3-$\mu$m feature (Table\,\ref{fit_data}), 
taking into account their error bars (shaded areas). The dashed line is an 
optical-depth spectrum for the Galactic YSO IRAS\,04016+2610 from 
\citet{zasowski09} showing water ice and the additional ice feature at 
6.8\,$\mu$m (spectrum scaled by a factor 1.5). Both LMC objects clearly show the
presence of water ice and SAGE1C\,J052546.49$-$661411.3 also shows the 
6.8-$\mu$m feature (see text).}
\end{center}
\end{figure}

Figure\,\ref{water_ice} shows the spectra for the two objects in the sample for 
which we could identify the H$_2$O ice absorption band at 6\,$\mu$m: 
IRAS\,05240\,$-$6809 and SAGE1C\,J052546.49$-$661411.3. None of these 
spectra show obvious PAH emission but IRAS\,05240\,$-$6809 shows unresolved H$_2$
emission. These YSOs show two of the largest CO$_2$ ice column densities 
($\ga$\,9\,$\times$\,10$^{17}$\,cm$^{-2}$, Table\,\ref{fit_data}), and thus 
should also have the largest water ice column densities (see next section). 
IRAS\,F04532$-$6709 shows similarly large CO$_2$ column density, but we cannot 
detect the 6-$\mu$m water ice feature which is probably hidden by PAH emission. 

We also show in Figure\,\ref{water_ice} typical laboratory ice spectra 
\citep[$T = 40$\,K,][]{hudgins93} scaled to match published water column
densities measured for the 3-$\mu$m feature (Table\,\ref{fit_data}) and taking
into account the measurement uncertainties (shaded areas in the figure). We 
have not made any allowance for the variable contribution of other ice species 
to the water ice band as described above. It is immediately obvious that, in 
particular between 6 and 6.5\,$\mu$m, the observed features are not deep enough 
to match the published water column densities. This suggests a deficiency in our 
continuum determination in this region. The figure also shows an example of a 
Galactic YSO from \citet{zasowski09} for the complete 5\,$-$\,7\,$\mu$m range, 
highlighting the contribution of the 6.8-$\mu$m ice band. The spectrum of 
SAGE1C\,J052546.49$-$661411.3 does indeed show this feature, likely attributed to
NH$_4^+$ \citep{boogert08}.

\section{Discussion}

In this section we compare in more detail the properties of ices in the LMC YSOs
to Galactic samples and interpret our results in terms of the star formation
environment in the LMC.

\subsection{Observed properties of the CO$_2$ ice profiles}

As we have seen in Section\,4.4, a large fraction of CO$_2$ ice is locked in 
a water-rich ice matrix, with a possible contribution from CO$_2$ ice in a 
CO-rich layer. Figure\,\ref{water_rich} compares the fraction of CO$_2$ ice in
the water-rich component for the LMC sample and Galactic samples from the
literature: high-luminosity YSOs and background sources\footnotemark
\citep{gerakines99,whittet09} and low-luminosity YSOs \citep{pontoppidan08}. 
\footnotetext{Background objects are bright sources that coincidentally sit 
behind the molecular cloud but are not associated with it in any way. They
represent quiescent sightlines.}
The column densities for this component are derived using the polar\,+\,apolar 
modelling approach (Section\,4.3.2) for all samples. In the Galaxy, luminous 
YSOs and background sources display a larger fraction of CO$_2$ ice in the 
water-rich component ($\sim$\,85\%) when compared to low-luminosity sources 
($\sim$\,70\%). The typical fraction for the LMC objects is $\sim$\,60\%. This 
suggests that, even though both in the LMC and in the Galaxy the water-rich 
component is the major contributor to CO$_2$ ice, its fraction might be smaller 
for the LMC sources. However, we note that the LMC sample is small, and 
systematic uncertainties could bring the samples into better agreement. Even
though a similar profile decomposition was used to isolate the water-rich
contribution, there are differences in the modelling approach (e.g., number and
properties of the components considered). 

A contribution from pure-CO$_2$ ice is also needed to fit the majority of the
profiles. Whether this contribution arises from inclusions in annealed ices or a 
pure-CO$_2$ layer resulting from CO desorption is less obvious. If CO$_2$ forms 
in either water- and/or CO-rich ice mixtures as has been widely suggested, then 
the presence of pure CO$_2$ diagnoses thermal processing. Therefore, our analysis
of the morphology (i.e. shape and composition) of the 15.2-$\mu$m CO$_2$ ice 
profiles of a sample of embedded YSOs in the LMC reveals only marginal 
differences when compared to Galactic YSOs. In the next section we discuss the 
concentration of CO$_2$ ice in relation to water ice. 

\subsection{CO$_2$ ice abundance in LMC}

Using compiled column-density measurements (Figure\,\ref{ratio}), we recompute 
the Galactic $N({\rm CO}_2)/N({\rm H_2O})$ ratio. Filled and open circles are for
Galactic YSOs and background sources, respectively. Measurements for LMC YSOs are
shown as star symbols. The YSOs in the \citet{gerakines99} sample are luminous, 
while those in the \citet{nummelin01}, \citet{pontoppidan08} and 
\citet{zasowski09} samples are low-luminosity objects. The \citet{knez05} and 
\citet{whittet07,whittet09} samples are exclusively of background sources.

Measurements of the column density of CO$_2$ ice are for either the 15.2-$\mu$m 
bending mode \citep{gerakines99,knez05,pontoppidan08,zasowski09,whittet09} or 
the 4.3-$\mu$m stretching mode \citep{nummelin01,shimonishi08}. Measurements of 
H$_2$O ice are for the 3-$\mu$m stretching mode, except for \citet{zasowski09},
who used the 6-$\mu$m bending mode. They did not account for the contribution of 
other ice species to the 6-$\mu$m optical depth. Therefore, they have probably
overestimated the column density of water ice. To compare them with the other 
samples, we need to scale them accordingly. \citet{boogert08} compared the column
densities of water ice at 3 and 6\,$\mu$m, and they found that the 6-$\mu$m value
is on average higher than the 3-$\mu$m value by a factor of 1.74 (the median of 
49 sources; the standard deviation is 0.62). The adjusted column densities of
\citet{zasowski09} appear in Figure\,\ref{ratio}. We have also converted their
column densities for CO$_2$ ice (computed using 
A\,=\,1.5\,$\times$\,10$^{-17}$\,cm\,molecule$^{-1}$ ) to use the same band 
strength as the other measurements discussed here 
(A\,=\,1.1\,$\times$\,10$^{-17}$\,cm\,molecule$^{-1}$).

Figure\,\ref{ratio} shows the strong correlation between $N({\rm CO}_2)$ and 
$N({\rm H_2O})$. The median value for the number density ratio is 
$N({\rm CO}_2)/N({\rm H_2O})$\,$\sim$\,0.2 for luminous Galactic YSOs and 
background sources \citep{gerakines99,whittet07,knez05,whittet09}. For 
low-luminosity Galactic YSOs, that ratio is 
$N({\rm CO}_2)/N({\rm H_2O})$\,$\sim$\,0.3 
\citep{nummelin01,boogert04a,pontoppidan08} with a very large spread. 
\citet{zasowski09} reported a ratio of 0.16; however if the correction factor 
derived in the previous paragraph is applied to the measured water column
densities then that ratio becomes 0.28, fully consistent with the previous value.
In fact, the right-hand panels (b and d) in Figure\,\ref{ratio} show that, with 
this correction, the low-luminosity measurements are essentially 
indistinguishable for the different samples. Note that there is a larger scatter 
for the measurements of low-luminosity objects when compared to the 
high-luminosity and background sample. Therefore, Figure\,\ref{ratio} suggests a 
clear difference between the high-luminosity YSO and background samples, and the 
low-luminosity YSO samples \citep[see also][]{pontoppidan08}.  

\citet{shimonishi08} estimated a ratio 
$N({\rm CO}_2)/N({\rm H_2O})$\,$\sim$\,0.45\,$\pm$\,0.17 for five LMC YSOs. As 
mentioned previously, we revised some of Shimonishi's estimates downwards. For 
the seven LMC YSOs with both column-density measurements we estimate a median 
ratio $N({\rm CO}_2)/N({\rm H_2O})$\,=\,0.32\,$\pm$\,0.15 (1\,$\sigma$). A single
Shimonishi's measurement drives this large spread, it would otherwise be
$N({\rm CO}_2)/N({\rm H_2O})$\,=\,0.32\,$\pm$\,0.05 (1\,$\sigma$). Estimated 
total luminosities in our sample are 5\,$-$50\,$\times$\,10$^{3}$\,L$_{\odot}$, 
comparable to the massive YSO sample in the Galaxy. Figure\,\ref{ratio} (left 
panels) shows that if the high-luminosity Galactic and LMC samples are compared 
the $N({\rm CO}_2)/N({\rm H_2O})$ ratio is higher in the LMC, by a factor 1.6. It
is curious, however, that the ratio measured for the LMC objects is consistent 
with the {\em low-luminosity} Galactic sample. Note that column-density 
measurements do not depend on any modelling details but are prone to systematic 
uncertainties, namely in the continuum determination, that are generally 
difficult to quantify.

One should investigate the homogeneity of the samples discussed here. 
The LMC sample is restricted to very luminous YSOs, and it should, in terms of 
luminosity at least, be comparable to the sample of luminous Galactic YSOs 
observed with {\it ISO}. \spitzer\ studies of star formation concentrated on 
nearby star-forming regions, molecular clouds and isolated globules that form 
predominantly intermediate- and low-luminosity objects. These low-luminosity
Galactic samples include a variety of evolutionary stages from more embedded 
YSOs to Herbig AeBe, T Tauri and zero-age main-sequence stars. The different 
{\em timescales} involved in the formation of objects with different luminosity 
may leave an imprint on the observed ice chemistry. Furthermore, YSOs in the
LMC should in general appear more evolved due to the smaller dust column density 
in LMC objects when compared to Galactic objects. 

Galactic and LMC measurements sample different {\em spatial scales} 
(Section\,2), from a few parsecs in the LMC to milliparsecs in the Galaxy. If 
there is a gradient of ice abundances between the more immediate YSO cocoon and 
the wider molecular environment in which it is immersed \citep{pontoppidan06}, 
then the fact that ice properties are integrated over different spatial scales 
could affect global abundance ratios.

Such issues should be considered, since models of molecular clouds and 
their chemical evolution during the star formation process suggest that both 
temporal and radial variations of molecular concentrations are important 
\citep[e.g.,][]{lee04,rodgers03,garrod08,vanweeren09,hollenbach09}. \citet{lee04}
singles out CO abundances (and the balance of its gas and solid phases) as 
particularly critical in determining the chemistry of star-forming cores. Even 
though they did not explicitly discuss CO$_2$ formation, the evolution of the 
two molecules is obviously closely linked. The balance of water in the gas phase 
and in ice form is also controlled by time- and space-dependent effects 
\citep{hollenbach09}. 

\subsection{The star formation environment in the LMC}

The focus of this project is how the metal-poor star-formation environment in the
LMC influences the chemistry of the envelopes that surround YSOs. Thus we now 
look at what is known about the environment in molecular clouds in the LMC. 

The structure of Giant Molecular Clouds (GMCs) in metal-poor galaxies including 
the LMC is different from that in the Galaxy. Observations suggest that in 
metal-poor galaxies there are large reservoirs of H$_2$ that are not traced by 
CO \citep[e.g.,][]{israel97}. The magnitude of this H$_2$-to-CO excess was found
to depend strongly on both metallicity and strength of the radiation field. This 
has been interpreted as the result of selective photodissociation of CO in the 
outer parts of metal-poor GMCs \citep[e.g.,][]{israel97,bolatto99}: H$_2$ 
readily self-shields while CO is shielded from the photodissociating radiation 
mostly by dust, which is less abundant at low metallicities. In this case, CO 
emission would trace only the inner parts of metal-poor GMCs. If CO cores are 
smaller in the LMC, due to the harsher UV radiation field \citep[e.g.,][]
{welty06} and lower dust abundance \citep[e.g.,][]{bernard08} then we could 
naively expect CO$_2$ production to be reduced. A lower production rate of CO$_2$
could just result from the lower abundance of gas-phase CO observed in the LMC 
\citep[e.g.,][]{bel86,fukui01,pineda09}. It could also result from a smaller
density of dust grains that act as seeds for the surface chemistry. Both effects 
would then tend to reduce the value of $N({\rm CO}_2)/N({\rm H_2O})$. However as 
described above this is not what we observed in the LMC.

The formation of CO$_2$ ice in a wide variety of environments (e.g., both
quiescent sightlines and low- and high-luminosity YSOs) argues against the need 
for enhanced radiative processing. Still, laboratory experiments do show that UV
irradiation enhances CO$_2$ production \citep[e.g.,][]{dhendecourt86}, and models
of diffusive surface chemistry can produce high CO$_2$ densities at slightly 
elevated temperatures \citep[e.g.,][]{ruffle01}. The LMC has a stronger ambient 
UV radiation field when compared to the Galaxy \citep[by a factor of 1\,$-$10,][]
{bel86,welty06}. As a result of the stronger radiation field, the average  
ISM dust temperature is also generally higher in the LMC 
\citep[12\,$\la$\,$T_{\rm d}$\,$\la$\,35\,K,][]{bernard08} than in the Galaxy 
\citep[19\,$\la$\,$T_{\rm d}$\,$\la$\,25\,K,][]{reach96}  --- dust temperatures 
in cloud cores are lower than in the diffuse ISM. Both these effects could 
increase CO$_2$ production and increase $N({\rm CO}_2)/N({\rm H_2O})$, consistent
with what is observed. 

We should also consider the possibility that the formation of water ice is 
somewhat inhibited in the LMC. The largest column density of water ice measured 
so far in the LMC is $\sim$\,5\,$\times$\,10$^{18}$\,cm$^{-2}$ 
\citep{shimonishi08}. This is below the highest column densities measured towards
Galactic YSOs, 10\,$-$30\,$\times$\,10$^{18}$\,cm$^{-2}$ 
\citep{gibb04,pontoppidan08}. Due to the small number of LMC objects investigated
to date and possible selection effects this may not be a significant difference. 
Water ice seems to form rather easily in molecular clouds. However the 
A$_{\rm V}$-threshold for water ice formation ($A_{\rm V}$\,$\sim$\,3\,mag for 
typical Galactic molecular clouds) is affected by the strength of the incident 
radiation field and dust temperature \citep{hollenbach09}. Still, it is difficult
to see how water ice formation could be inhibited when CO$_2$ ice formation is 
not.

\section{Summary and final remarks}

We analyzed \spitzer-IRS spectra of a sample of embedded young stellar objects in
the LMC, obtained as part of the \spec\ Legacy program. We selected sources for 
this analysis based on the shape of their mid-IR continuum and the presence of 
the bending mode of CO$_2$ ice at 15.2\,$\mu$m; a total of 15 sources were 
included, with three CO$_2$ ice detections considered marginal only
($<$\,3-$\sigma$). Based on modelling of the objects' SEDs, we characterize the 
LMC YSOs as luminous and still in the early embedded stages of evolution. 

A large fraction of the spectra of the LMC YSOs show prominent PAH emission. This
does not necessarily mean that the sources are more evolved, ultracompact 
H\,{\sc ii} regions. The \spitzer-IRS spatial resolution combined with the
distance to the LMC implies that our spectra sample a relatively large fraction 
of the YSOs' environment (a few square parsecs) when compared to Galactic YSOs, 
and thus PAH emission frequently appears superimposed onto the spectra of 
embedded YSOs. This makes the identification of the 5$-$7\,$\mu$m ice band 
difficult. We detect water ice at 6\,$\mu$m for two YSOs and the 6.8-$\mu$m 
feature (likely due to ammonium) in a single object.

It is usually believed that CO$_2$ forms on icy mantles, in water-dominated 
(polar) and/or CO-dominated (apolar) ice matrices. In Galactic YSOs, the
water-rich component is the most abundant. These components exhibit a distinct 
morphology allowing a multi-component analysis of the observed profiles. The 
double-peaked structure found in the CO$_2$ ice profiles towards YSOs in the 
Galaxy is a signature of thermal processing, via which the CO$_2$ segregates from
the other ice constituents. There are two types of modelling usually found in the
literature: a dominant water-rich component is combined with either one or more 
apolar (CO- or CO$_2$-dominated) components, or with an annealed ice component 
(water, CO$_2$ and methanol in equal parts). Accordingly, we model the observed 
profiles of CO$_2$ ice towards LMC YSOs in these two ways, using databases of
laboratory ice profiles available in the literature. We find that at the 
lower-resolution of the IRS LL mode, a considerable amount of detail is lost from
the laboratory spectra. Thus our comparison of the two types of modelling is not 
conclusive. 

We find that the CO$_2$ ice towards LMC YSOs is also dominated by the water-rich
component, and they also require a contribution from pure CO$_2$ ice. Thus in 
terms of general morphology and composition the CO$_2$ ice towards LMC YSOs is 
similar to that observed towards Galactic YSOs. There are however two 
possible differences. First, there is a hint that the fraction of CO$_2$ ice 
locked into a water-rich matrix is smaller in the LMC. Furthermore, the column 
density ratio $N({\rm CO}_2)/N({\rm H_2O})$, available for some of the targets, 
seems larger in the LMC. 

\begin{figure}
\begin{center}
\includegraphics[scale=0.8]{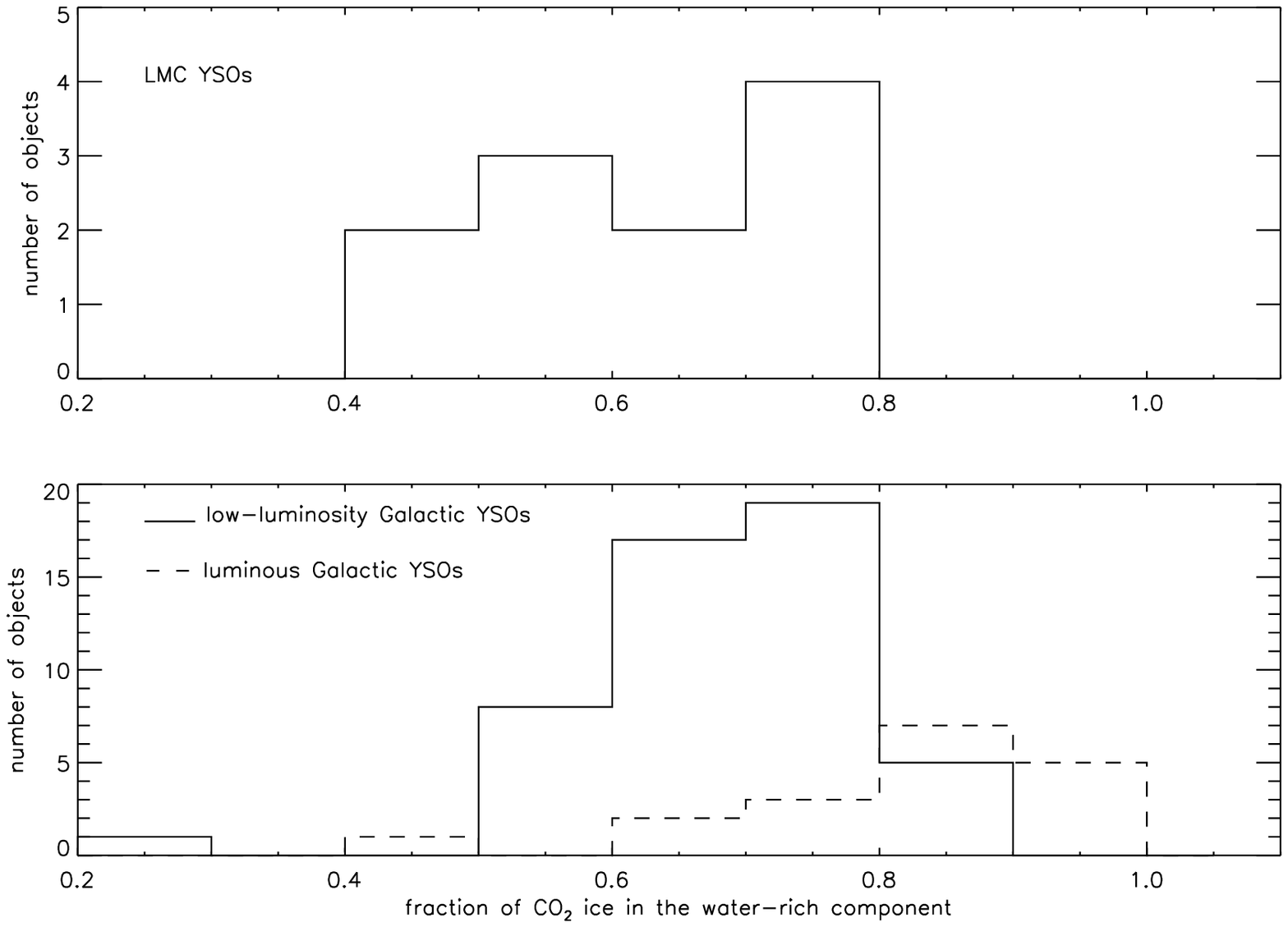}\\
\caption{Fraction of CO$_2$ locked in the water-rich (polar) component, for the 
LMC sample (top) and Galactic YSOs and background sources (bottom). Galactic data
come from \citet{gerakines99}, \citet{pontoppidan08}, and \citet{whittet09}.
\label{water_rich}}
\end{center}
\end{figure}

\begin{figure}
\begin{center}
\includegraphics[scale=0.8]{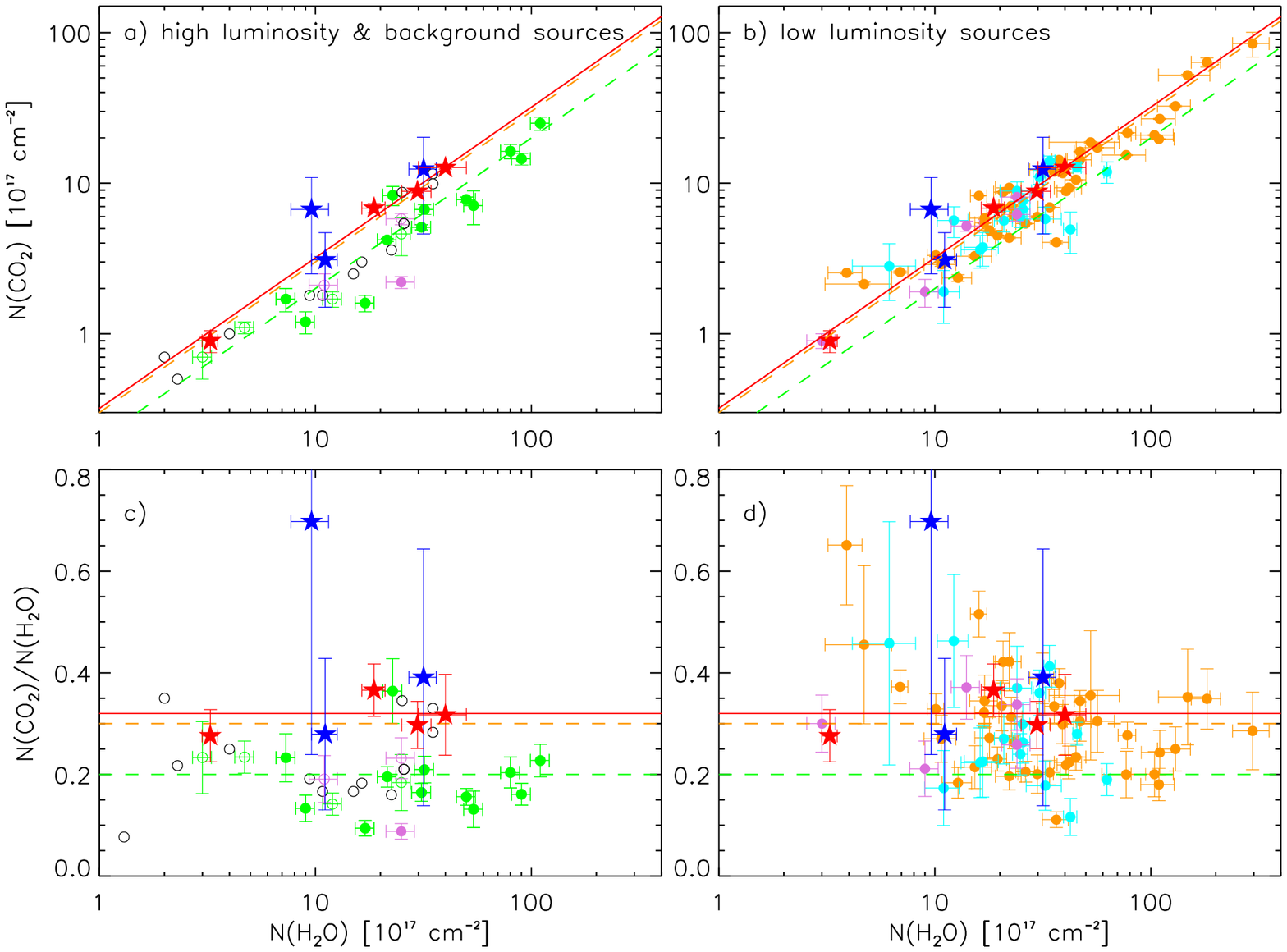}\\
\caption{CO$_2$ ice versus H$_2$O ice column densities. We compiled measurements
from the literature for Galactic sources (circles): \citet[][ green]
{gerakines99}, \citet[][ pink]{nummelin01}, \citet[][ cyan]{zasowski09}, 
\citet[][ gold]{pontoppidan08}, and \citet{whittet07,whittet09} and 
\citet{knez05} (black). Filled circles are YSOs while open circles are 
background sources (i.e. quiescent sightlines). Star symbols are the seven
LMC measurements: from \citet[][ blue stars]{shimonishi08} and our CO$_2$ ice 
measurements combined with literature H$_2$O ice measurements (red stars). The 
full lines are median ratios N(CO$_2$)/N(H$_2$O) for  Galactic high-luminosity 
and background sources ($\sim$\,0.2, green), Galactic low-luminosity sources 
($\sim$\,0.3, gold) and LMC sources ($\sim$\,0.32, red) --- see text for full 
explanation.\label{ratio}}
\end{center}
\end{figure}

One puzzling fact becomes noticeable when the available Galactic samples are 
separated into luminous and low-luminosity YSOs. The YSOs in our LMC sample are
luminous and thus should in principle be compared to the Galactic high-luminosity
sample. However, it is apparent from Figures \ref{water_rich} and
\ref{ratio} that the properties of the CO$_2$ profiles observed towards LMC YSOs
more resemble those of {\it low-luminosity Galactic YSOs} rather than those of 
the high-luminosity Galactic YSOs, both in terms of water-rich fraction and 
concentration ratio. Admittedly the size of the LMC sample is small thus these 
results are tentative for now. We discuss how the temporal and spatial evolution 
of the physical conditions in molecular clouds can affect the observed properties
of the different samples we compare.

We also discuss the implications of our results in terms of the star formation
conditions in molecular clouds in the LMC. The ISM of the LMC has lower
metallicity than the Galactic ISM ($Z_{\rm LMC}\sim 0.4 $\,Z$_\odot$). This 
affects the conditions in molecular clouds in the LMC in two parallel ways. On 
one hand there is less carbon and oxygen, and less dust. On the other hand, due 
to reduced shielding, the physical conditions in molecular clouds are harsher.

CO in the gas phase is less abundant, and the gas:dust ratio is larger by a 
factor 2 to 3 in the LMC than in the Galaxy. Observations also suggest that CO 
cores are smaller in the LMC, the result of selective photodissociation of CO in 
the outer parts of GMCs. If water ice abundance is unchanged, these effects would
conspire to reduce the ratio of CO$_2$ ice to water ice in the LMC. This is 
clearly not what we observe. On the other hand, the twined facts that the UV 
radiation field is stronger and the average dust temperature is higher in the LMC
could increase CO$_2$ production, which would be consistent with what we see
in our sample. 

Our results are tantalizing but not yet conclusive. We suggest that dust
temperature could be responsible for observed differences between the LMC and the
Galaxy. CO ice is more volatile than CO$_2$ and the CO profiles can be used as 
an indicator of thermal processing if different components correspond to
environments of different volatility. The suggestion is that CO ice is more
sensitive to the relatively small dust temperature differences mentioned 
previously. Furthermore, the possible increases in UV photolysis and thermal 
processing are related to metallicity. We have obtained \spitzer-IRS 
observations in the SMC and complementary ground-based data in both LMC and SMC, 
with the goal to incorporate CO ice profiles in our analysis, and to extend these
studies to the even more metal-poor environment of the SMC.

\acknowledgments We thank the anonymous referee for his/her comments. This work 
is based on observations made with the \spitzer\ Space Telescope, which is 
operated by the Jet Propulsion Laboratory, California Institute of Technology 
under a contract with NASA. This research has made use of NASA's Astrophysics 
Data System Bibliographic Services. J. M. Oliveira acknowledgs support from the 
UK's STFC. SAGE and SAGE-Spec research by M. Meixner at STScI has been funded by 
NASA/\spitzer\ grants 12755598, 1282638 and 1310534 and NASA NAG5-12595.

{\it Facilities:} \facility{Spitzer (IRAC)}, \facility{Spitzer (IRS)}

\clearpage

\begin{deluxetable}{lccccccccccccccccl}
\tabletypesize{\tiny}
\rotate
\tablecaption{Sample of embedded LMC YSOs, with groundbased optical and 
near-IR photometry. $UBVI$ and $JHK$ fluxes are from the MCPS and IRSF catalogs,
respectively \citep{zaritsky04,kato07}.
\label{targets1}}
\tablewidth{0pt}
\tablehead{
\colhead{ID} & \colhead{RA} & \colhead{DEC} &
F$_{U}$ & 
F$_{B}$ &
F$_{V}$ &
F$_{I}$ &
F$_{J}$ &
F$_{H}$ &
F$_{K}$ \\
&($^{h\,\,m\,\,s}$)
&($^{d\,\,m\,\,s}$)&
($\mu$Jy) &
($\mu$Jy) &
($\mu$Jy) &
($\mu$Jy) &
(mJy) &
(mJy) &
(mJy) 
}
\startdata
IRAS\,04514$-$6931\tablenotemark{a}&04:51:11.47&$-$69:26:46.9& 58.6\,$\pm$\,10.0       &\llap{1}79.4\,$\pm$\,18.7&\llap{3}24.4\,$\pm$\,40.7  &\llap{1}94.3\,$\pm$\,19.5&0.168\,$\pm$\,0.006&0.396\,$\pm$\,0.018& 0.730\,$\pm$\,0.067\\
SSTISAGE1C\,J045228.65$-$685451.3  &04:52:28.65&$-$68:54:51.3&  5.0\,$\pm$\, 1.8       & 26.4\,$\pm$\, 2.4       & 61.5\,$\pm$\, 4.6         &\llap{1}11.1\,$\pm$\, 7.7&0.068\,$\pm$\,0.007&0.275\,$\pm$\,0.020& 0.748\,$\pm$\,0.041\\
IRAS\,F04532$-$6709                &04:53:11.03&$-$67:03:56.0& 34.2\,$\pm$\, 3.7       & 56.6\,$\pm$\, 4.1       &\llap{1}07.6\,$\pm$\,22.1  &\nodata                  &0.112\,$\pm$\,0.009&0.190\,$\pm$\,0.018& 0.323\,$\pm$\,0.027\\
SSTISAGE1C\,J050354.56$-$671848.5  &05:03:54.56&$-$67:18:47.7& 50.1\,$\pm$\,10.1       & 55.9\,$\pm$\, 7.9       & 53.7\,$\pm$\, 4.5         & 61.0\,$\pm$\, 5.2       &0.475\,$\pm$\,0.009&2.034\,$\pm$\,0.019& 6.658\,$\pm$\,0.061\\
SSTISAGE1C\,J051347.73$-$693505.1  &05:13:47.79&$-$69:35:05.1&\nodata                  &\nodata                  &\nodata                    &\nodata                  &\nodata            &\nodata	       & 0.389\,$\pm$\,0.037\\
SSTISAGE1C\,J051449.41$-$671221.5  &05:14:49.41&$-$67:12:21.5&  5.7\,$\pm$\, 2.5       & 14.0\,$\pm$\, 1.5       & 19.7\,$\pm$\, 4.7         & 73.8\,$\pm$\, 4.0       &0.224\,$\pm$\,0.008&1.112\,$\pm$\,0.021& 4.811\,$\pm$\,0.133\\
IRAS\,05240$-$6809                 &05:23:51.14&$-$68:07:12.4&\nodata                  & 68.2\,$\pm$\,10.4       &\llap{7}80.4\,$\pm$\,108\rlap{.5}&\nodata            &5.936\,$\pm$\,0.711&6.021\,$\pm$\,0.388&\llap{1}3.158\,$\pm$\,0.727\\
IRAS\,05246$-$7137                 &05:23:53.95&$-$71:34:44.0&\nodata                  &\nodata                  &\nodata                    &\nodata                  &0.055\,$\pm$\,0.004&0.484\,$\pm$\,0.018& 3.966\,$\pm$\,0.037\\
MSX\,LMC\,464                      &05:24:13.30&$-$68:29:59.0&\llap{1}43.6\,$\pm$\,12.0&\llap{1}92.8\,$\pm$\,18.3&\llap{2}55.1\,$\pm$\,23.7  &\nodata                  &1.765\,$\pm$\,0.032&6.909\,$\pm$\,0.127&\llap{2}0.231\,$\pm$\,0.373\\
SSTISAGE1C\,J052546.49$-$661411.3  &05:25:46.52&$-$66:14:11.3&  6.0\,$\pm$\, 1.4       & 15.5\,$\pm$\, 1.7       & 34.2\,$\pm$\, 2.0         & 58.8\,$\pm$\, 6.6       &0.110\,$\pm$\,0.007&0.462\,$\pm$\,0.017& 3.254\,$\pm$\,0.060\\
IRAS\,05328$-$6827\tablenotemark{a}&05:32:38.59&$-$68:25:22.2&\nodata                  &\nodata                  &\nodata                    &\nodata                  &0.191\,$\pm$\,0.007&0.028\,$\pm$\,0.418& 9.111\,$\pm$\,0.168\\
MSX\,LMC\,1786\tablenotemark{a}    &05:37:28.07&$-$69:08:48.0&\nodata                  & 18.4\,$\pm$\, 2.4       & 27.6\,$\pm$\, 3.7         &\nodata                  &0.254\,$\pm$\,0.012&0.611\,$\pm$\,0.034& 1.180\,$\pm$\,0.076\\
SSTISAGE1C\,J054059.29$-$704402.6  &05:40:59.29&$-$70:44:02.6&\nodata                  & 31.1\,$\pm$\, 6.4       & 26.6\,$\pm$\, 3.8         & 31.9\,$\pm$\, 3.5       &\nodata            &0.289\,$\pm$\,0.032& 0.780\,$\pm$\,0.044\\
IRAS\,05421$-$7116\tablenotemark{b}&05:41:25.08&$-$71:15:32.7& 50.5\,$\pm$\, 4.8       &120.8\,$\pm$\, 5.5       & 57.4\,$\pm$\, 7.8         &\nodata                  &0.423\,$\pm$\,0.026&0.690\,$\pm$\,0.044& 1.285\,$\pm$\,0.058\\
IRAS\,05452$-$6924                 &05:44:50.23&$-$69:23:04.7& 12.9\,$\pm$\, 2.5       & 23.5\,$\pm$\, 1.5       & 35.0\,$\pm$\, 2.3         & 78.5\,$\pm$\, 4.2       &0.181\,$\pm$\,0.007&0.321\,$\pm$\,0.012& 0.671\,$\pm$\,0.025\\
\enddata
\tablenotetext{a}{sources from the \spitzer\ archive; all other sources are from
the \spec\ program.}
\tablenotetext{b}{$JHK$ fluxes from the 2MASS catalog \citep{cutri03}.}
\end{deluxetable}

\begin{deluxetable}{lccccccccl}
\tabletypesize{\tiny}
\rotate
\tablecaption{Sample of embedded LMC YSOs, with \spitzer\ IRAC and MIPS 
photometry from the SAGE catalog \citep{meixner06}. The last column provides
alternative identifications and known classifications from the literature.
\label{targets2}}
\tablewidth{0pt}
\tablehead{
\colhead{ID} & 
F$_{3.6\,\mu{\mathrm m}}$ &
F$_{4.5\,\mu{\mathrm m}}$ &
F$_{5.8\,\mu{\mathrm m}}$ &
F$_{8\,\mu{\mathrm m}}$ &
\multicolumn{2}{c}{F$_{24\,\mu{\mathrm m}}^b$} &
F$_{70\,\mu{\mathrm m}}$ &
F$_{160\,\mu{\mathrm m}}$& 
\colhead{Comments}\\
&(mJy)&
(mJy) &
(mJy) &
(mJy) &
\multicolumn{2}{c}{(mJy)} &
(mJy) &
(mJy) &
}
\startdata
IRAS\,04514$-$6931\tablenotemark{a}& 5.72\,$\pm$\,0.494       &\llap{1}8.90\,$\pm$\,0.547 &\llap{5}2.70\,$\pm$\,1.529 & 96.0\,$\pm$\,6.386	 &\llap{2}538.0\,$\pm$\,16.590&\llap{2}538.0\,$\pm$\,16.590&\llap{1}5820\,$\pm$\,127.40 &1257\,$\pm$\,336.70 &MSX\,LMC\,1200\\
SSTISAGE1C\,J045228.65$-$685451.3  & 3.21\,$\pm$\,0.173       &  6.05\,$\pm$\,0.169       &\llap{1}0.80\,$\pm$\,0.288 & 23.0\,$\pm$\,0.514	 & 198.3\,$\pm$\, 1.364       & 217.4\,$\pm$\, 1.064	   & 1435\,$\pm$\, 24.24	&2793\,$\pm$\, 81.96 &YSO (1)\\
IRAS\,F04532$-$6709                & 2.03\,$\pm$\,0.350       &  4.00\,$\pm$\,0.171       &  9.10\,$\pm$\,0.381       & 16.9\,$\pm$\,0.880       & 391.3\,$\pm$\, 2.158       & 350.1\,$\pm$\, 2.214       & 2454\,$\pm$\, 29.70        &3815\,$\pm$\,110.10 &\\
SSTISAGE1C\,J050354.56$-$671848.9  &24.50\,$\pm$\,0.475       &\llap{3}5.60\,$\pm$\,0.514 &\llap{4}5.40\,$\pm$\,0.869 & 64.5\,$\pm$\,1.579	 & 180.6\,$\pm$\, 4.059       & 307.2\,$\pm$\, 2.703	   & 7374\,$\pm$\, 72.46	&1091\,$\pm$\,385.10 &YSO (1)\\
SSTISAGE1C\,J051347.73$-$693505.1  &\nodata                   &  3.13\,$\pm$\,0.158       &  8.25\,$\pm$\,0.417       & 17.6\,$\pm$\,0.830       & 372.5\,$\pm$\, 1.899       & 276.4\,$\pm$\, 2.949       &\nodata                     &\nodata             &YSO (1)\\
SSTISAGE1C\,J051449.41$-$671221.5  &\llap{1}9.20\,$\pm$\,0.586&\llap{3}4.30\,$\pm$\,0.651 &\llap{5}1.20\,$\pm$\,0.939 & 71.3\,$\pm$\,1.238	   & 281.7\,$\pm$\, 1.634     & 294.8\,$\pm$\, 1.674       & 1243\,$\pm$\, 19.92	  &\nodata	       &YSO (1), ST4 (2)\\
IRAS\,05240$-$6809                 &\llap{2}2.50\,$\pm$\,0.685&\llap{3}3.40\,$\pm$\,0.529 &\llap{4}2.20\,$\pm$\,0.860 & 56.8\,$\pm$\,0.797	   &\llap{1}216.0\,$\pm$\, 6.113&\llap{1}270.0\,$\pm$\, 7.630& 3962\,$\pm$\, 42.34	  &1384\,$\pm$\, 91.71 &YSO (1), ST7 (2)\\
IRAS\,05246$-$7137                 &\llap{3}4.50\,$\pm$\,0.694&\llap{6}2.50\,$\pm$\,1.799 &\llap{9}2.90\,$\pm$\,1.672 &\llap{1}19.0\,$\pm$\,1.877& 342.6\,$\pm$\, 1.588       & 348.2\,$\pm$\, 2.042	     & 1490\,$\pm$\, 19.09	  &\nodata	&YSO (1), MIPS-SED (3)\\
MSX\,LMC\,464                      &\llap{6}3.60\,$\pm$\,1.890&\llap{8}9.50\,$\pm$\,2.576 &\llap{12}2.00\,$\pm$\,2.684&\llap{1}75.0\,$\pm$\,4.752& 868.0\,$\pm$\, 4.333       & 668.0\,$\pm$\, 4.412	   & 6741\,$\pm$\, 58.81	&1961\,$\pm$\,421.60 &H\,{\sc ii}? (4)\\
SSTISAGE1C\,J052546.49$-$661411.3  &\llap{3}0.00\,$\pm$\,0.668&\llap{5}6.60\,$\pm$\,1.554 &\llap{7}9.80\,$\pm$\,1.430 &\llap{1}15.0\,$\pm$\,1.958& 504.7\,$\pm$\, 3.381             & 498.4\,$\pm$\, 3.021      & 1417\,$\pm$\, 34.56	  &17510\,$\pm$\,363.90 &YSO (1), ST3 (2)\\
IRAS\,05328$-$6827\tablenotemark{a}&\llap{7}7.24\,$\pm$\,3.294&\llap{14}8.20\,$\pm$\,3.771&\llap{21}2.90\,$\pm$\,3.233&\llap{2}61.7\,$\pm$\,7.766&\llap{1}070.0\,$\pm$\, 5.392& 919.3\,$\pm$\, 5.637       & 3052\,$\pm$\, 63.57        &1113\,$\pm$\,230.80 &YSO (5), MIPS-SED (3)\\
MSX\,LMC\,1786\tablenotemark{a}    &\llap{1}1.20\,$\pm$\,0.500&\llap{3}4.20\,$\pm$\,0.784 &\llap{8}8.10\,$\pm$\,1.782 &\llap{1}42.0\,$\pm$\,2.200&\llap{1}444.0\,$\pm$\, 9.927& 906.0\,$\pm$\,11.460	     & 4139\,$\pm$\,231.90	 &2002\,$\pm$\,636.10 &30\,Dor$-$17, MIPS-SED (3)\\
SSTISAGE1C\,J054059.29$-$704402.6  & 4.51\,$\pm$\,0.202       &\llap{1}0.30\,$\pm$\,0.297 &\llap{1}7.20\,$\pm$\,0.480 & 29.5\,$\pm$\,0.581	   & 209.7\,$\pm$\, 1.055	& 216.0\,$\pm$\, 1.394       &  945\,$\pm$\, 12.78	  &\nodata	       &YSO (1)\\
IRAS\,05421$-$7116                 & 3.70\,$\pm$\,0.272       &  4.02\,$\pm$\,0.120       &  7.55\,$\pm$\,0.311       & 13.3\,$\pm$\,0.949       & 331.4\,$\pm$\, 1.770       & 329.0\,$\pm$\, 2.454       & 2040\,$\pm$\, 22.66        &\nodata             &YSO (1)\\
IRAS\,05452$-$6924                 & 2.36\,$\pm$\,0.081       &  3.62\,$\pm$\,0.114       &  8.58\,$\pm$\,0.176       & 19.4\,$\pm$\,0.406       & 674.9\,$\pm$\, 3.158       & 659.3\,$\pm$\, 3.609       & 2967\,$\pm$\, 47.63        &\nodata             &YSO (1)\\
\enddata
\tablenotetext{a}{sources from the \spitzer\ archive; all other sources are from
the SAGE-Spec program. $^b$ two epochs of 24-$\mu$m photometry are available.}
\tablerefs{(1) \citet{whitney08}; (2) \citet{shimonishi08}; (3) \citet{vanloon09}; (4)
\citet{kastner08}; (5) \citet{vanloon05a}.}
\end{deluxetable}

\begin{figure}
\begin{center}
\includegraphics[scale=0.85]{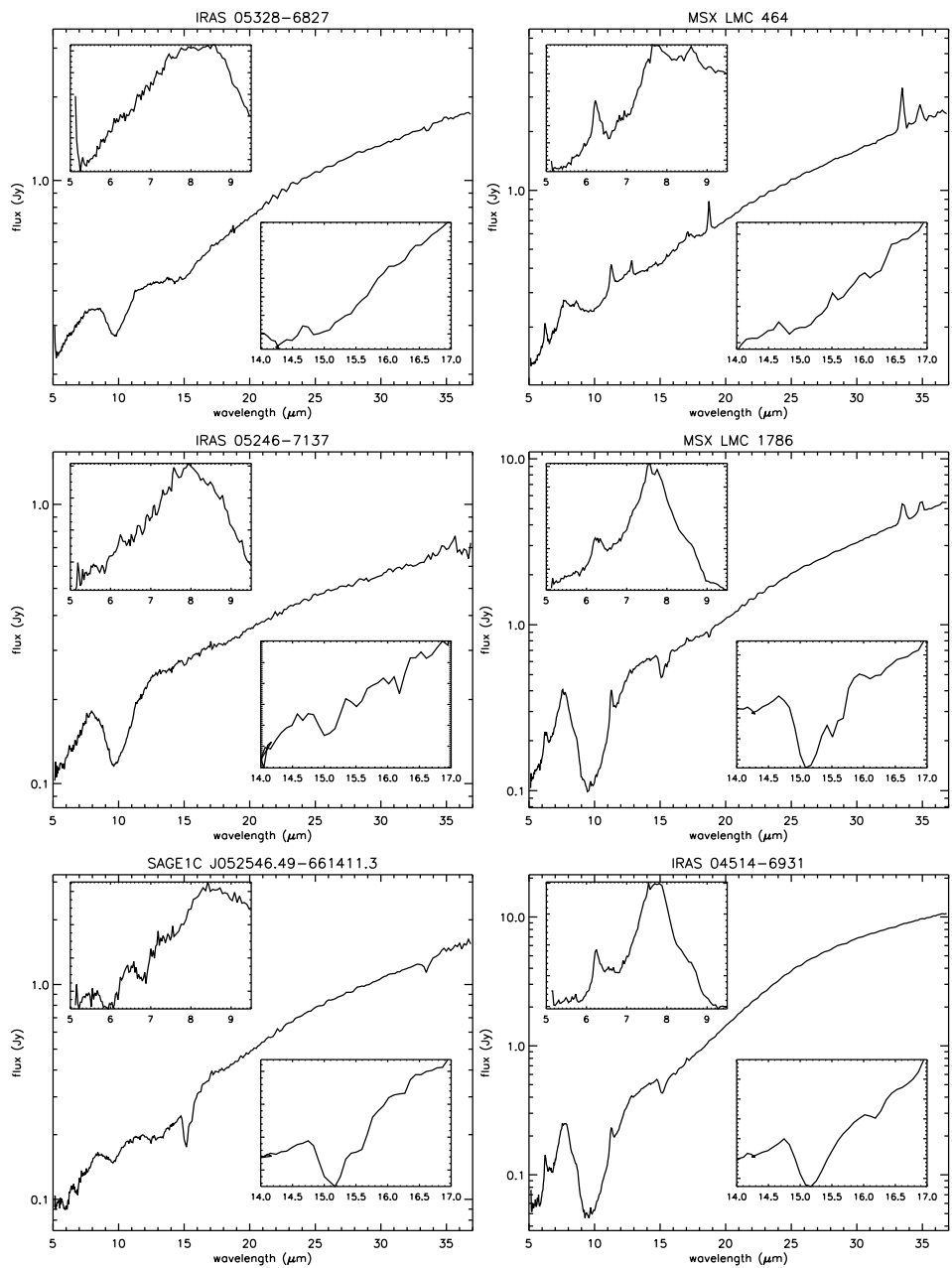}
\caption{\spitzer-IRS spectra of the 15 sources discussed in this work. From 
left to right and top to bottom, sources are shown in decreasing order of their
measured IRAC flux at 5.8\,$\mu$m. The insets in each graph show the spectral 
regions where ice bands can be found, namely the CO$_2$ ice feature at 
15.2\,$\mu$m (lower right corner) and the 5\,$-$\,7\,$\mu$m complex that 
includes the water ice O$-$H bending mode at 6\,$\mu$m (upper left corner). 
\label{spectra}}
\end{center}
\end{figure}

\begin{figure}
\ContinuedFloat
\begin{center}
\includegraphics[scale=0.85]{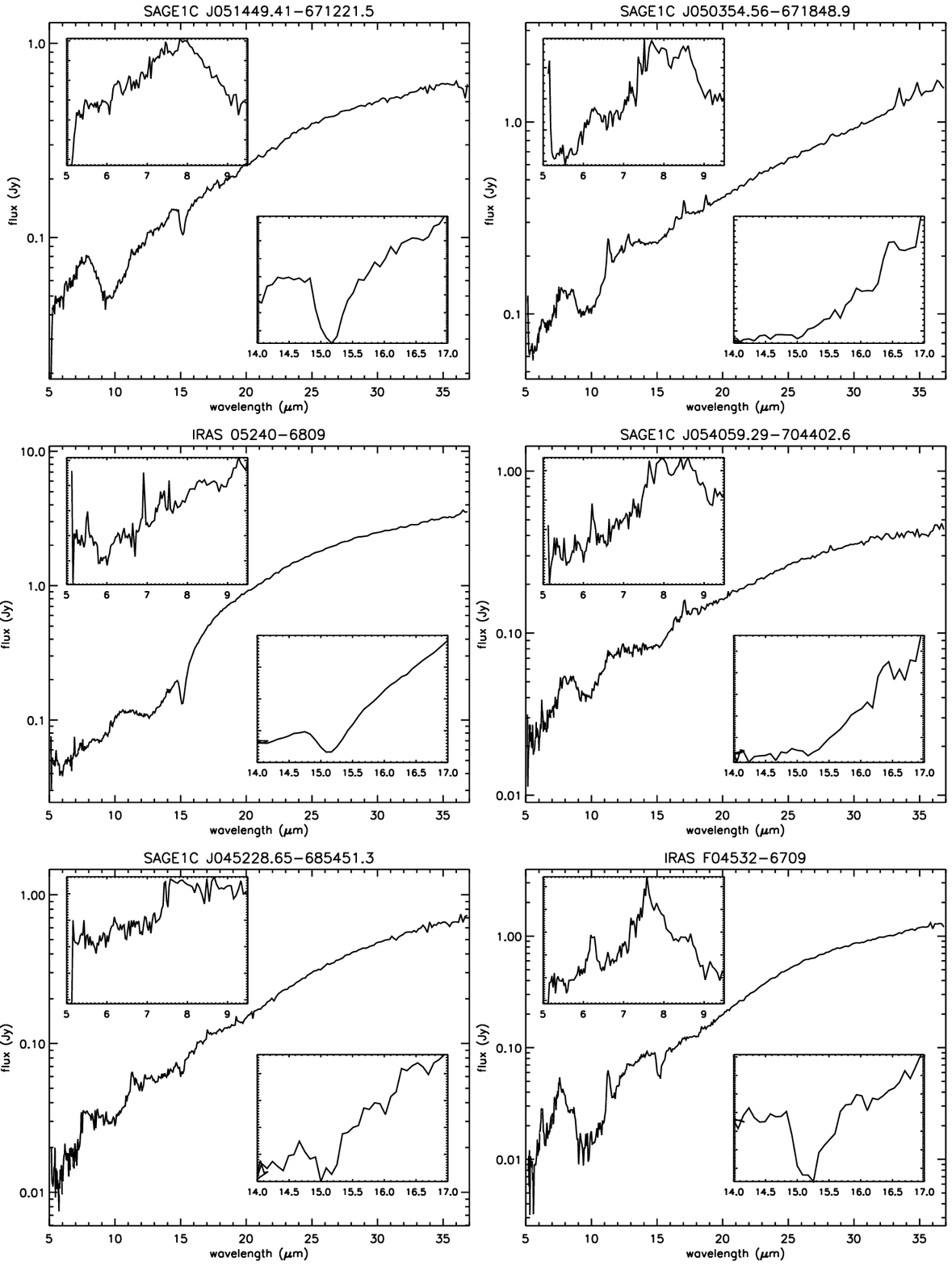}
\caption{Continued.}
\end{center}
\end{figure}

\begin{figure}
\ContinuedFloat
\begin{center}
\includegraphics[scale=0.85]{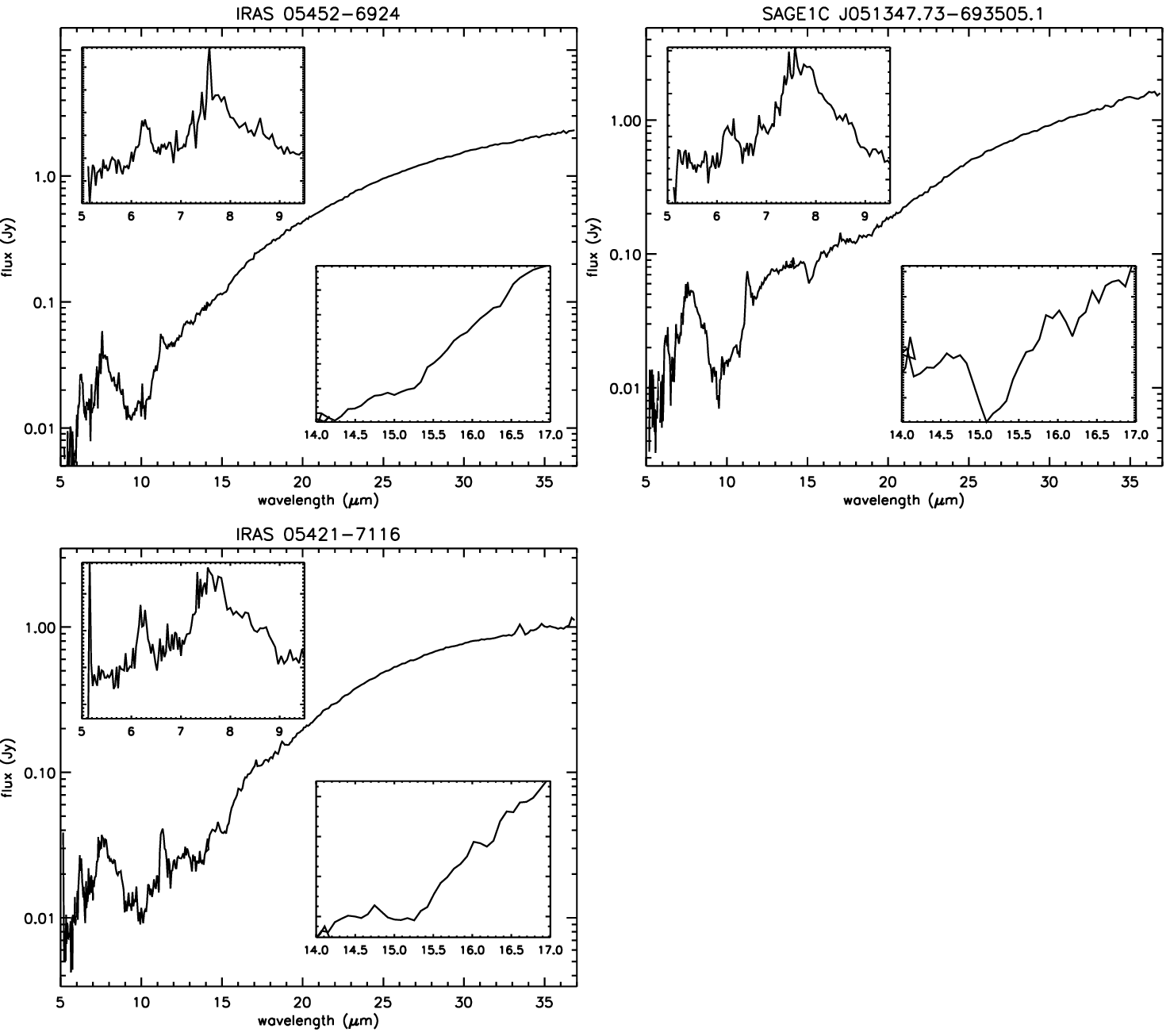}
\caption{Continued.}
\end{center}
\end{figure}

\begin{figure}
\begin{center}
\includegraphics[scale=0.68]{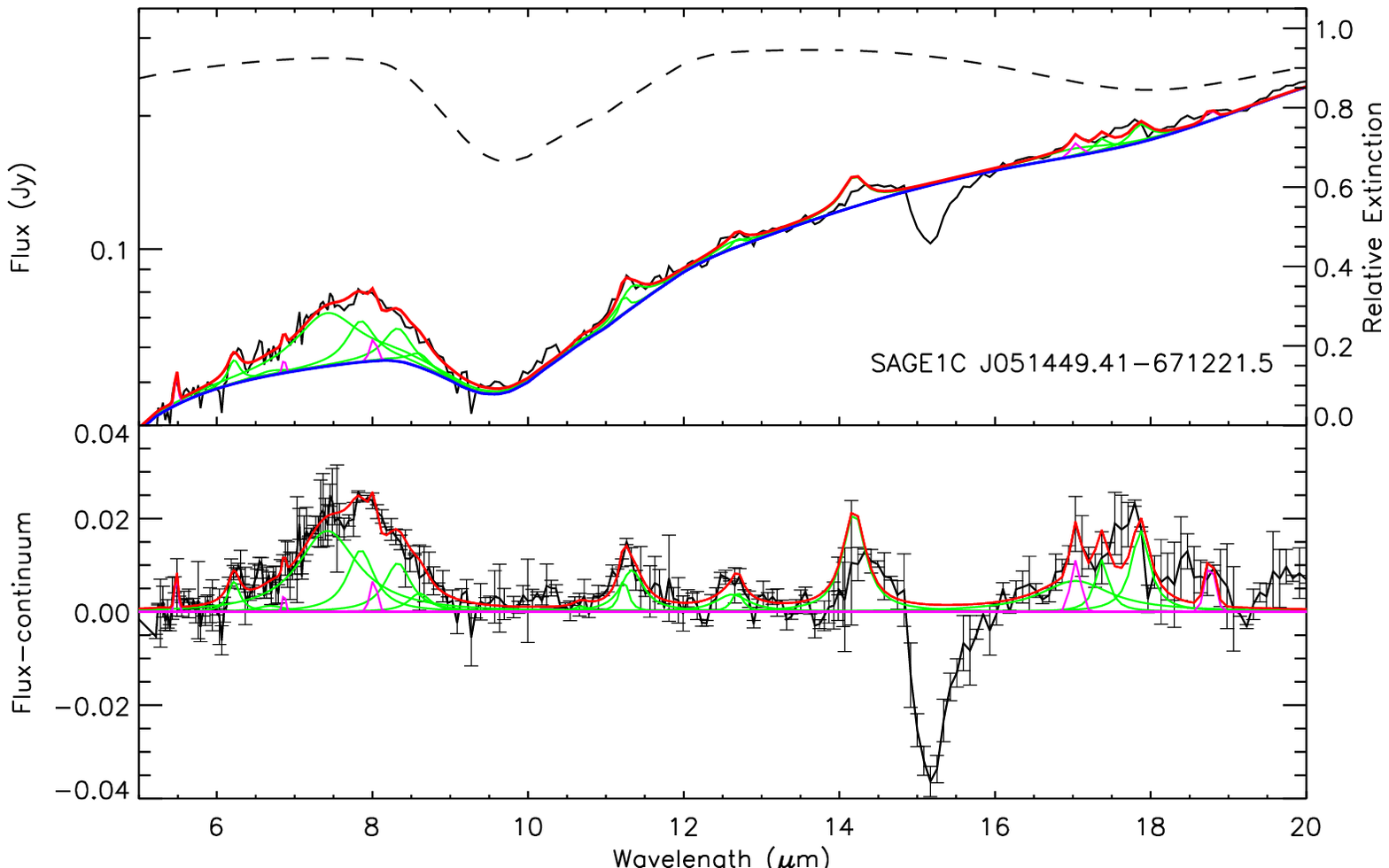}\\
\includegraphics[scale=0.68]{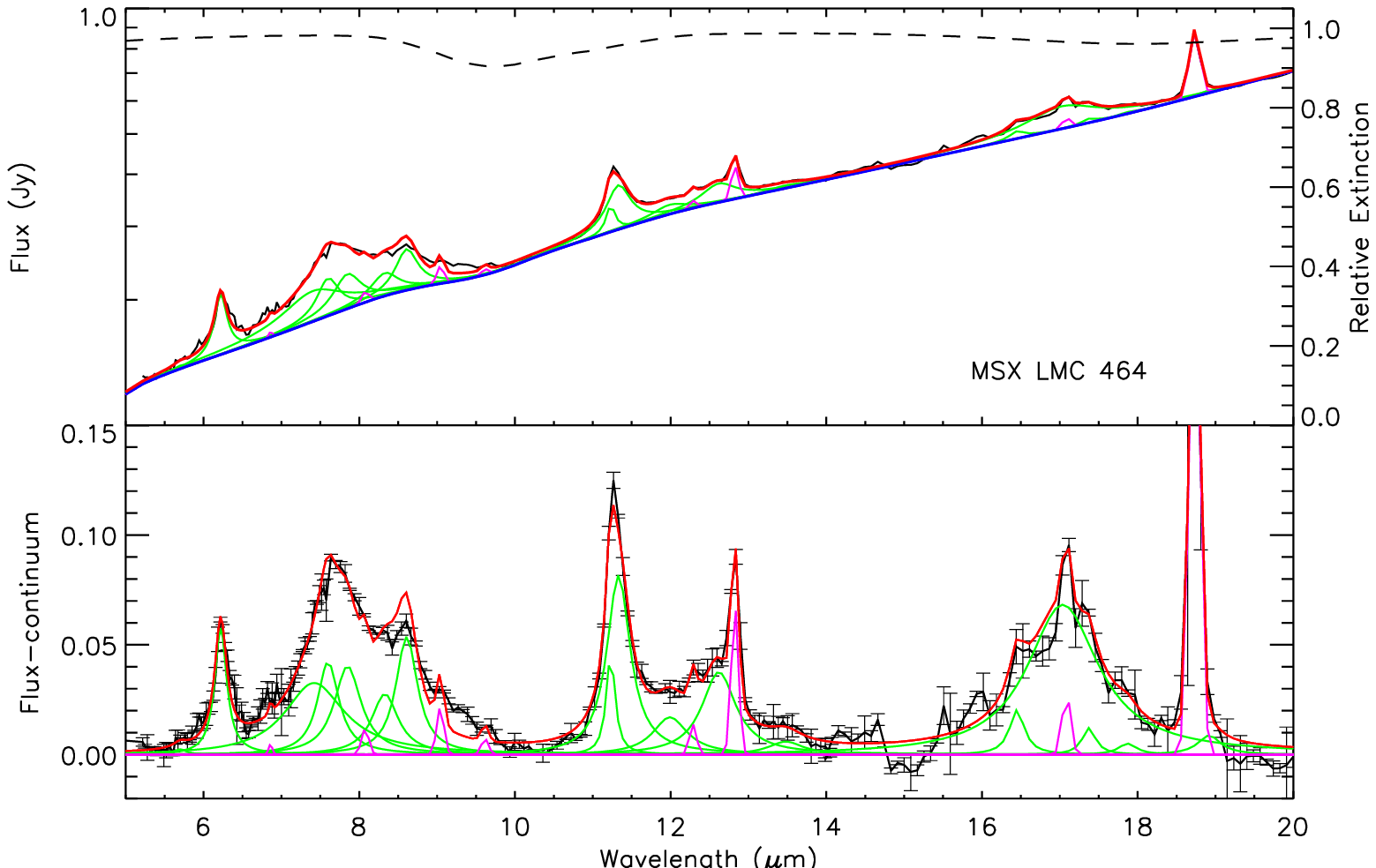}\\
\includegraphics[scale=0.68]{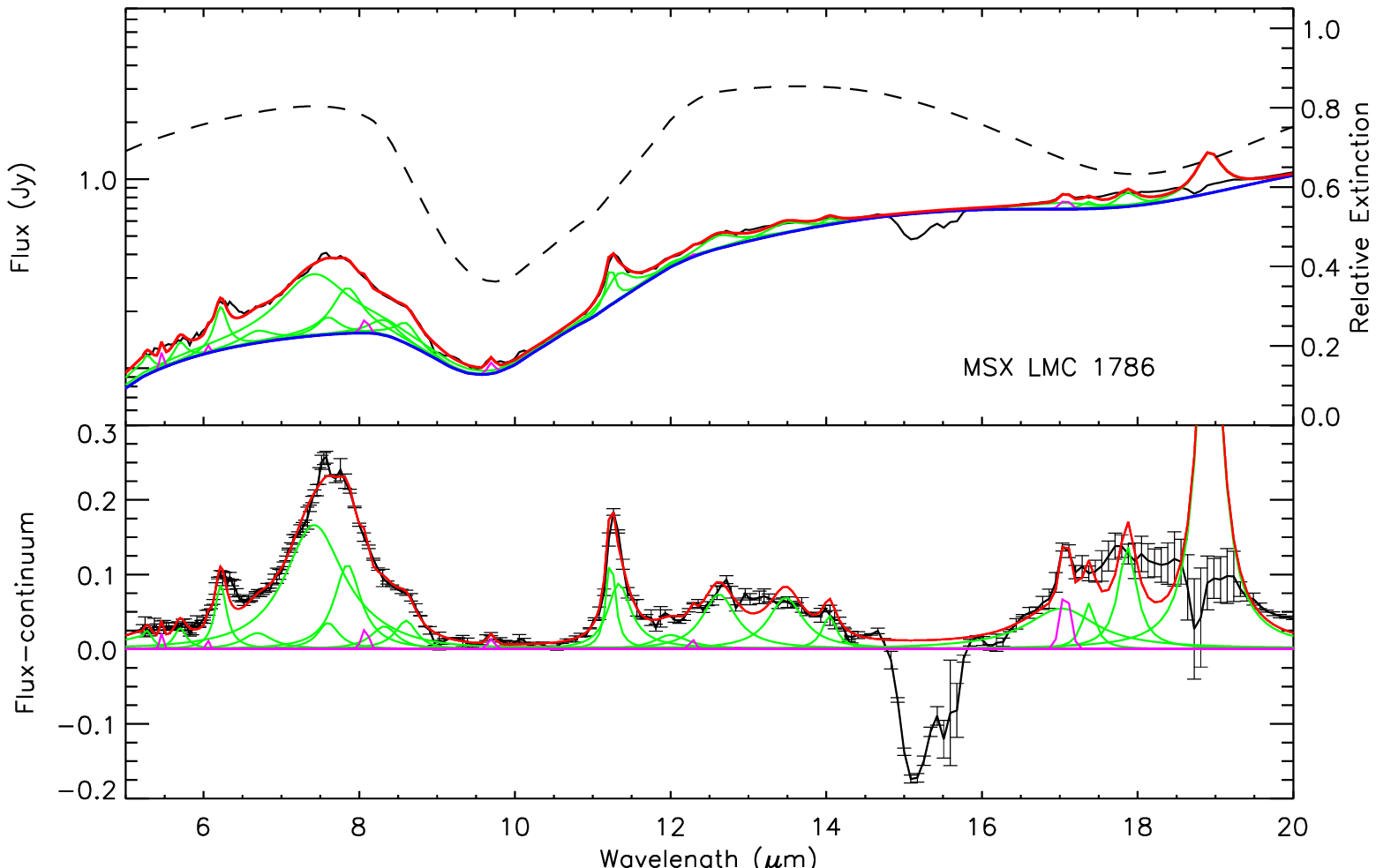}
\caption{Fits to YSO spectra using PAHFIT: spectra with all fitted components
(upper panels) and continuum-subtracted spectra (lower panels). The total fit is
shown (red) as well as the individual components of the fit: dust continuum 
(blue), PAH emission (green) and atomic and molecular emission lines (magenta) 
--- in some cases there is a very small contribution from starlight. Also shown
is the silicate dust relative extinction as fitted by PAHFIT (right-hand axis).
\label{PAHFIT}}
\end{center}
\end{figure}

\begin{figure}
\ContinuedFloat
\begin{center}
\includegraphics[scale=0.7]{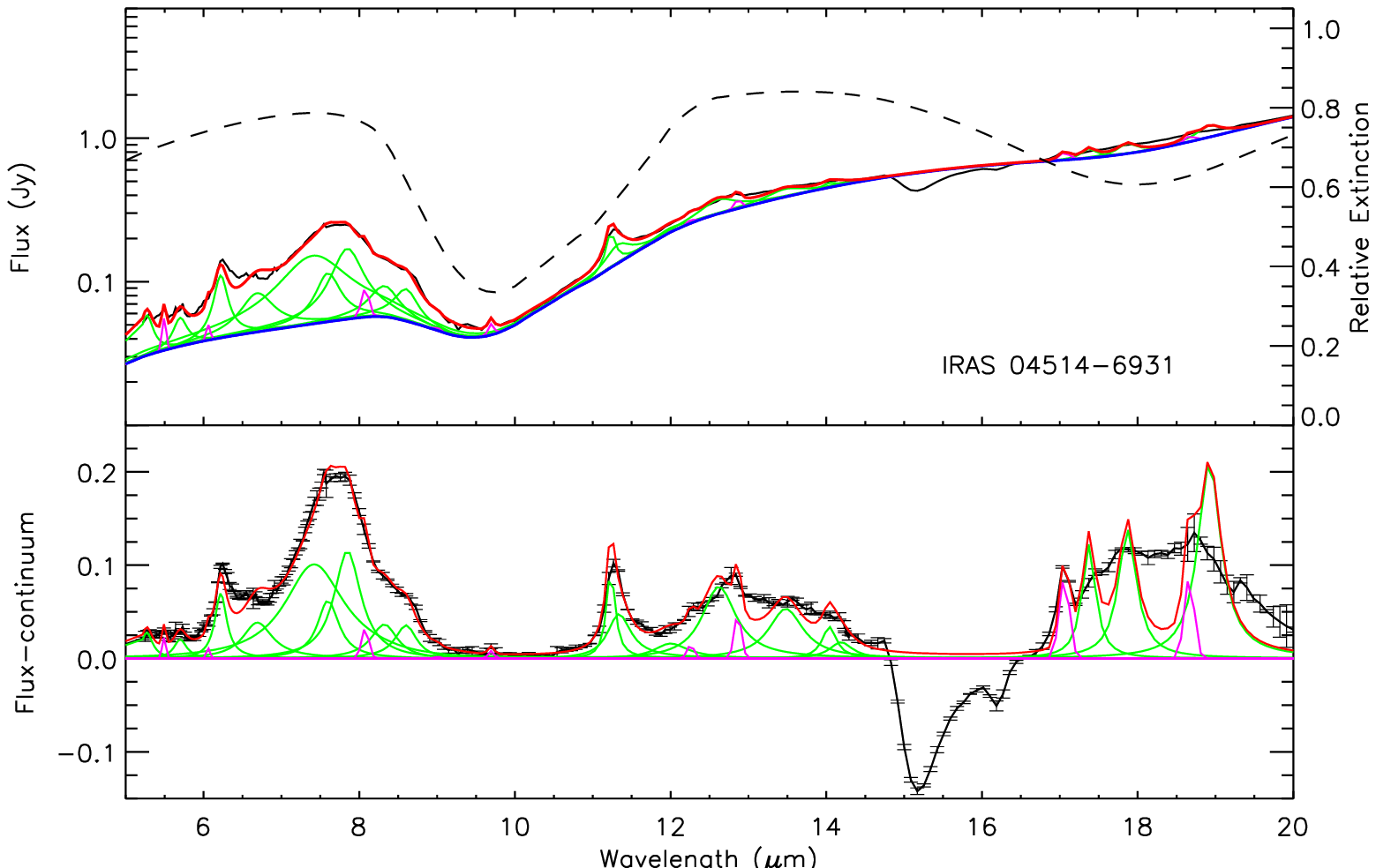}\\
\includegraphics[scale=0.7]{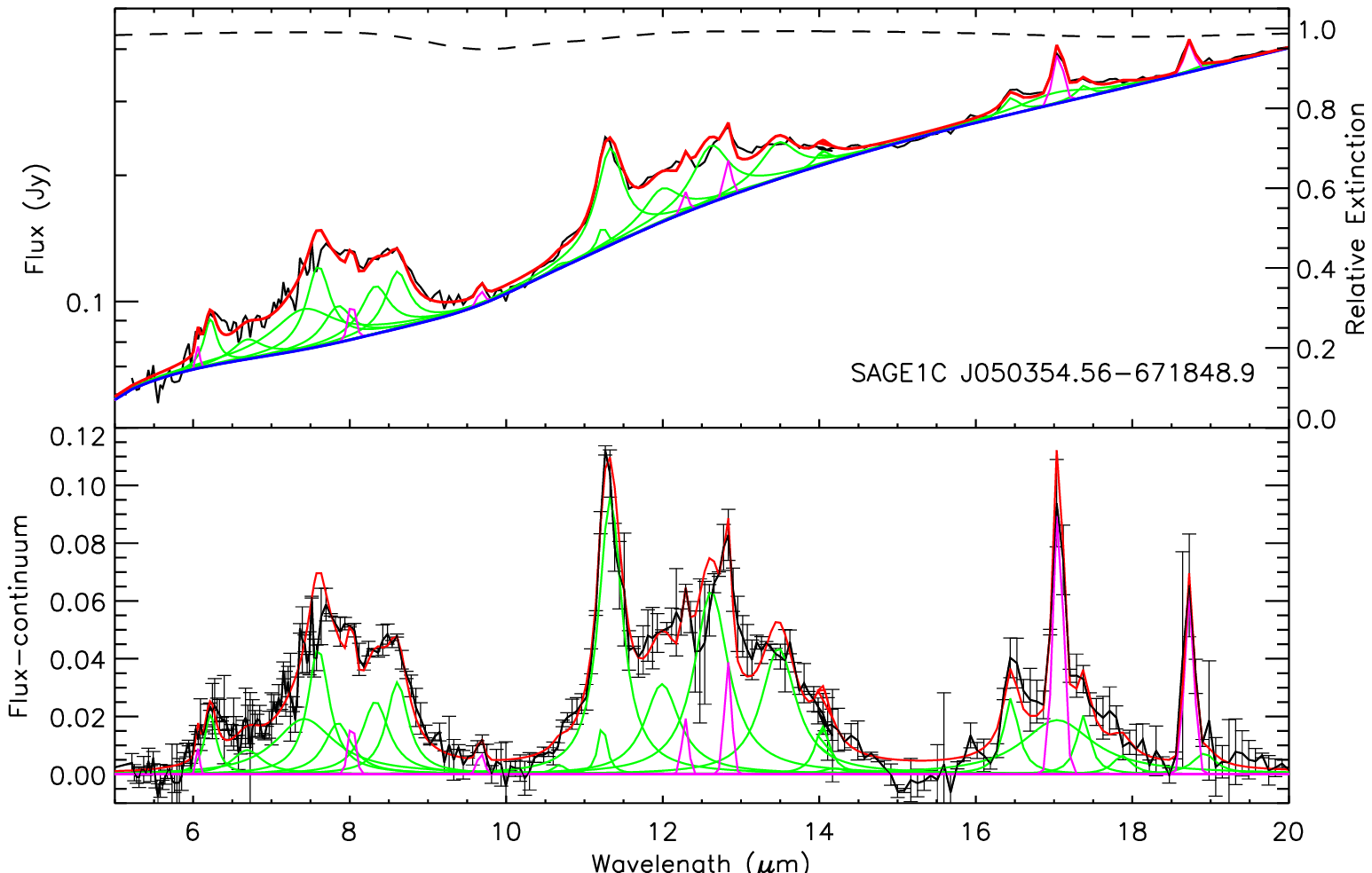}\\
\includegraphics[scale=0.7]{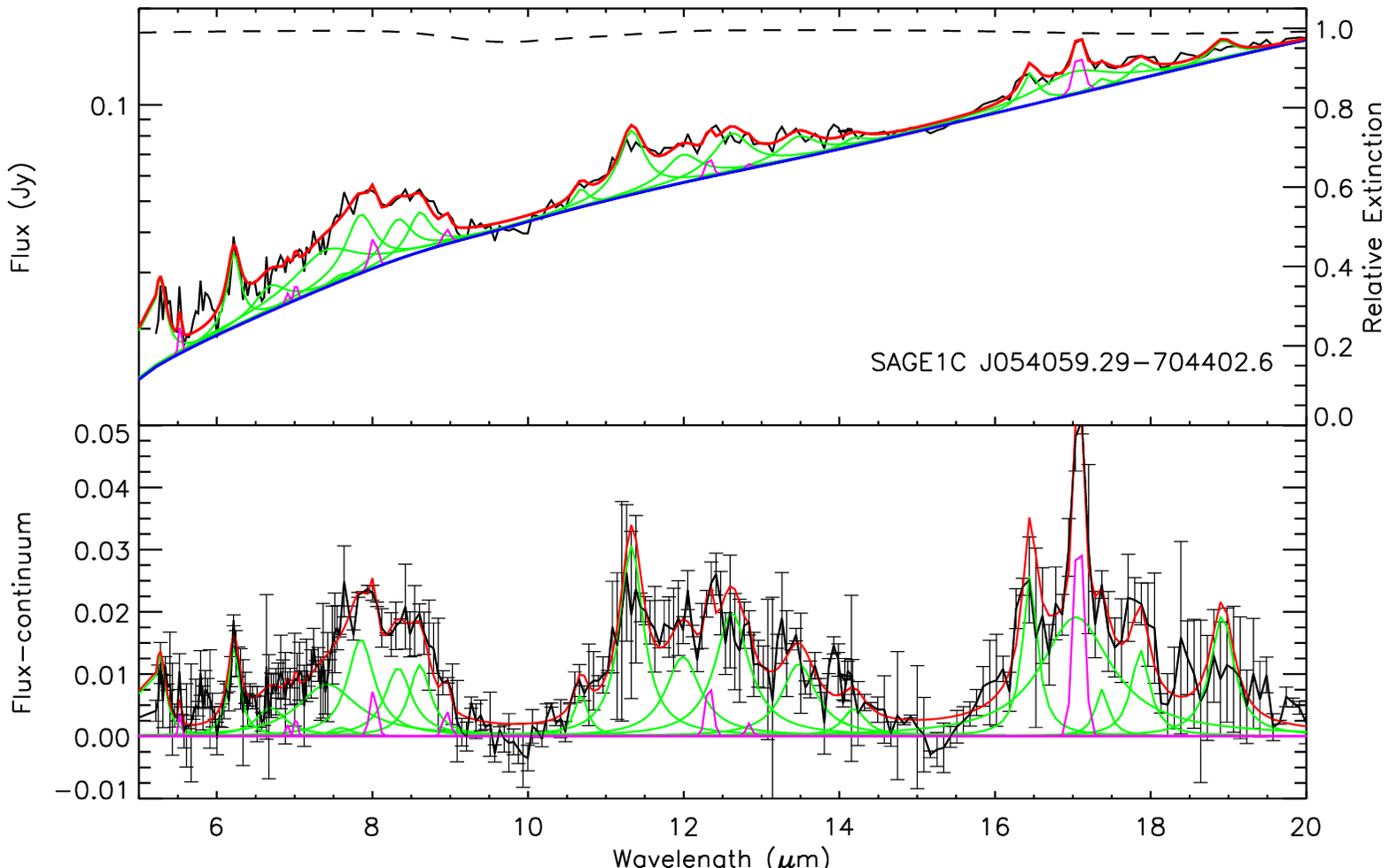}
\caption{Continued.}
\end{center}
\end{figure}

\begin{figure}
\ContinuedFloat
\begin{center}
\includegraphics[scale=0.7]{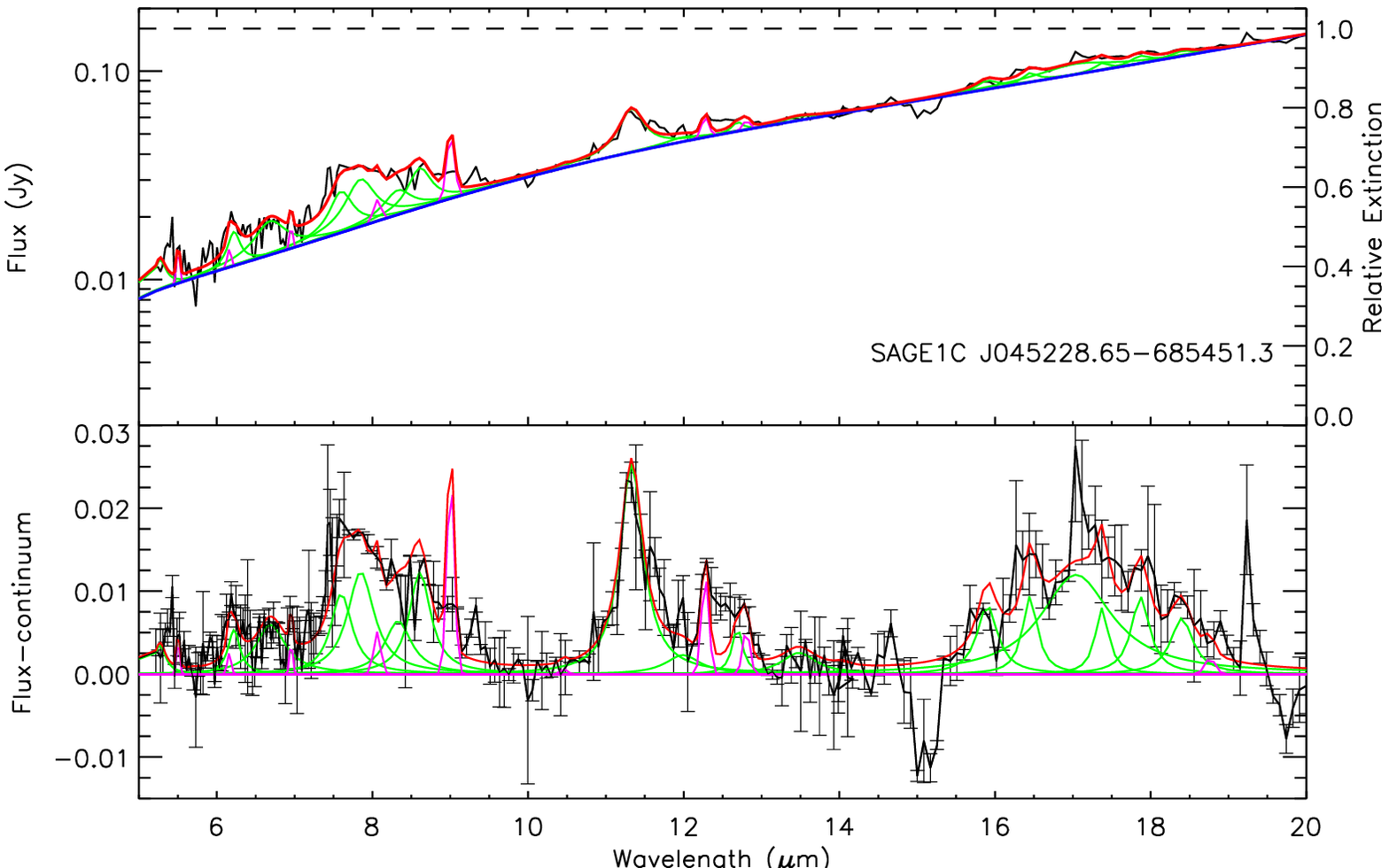}\\
\includegraphics[scale=0.7]{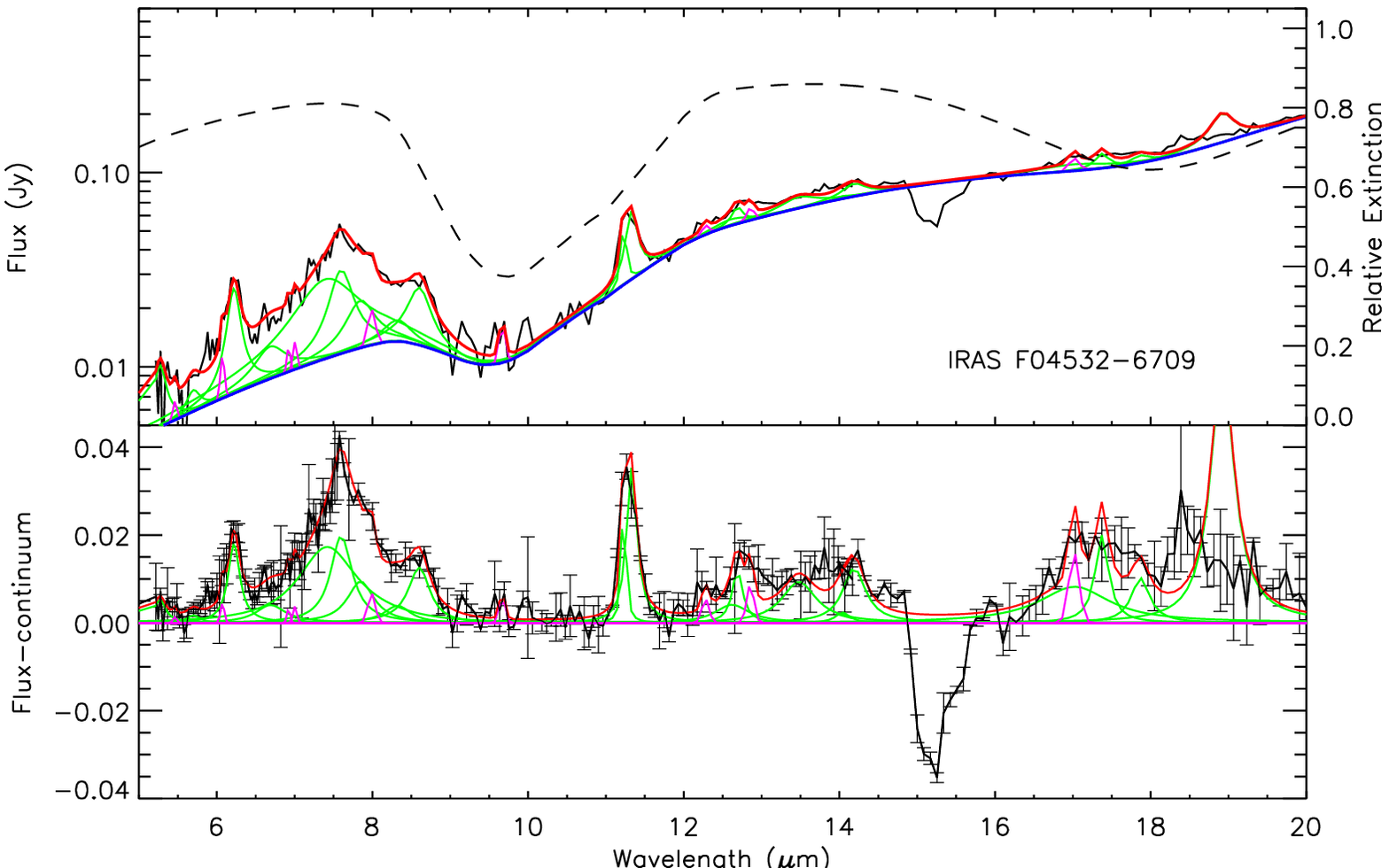}\\
\includegraphics[scale=0.7]{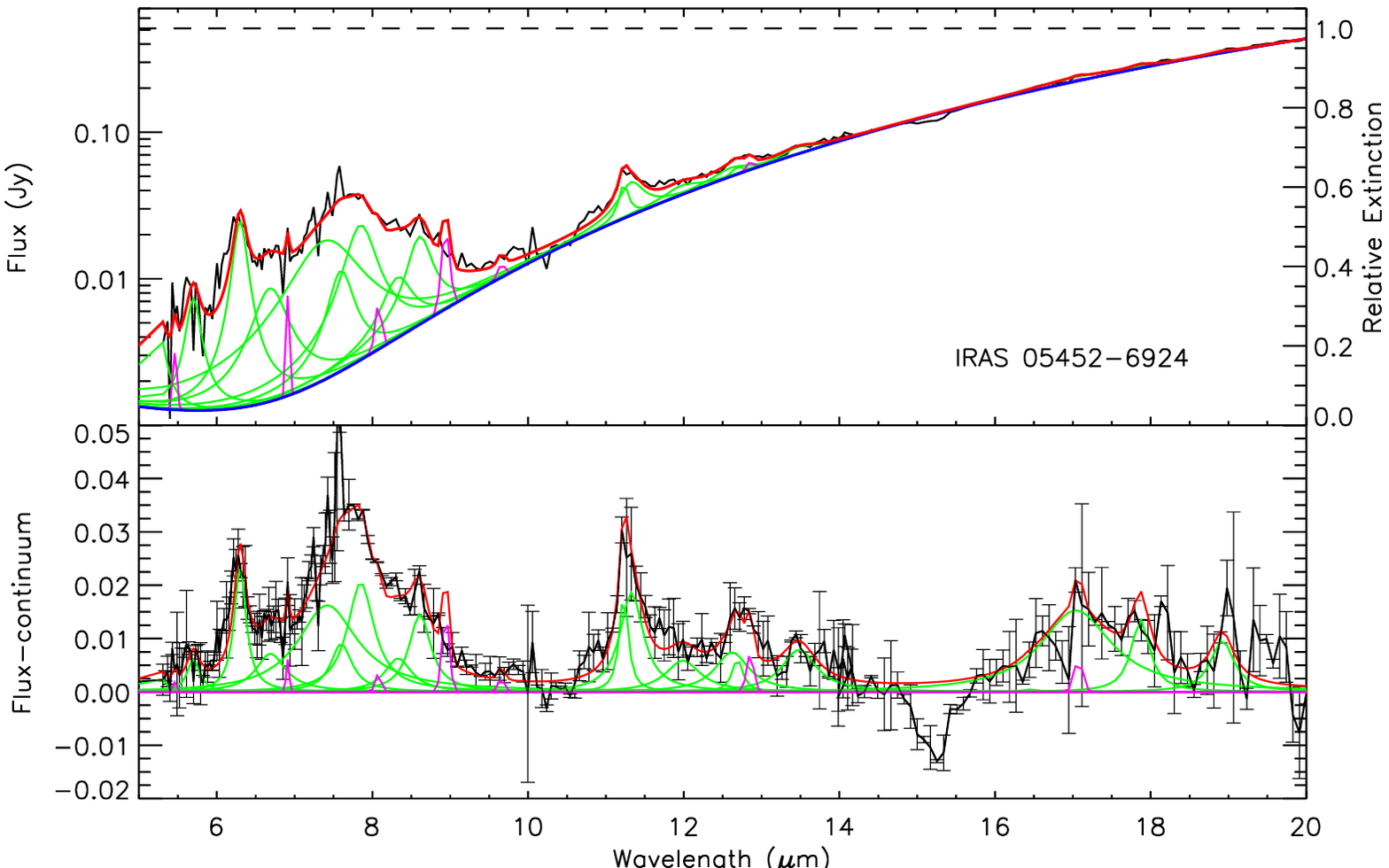}
\caption{Continued.}
\end{center}
\end{figure}

\begin{figure}
\ContinuedFloat
\begin{center}
\includegraphics[scale=0.7]{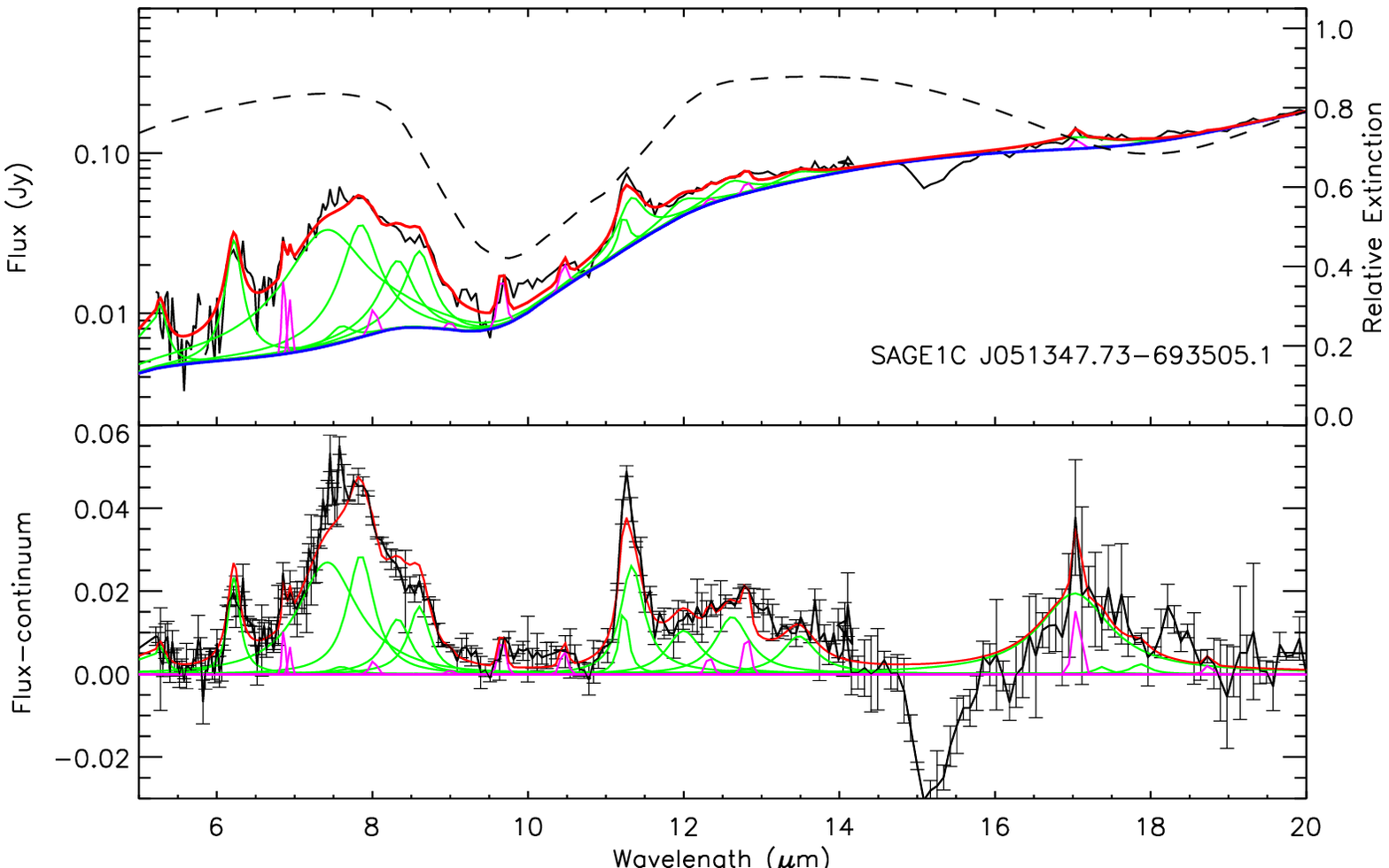}\\
\includegraphics[scale=0.7]{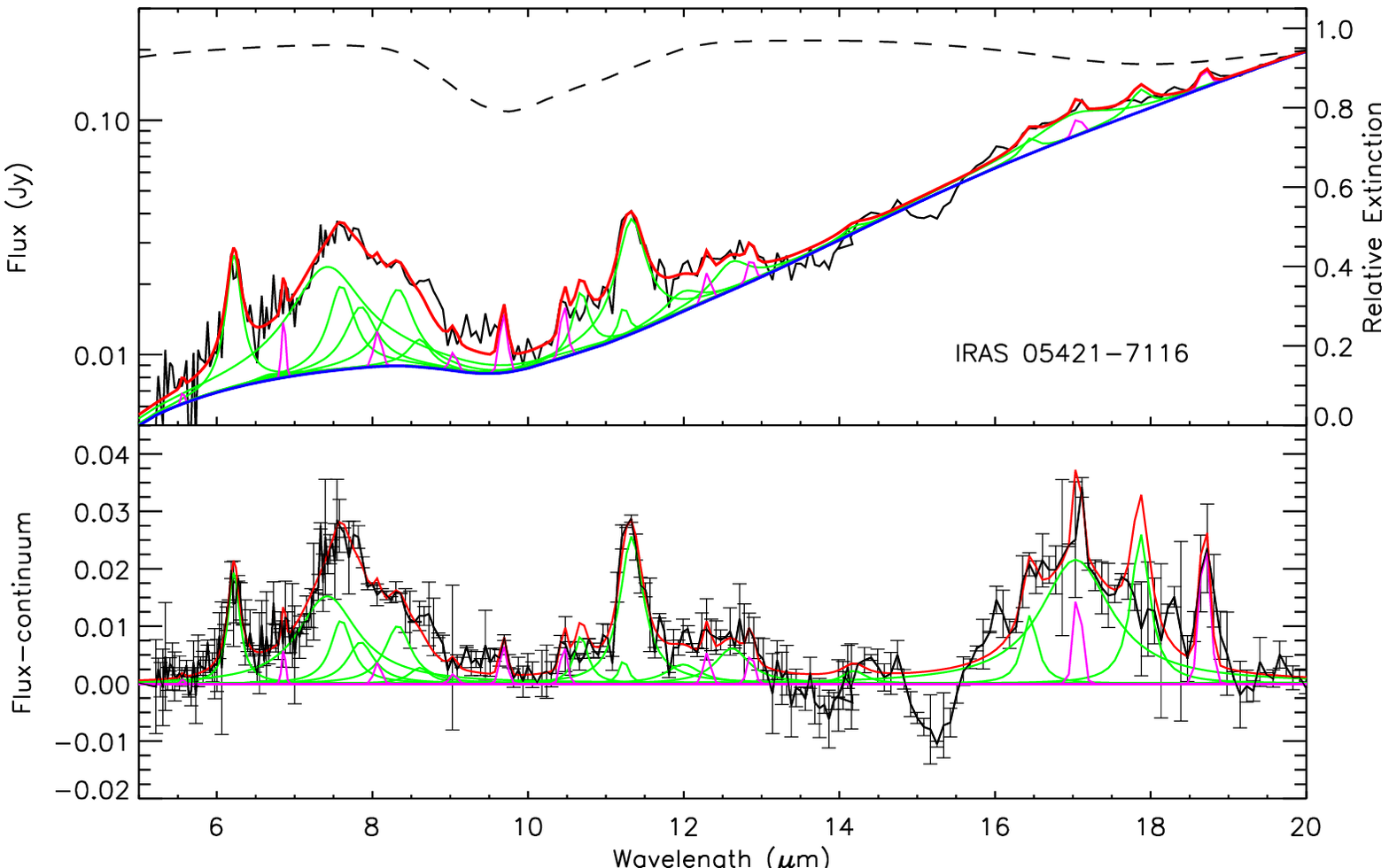}
\caption{Continued.}
\end{center}
\end{figure}

\begin{deluxetable}{lcc}
\tabletypesize{\scriptsize}
\tablecaption{Laboratory spectra used in the CO$_2$ ice fits from
\citet[][ E97]{ehrenfreund97} and \citet[][ W09]{white09}.
\label{lab}}
\tablewidth{0pt}
\tablehead{
mixture & temperature & reference \\
        &     (K)     &            }
\startdata
\multicolumn{3}{c}{H$_2$O-rich polar ices}\\
\hline
H$_2$O:CO$_2$\,=\,100:14 & 10 & E97\\
H$_2$O:CO$_2$:CO\,=\,100:20:3 & 20 & E97\\
\hline
\multicolumn{3}{c}{CO- or CO$_2$-rich apolar ices}\\
\hline
CO:CO$_2$\,=\,100:4 & 10 &E97\\
CO:CO$_2$\,=\,100:8 & 10 &E97\\
CO:CO$_2$\,=\,100:16 & 10 &E97\\
CO:CO$_2$\,=\,100:21 & 10 &E97\\
CO:CO$_2$\,=\,100:23 & 10 &E97\\
CO:CO$_2$\,=\,100:26 & 10 &E97\\
CO:CO$_2$\,=\,100:70 & 10 &E97\\
H$_2$O:CO:CO$_2$\,=\,1:50:56$^{a}$ & 10 &E97\\
CO:O$_2$:CO$_2$\,=\,100:50:4 & 10 &E97\\
CO:O$_2$:CO$_2$\,=\,100:50:8 & 10 &E97\\
CO:O$_2$:CO$_2$\,=\,100:50:16 & 10 &E97\\
CO:O$_2$:CO$_2$\,=\,100:50:21 & 10 &E97\\
CO:O$_2$:CO$_2$\,=\,100:50:32 & 10 &E97\\
pure CO$_2$&10&E97\\
pure CO$_2$&50&E97\\
pure CO$_2$&80&E97\\
\hline
\multicolumn{3}{c}{annealed ices}\\
\hline
H$_2$O:CH$_3$OH:CO$_2$&5, 10,..., 135$^{b}$&W09
\enddata
\tablenotetext{a}{This mixture contains minimal traces of water.}
\tablenotetext{b}{5-K temperature steps.}
\end{deluxetable}

\begin{figure}
\begin{center}
\includegraphics[scale=0.6]{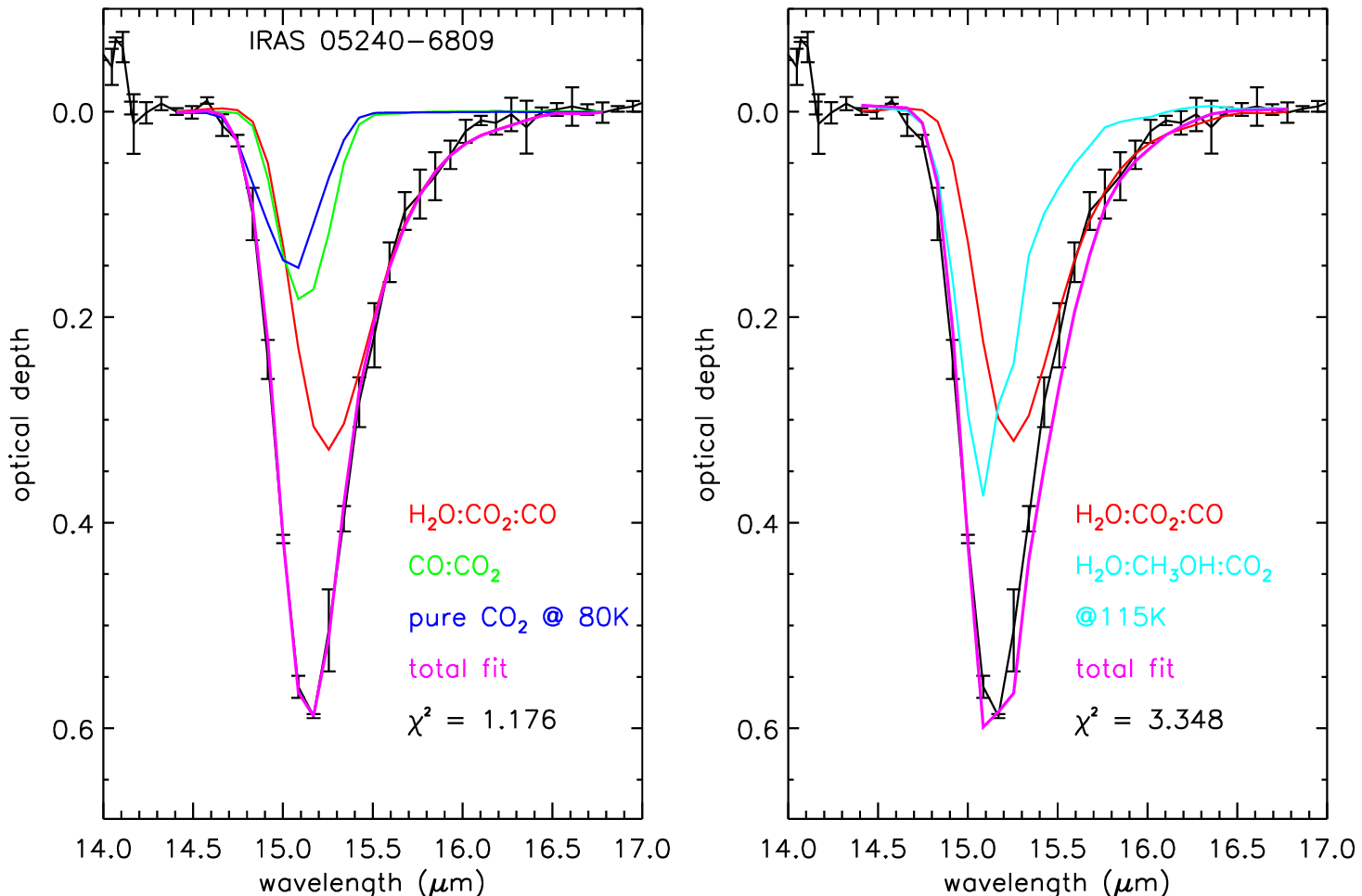}\\
\includegraphics[scale=0.6]{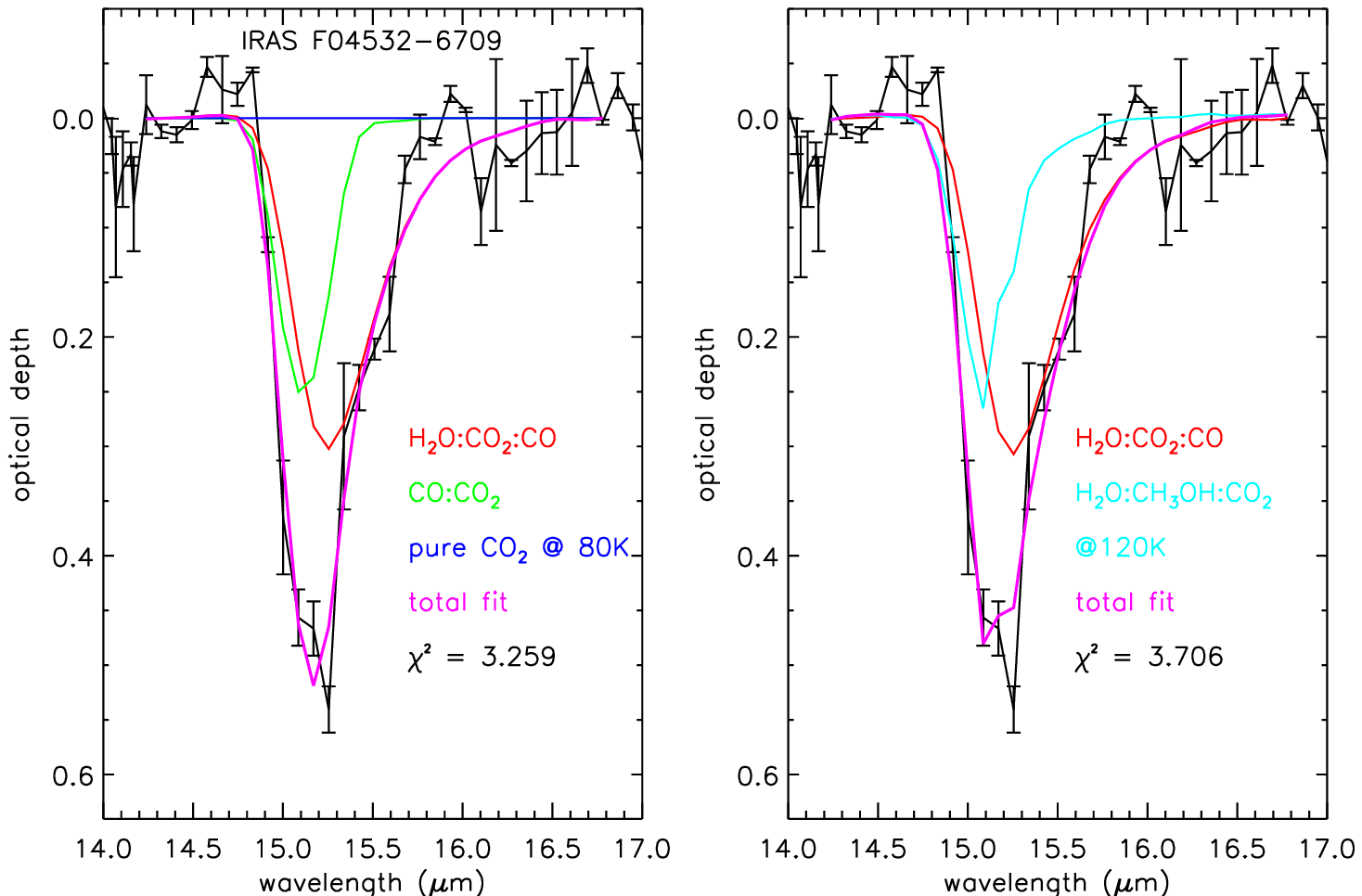}\\
\includegraphics[scale=0.6]{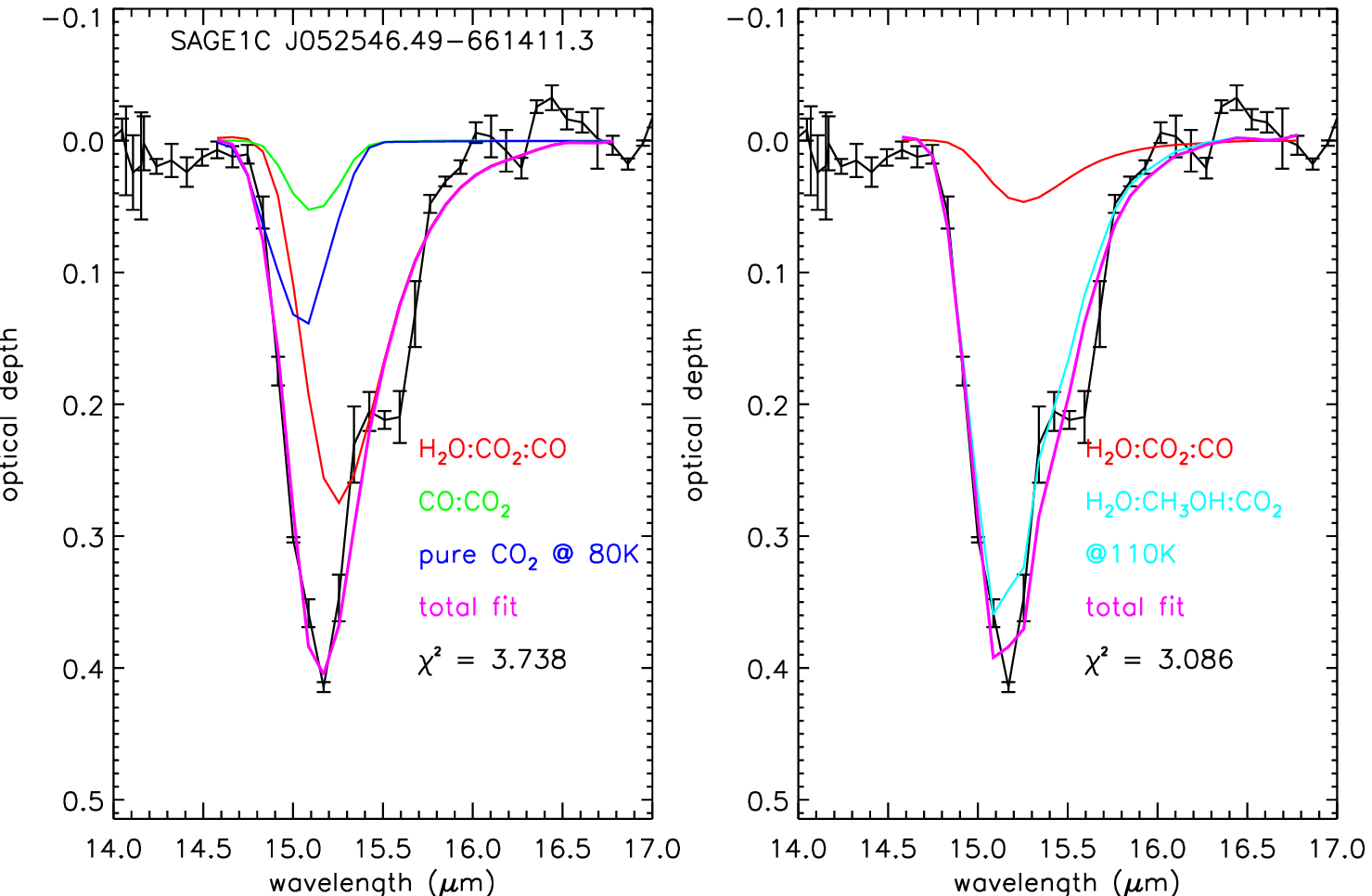}
\caption{CO$_2$ profiles and best fits to the observed profiles (magenta), using
a combination of either polar and apolar ices (left) or polar and annealed ices 
(right) --- see text for explanation. The different components are color-coded 
as follows: water-rich (red), CO-rich (green), pure CO$_2$ (blue) and annealed 
(cyan). Reduced $\chi^2$ values are also given. Objects are plotted in order of
decreasing total column density from top to bottom (Table\,\ref{fit_data}).
\label{ice_fits}}
\end{center}
\end{figure}

\begin{figure}
\begin{center}
\ContinuedFloat
\includegraphics[scale=0.6]{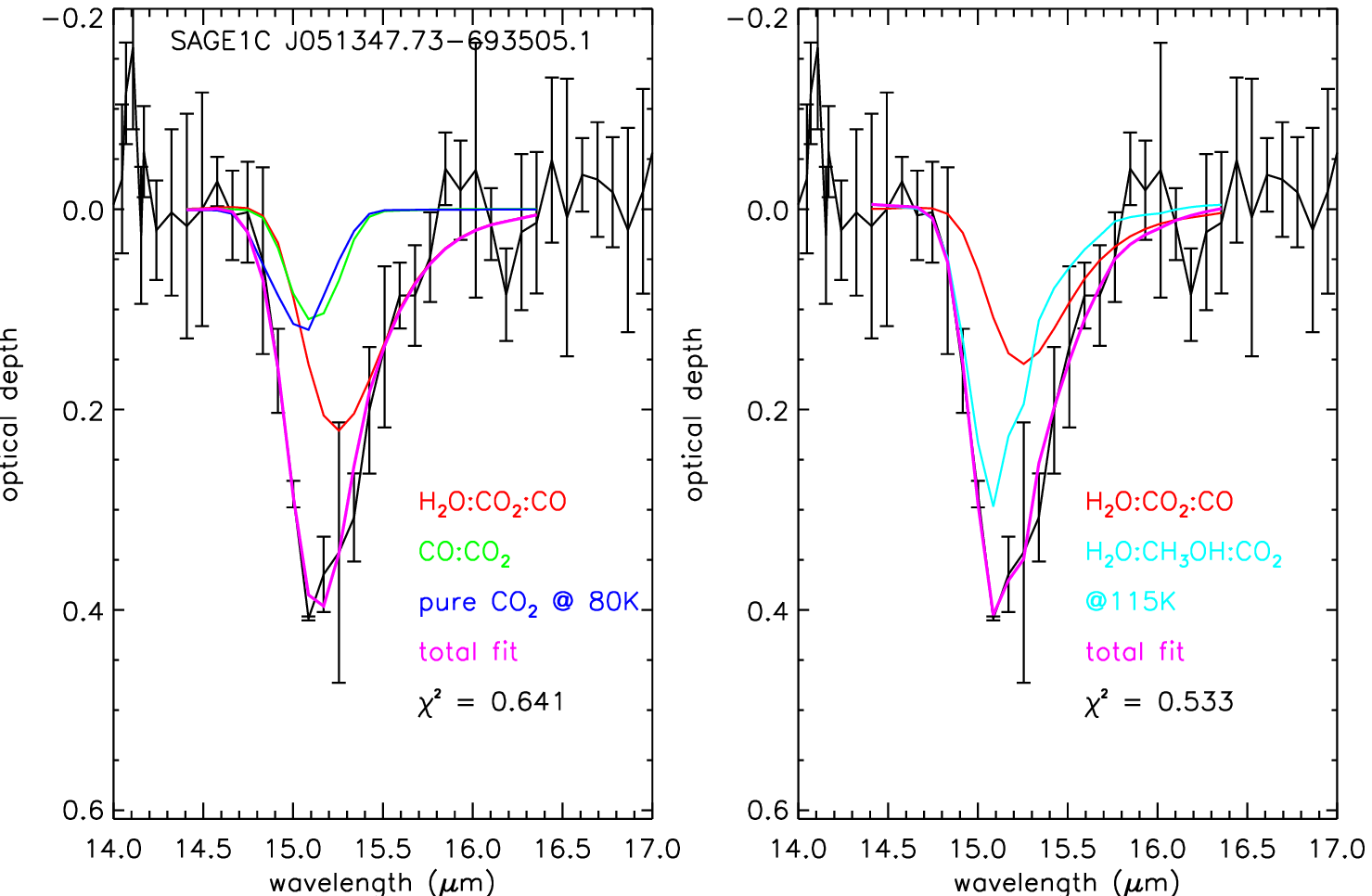}
\includegraphics[scale=0.6]{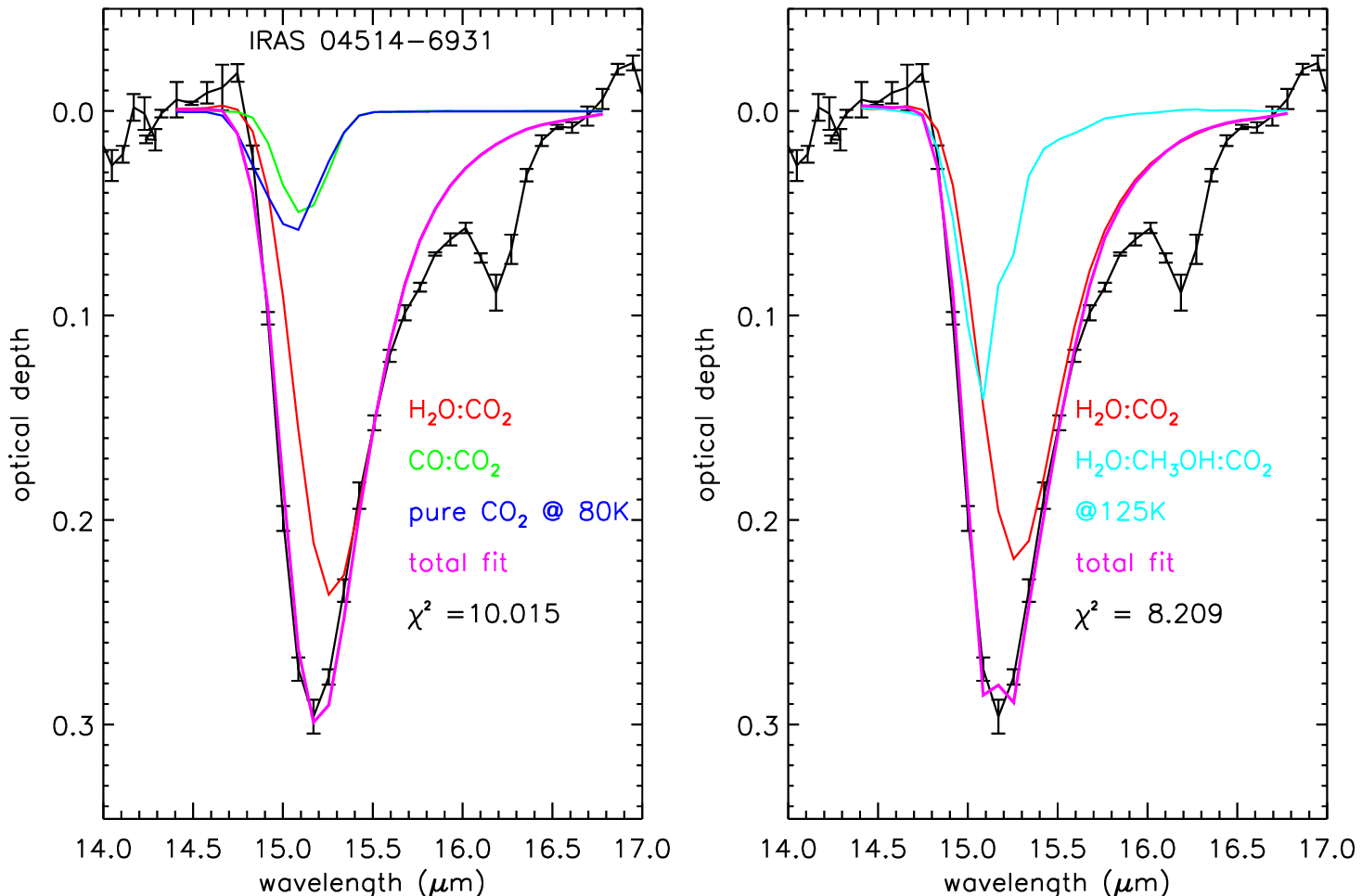}
\includegraphics[scale=0.6]{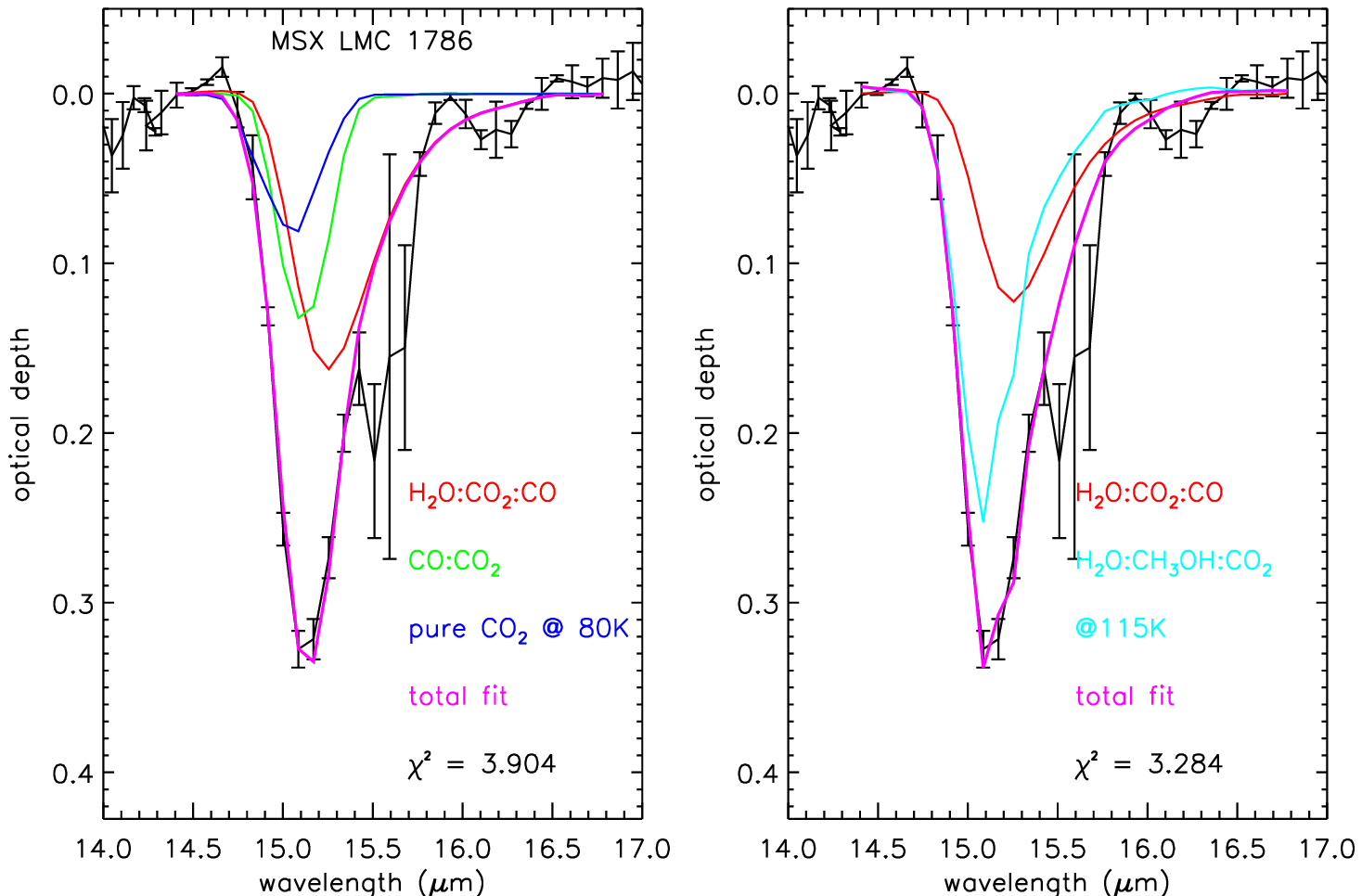}
\caption{Continued.}
\end{center}
\end{figure}

\begin{figure}
\begin{center}
\ContinuedFloat
\includegraphics[scale=0.6]{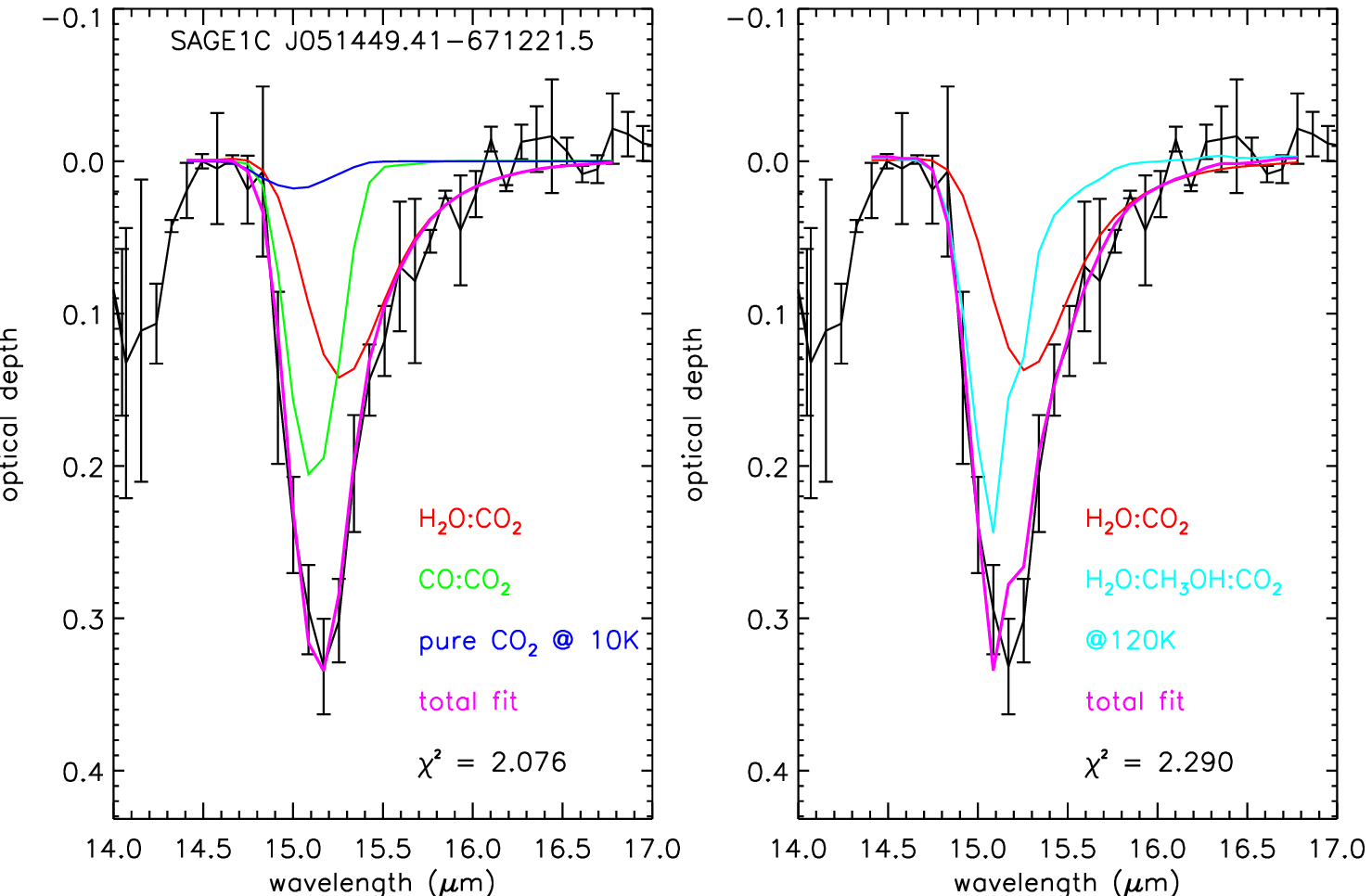}
\includegraphics[scale=0.6]{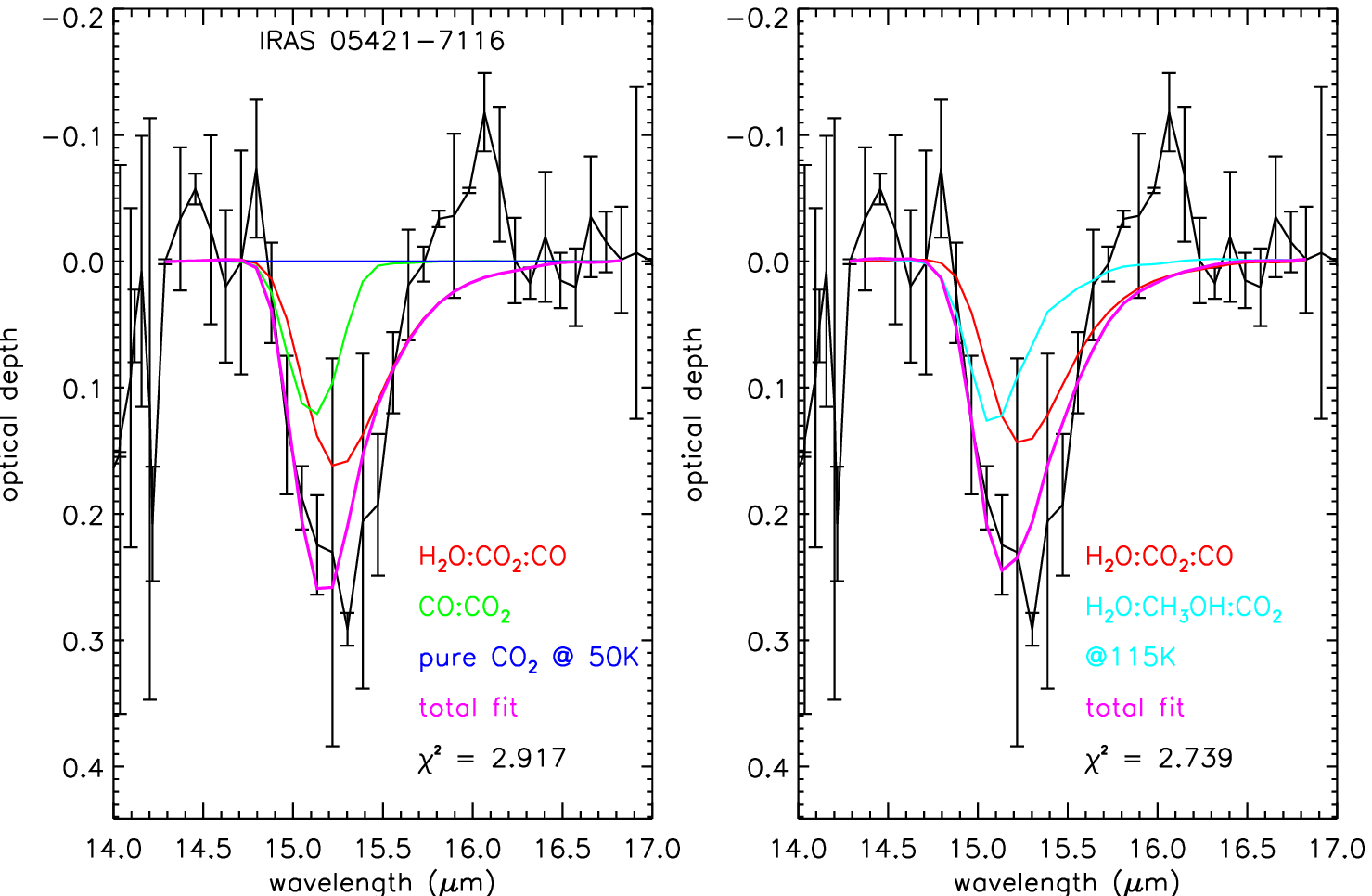}
\includegraphics[scale=0.6]{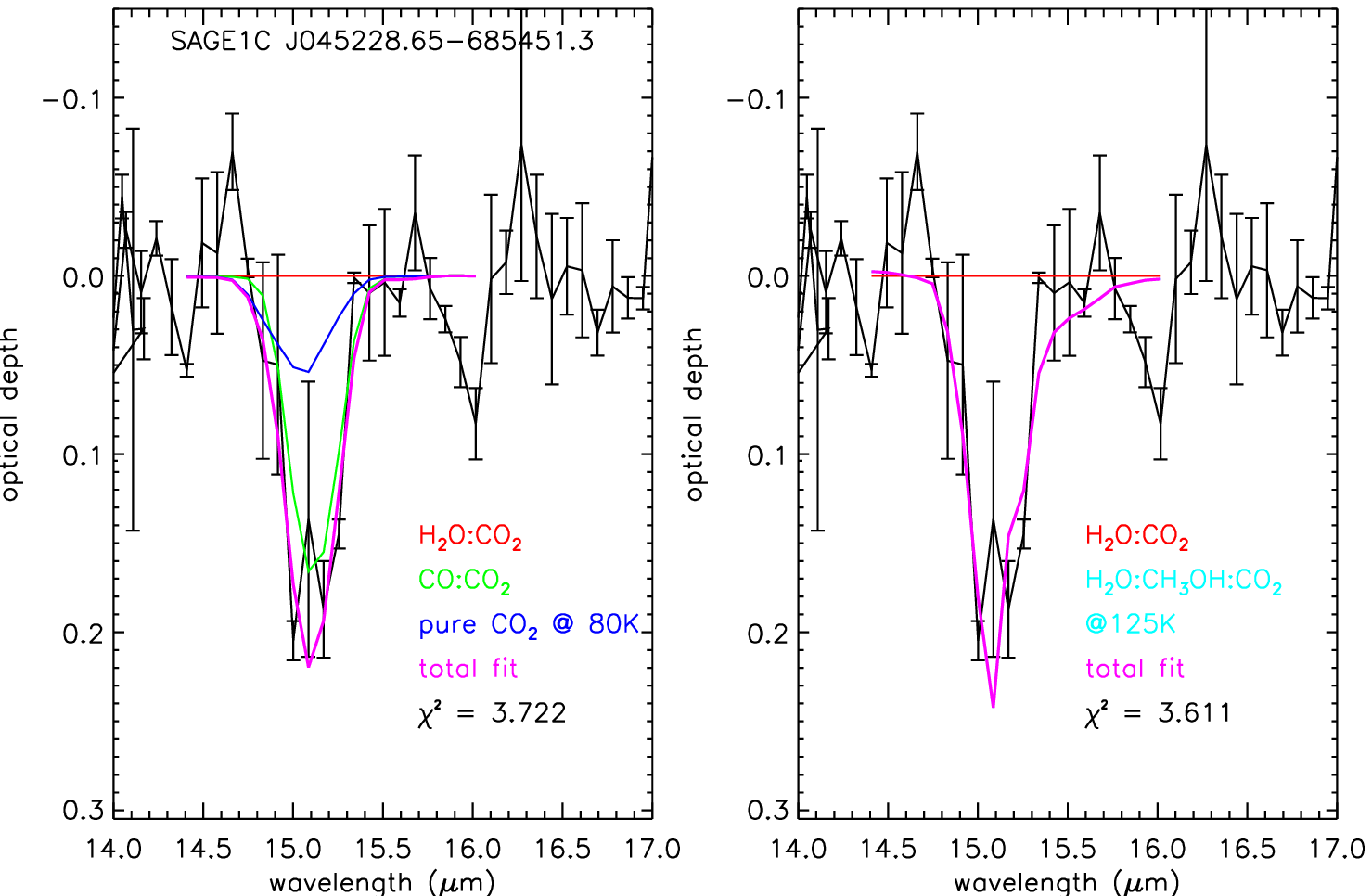}
\caption{Continued.}
\end{center}
\end{figure}

\begin{figure}
\begin{center}
\ContinuedFloat
\includegraphics[scale=0.6]{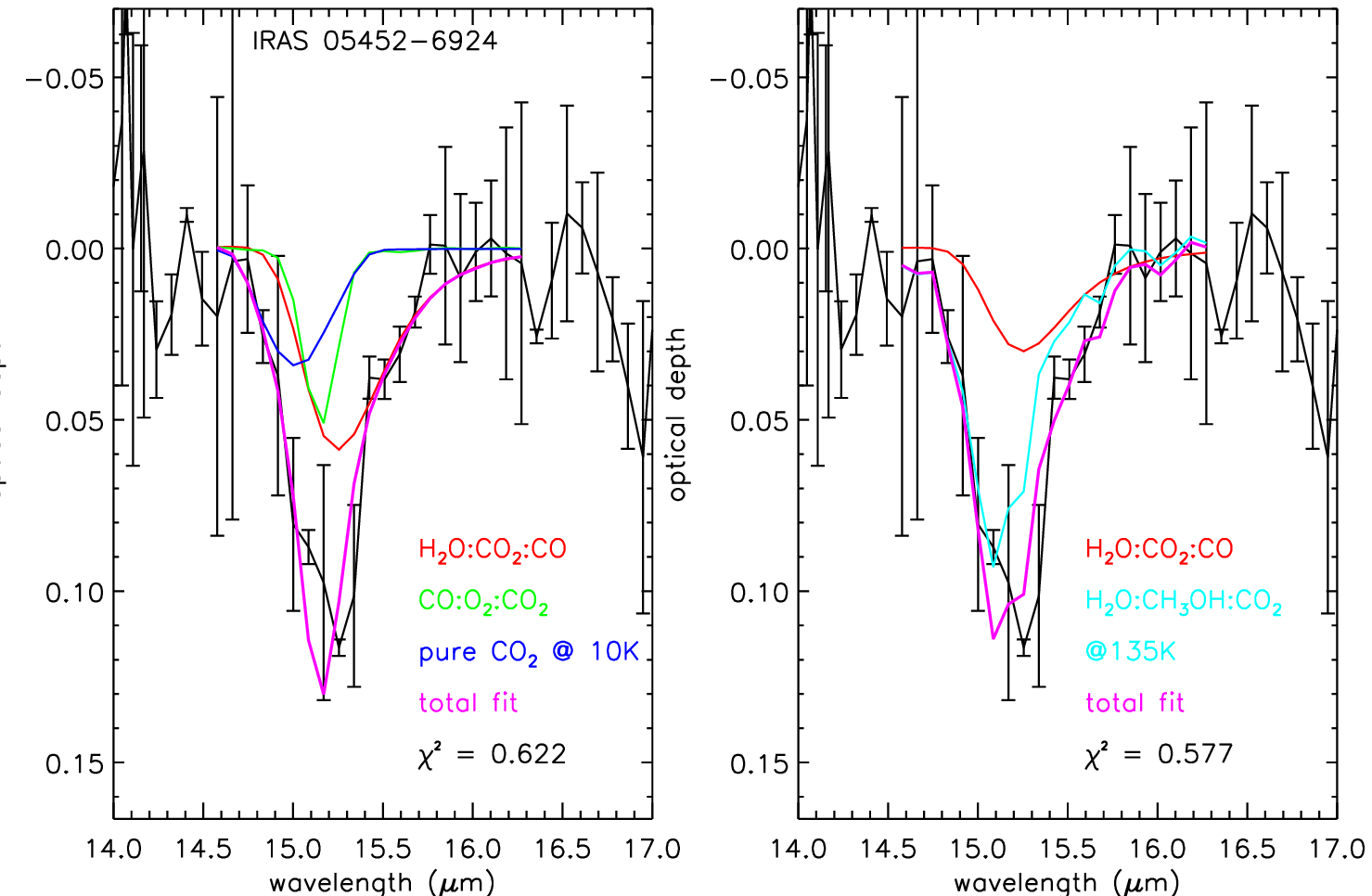}
\includegraphics[scale=0.6]{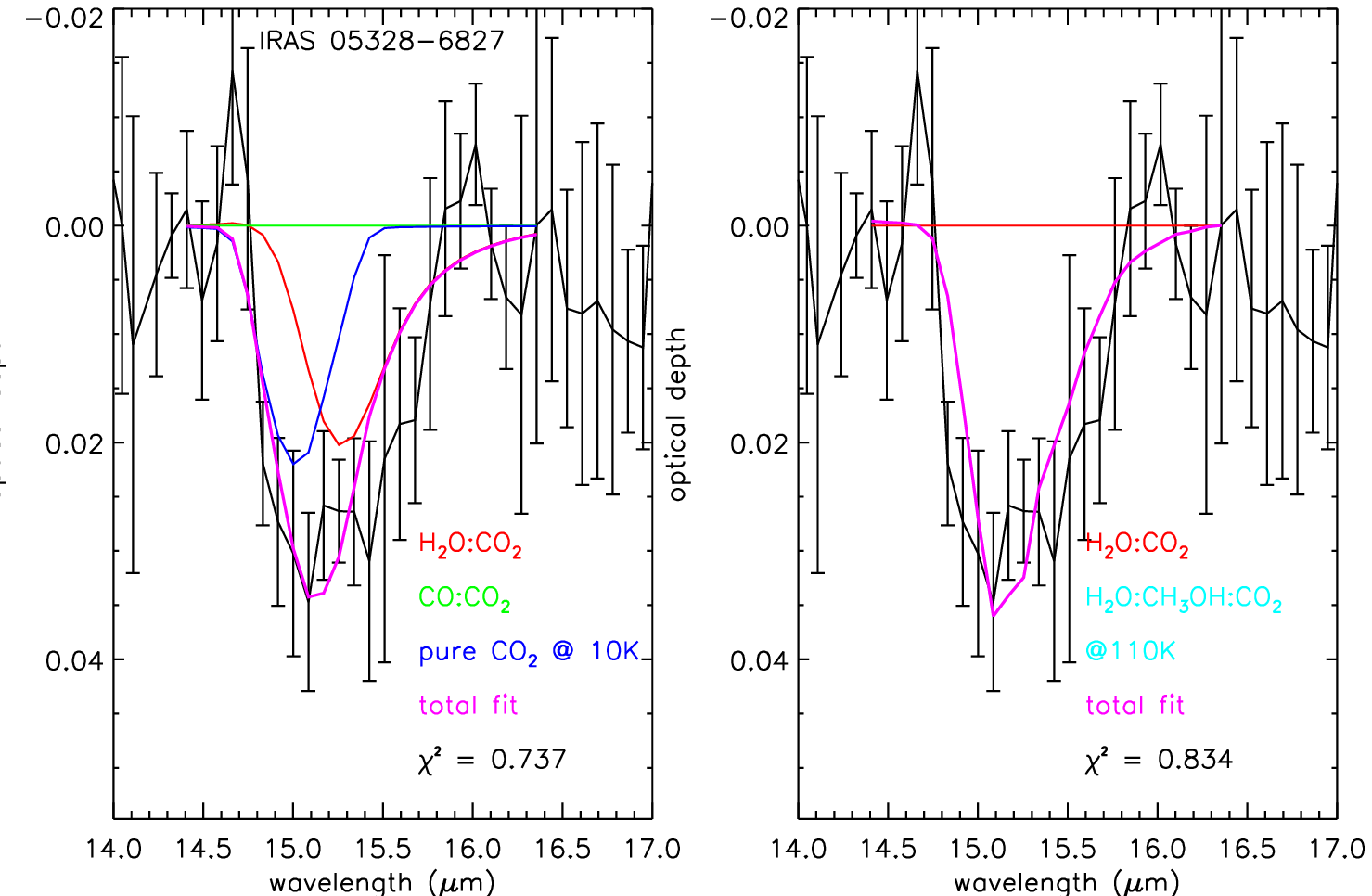}
\includegraphics[scale=0.6]{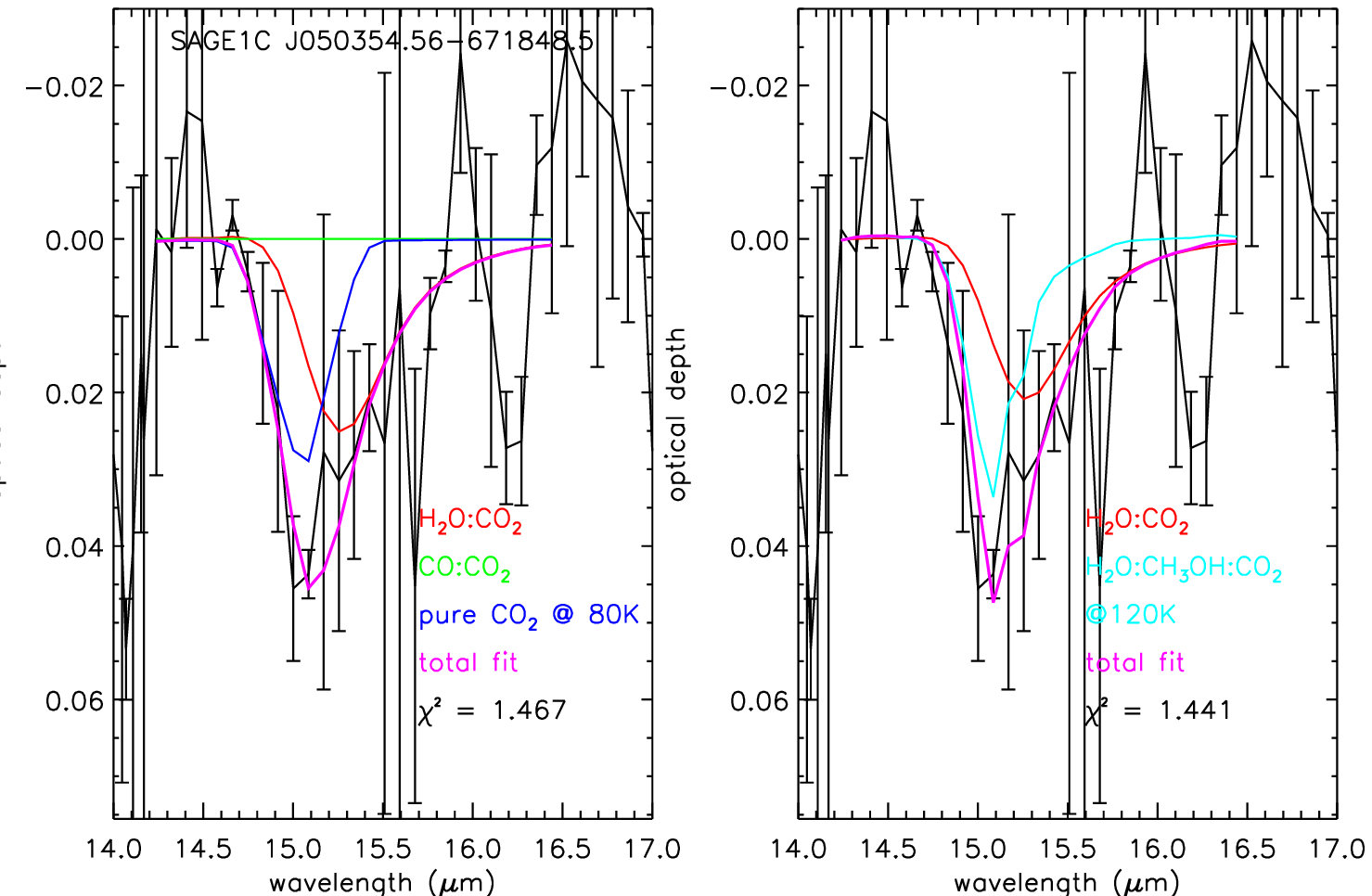}
\caption{Continued.}
\end{center}
\end{figure}

\clearpage
\begin{deluxetable}{l|cr|ccccc|ccccc|c|ll}
\tabletypesize{\tiny}
\rotate
\tablecaption{Column densities and model fit information for the observed
15.2-$\mu$m CO$_2$ features. Uncertainties are statistical and do not include 
those associated with the continuum determination. Fit components are water-rich 
(polar), CO-rich (apolar1), pure-CO$_2$ (apolar2) and annealed ices. For each 
component percentages (in terms of column density) are given. For the 
polar\,+\,apolar ice fits the best-fit CO$_2$:CO ratio is provided; for the 
polar\,+\,annealed fits the best-fit annealed temperature is given. $\chi^2$ 
values for the fits are listed. Additional ice column densities are provided in 
the last column for the 4.3-$\mu$m feature of CO$_2$ and 3-$\mu$m feature of 
water ice, either from the literature \citep{shimonishi08} or recomputed using 
published spectra \citep{shimonishi08,vanloon05a} --- see text and table notes.
\label{fit_data}}
\tablewidth{0pt}
\tablehead{\\[-0.6cm]
&&&\multicolumn{5}{|c|}{polar\,+\,apolar fit}&
\multicolumn{5}{|c|}{polar\,+\,annealed fit}&&\\
\multicolumn{1}{l|}{ID}&\multicolumn{2}{c|}{N(CO$_2$)$^a_{15 \mu m}$}&polar&apolar1&apolar2&
CO$_2$:CO&$\chi^2$&polar&annealed&T&T range&$\chi^2$&\multicolumn{1}{|l|}{best fit}
&N(CO$_2$)$_{4.3 \mu m}$&N(H$_2$O)$_{3 \mu m}$\\
&10$^{17}$\,cm$^{-2}$&$\sigma$&\%&\%&\%&&&\%&\%&K&K&&&\multicolumn{2}{c}{10$^{17}$\,cm$^{-2}$}
}
\startdata
\\[-0.7cm]
           IRAS\,05240$-$6809&\llap{1}2.69\,$\pm$\,0.28&\llap{4}4.8&60\,$\pm$\, 2&20\,$\pm$\, 2&19\,$\pm$\, 2&$\ga$\,1.12      &        1.18&54\,$\pm$\, 1& 46\,$\pm$\, 1&115&110\,$-$\,120&3.35&apolar&\llap{5}5.3$^{+40}_{-30}$&$\sim$\,40$^{b}$\\
          IRAS\,F04532$-$6709&        9.91\,$\pm$\,0.46&\llap{2}1.4&66\,$\pm$\, 2&34\,$\pm$\, 2&\nodata      &$\ga$\,1.12      &        3.26&66\,$\pm$\, 3& 34\,$\pm$\, 3&120&115\,$-$\,130&3.71&apolar\\
SAGE1C\,J052546.49$-$661411.3&        8.83\,$\pm$\,0.18&\llap{4}9.9&68\,$\pm$\, 4& 8\,$\pm$\, 4&24\,$\pm$\, 3&$\ga$\,1.12      &        3.74&12\,$\pm$\, 4& 88\,$\pm$\, 4&110&110\,$-$\,115&3.09&annealed&\llap{1}5.5$^{+13.8}_{-9.4}$&\llap{2}9.7$^{+4.5}_{-4.6}$\\
SAGE1C\,J051347.73$-$693505.1&        8.37\,$\pm$\,0.93&        9.0&59\,$\pm$\,14&18\,$\pm$\,14&23\,$\pm$\,13&$\ga$\,1.12      &        0.64&42\,$\pm$\, 7& 58\,$\pm$\, 7&115&100\,$-$\,130&0.53&\nodata\\
           IRAS\,04514$-$6931&        7.99\,$\pm$\,0.08&\llap{9}8.3&78\,$\pm$\, 3& 9\,$\pm$\, 3&13\,$\pm$\, 2&0.70\,$\pm$\,0.47&\llap{1}0.01&73\,$\pm$\, 1& 27\,$\pm$\, 1&125&110\,$-$\,130&8.21&annealed\\
              MSX\,LMC\,1786 &        7.82\,$\pm$\,0.47&\llap{1}6.6&54\,$\pm$\, 3&27\,$\pm$\, 4&19\,$\pm$\, 3&$\ga$\,1.12      &        3.90&40\,$\pm$\, 2& 60\,$\pm$\, 2&115&110\,$-$\,125&3.28&annealed\\
SAGE1C\,J051449.41$-$671221.5&        6.84\,$\pm$\,0.47&\llap{1}4.7&51\,$\pm$\, 6&44\,$\pm$\, 6& 5\,$\pm$\, 8&$\ga$\,1.12      &        2.08&49\,$\pm$\, 3& 51\,$\pm$\, 3&120&110\,$-$\,130&2.29&\nodata&8.2$^{+4.3}_{-3.7}$&\llap{1}8.7$^{+2.3}_{-2.3}$\\
           IRAS\,05421$-$7116&        4.95\,$\pm$\,0.86&        5.8&69\,$\pm$\, 6&31\,$\pm$\, 6&\nodata      &$\ga$\,1.12      &        2.92&60\,$\pm$\, 8& 40\,$\pm$\, 8&115&110\,$-$\,130&2.74&\nodata\\
SAGE1C\,J045228.65$-$685151.3&        2.65\,$\pm$\,0.44&        6.0& \nodata     &73\,$\pm$\,22&27\,$\pm$\,22&0.70\,$\pm$\,0.57&        3.72&\nodata      &\llap{1}00    &125&115\,$-$\,130&3.61&\nodata\\
           IRAS\,05452$-$6924&        2.36\,$\pm$\,0.42&        5.7&55\,$\pm$\,14&20\,$\pm$\,20&25\,$\pm$\,10&0.10\,$\pm$\,0.29&        0.62&29\,$\pm$\, 9& 71\,$\pm$\, 9&135&135\,$-$\,135&0.58&\nodata\\
           IRAS\,05328$-$6827&        0.90\,$\pm$\,0.15&        5.9&55\,$\pm$\, 8&\nodata      &45\,$\pm$\, 8&\nodata          &        0.74&\nodata      &\llap{1}00    &110&110\,$-$\,110&0.83&\nodata&&3.26$^{+0.27}_{-0.27}$$^{c}$\\
SAGE1C\,J050354.56$-$671848.5&        0.87\,$\pm$\,0.22&        4.0&50\,$\pm$\, 8&\nodata      &50\,$\pm$\, 8&\nodata          &        1.47&46\,$\pm$\,10& 54\,$\pm$\,10&120&110\,$-$\,130&1.44&\nodata\\
\hline
SAGE1C\,J054059.29$-$704402.6&        0.63\,$\pm$\,0.77&0.8&\multicolumn{5}{c|}{\nodata}&\multicolumn{5}{c|}{\nodata}&\nodata\\
           IRAS\,05246$-$7137&        0.46\,$\pm$\,0.18&2.5&\multicolumn{5}{c|}{\nodata}&\multicolumn{5}{c|}{\nodata}&\nodata\\
                MSX\,LMC\,464&        0.45\,$\pm$\,0.17&2.7&\multicolumn{5}{c|}{\nodata}&\multicolumn{5}{c|}{\nodata}&\nodata\\
\enddata
\tablenotetext{a}{assuming 
A\,=\,1.1\,$\times$\,10$^{-17}$\,cm\,molecule$^{-1}$ \citep{gerakines95} and
integrating between 14.4 and 16.8\,$\mu$m.}
\tablenotetext{b}{estimated from Figure\,1 by \citet{shimonishi08}. $^{\rm c}$ 
new estimate using the spectrum published by \citet{vanloon05a}.} 
\end{deluxetable}

\end{document}